\documentclass[longauth]{aa} 
\usepackage{graphicx}
\usepackage{txfonts}
\usepackage{natbib}
\usepackage{amsmath}
\usepackage{xcolor}
\usepackage[colorlinks=true,linkcolor=blue,citecolor=blue,urlcolor=blue]{hyperref}
\usepackage[normalem]{ulem}

\defcitealias{Jolly2026}{Paper~I}
\begin{document}

\newcommand{\noema}{NOEMA$^\text{3D}$}
\newcommand{\ci}{[C\,\textsc{i}](1$-$0)}
\newcommand{\co}{CO(4$-$3)}
\newcommand{\HST}{\textit{HST}}
\newcommand{\JWST}{\textit{JWST}}
\newcommand{\firstgroup}{group-1}
\newcommand{\secondgroup}{group-2}

  \title{NOEMA$^\text{3D}$: Extended CO, [C\,I] and dust in massive star-forming main-sequence galaxies at cosmic noon}

    \author{  Jianhang Chen \inst{1}\thanks{ \email{jhchen@mpe.mpg.de, cjhastro@gmail.com}} 
    \and
              Linda J. Tacconi \inst{1} \and
            Reinhard Genzel \inst{1,2} \and
             Roberto Neri \inst{3} \and
             Karl Schuster \inst{3} \and 
             Natascha~M.~F{\"o}rster~Schreiber \inst{1} \and
             Jean-Baptiste~Jolly \inst{1} \and
             Stavros~Pastras \inst{1,4} \and
             Letizia~Scaloni \inst{5,6} \and
             Giulia~Tozzi \inst{1} \and
              Capucine~Barfety \inst{1} \and
              Alberto~Bolatto \inst{7,8} \and
           Andreas~Burkert\inst{1,9} \and
            Françoise~Combes\inst{10} \and
             Pierre~Cox\inst{11} \and
             Ric~Davies \inst{1}\and
             Frank~Eisenhauer \inst{1,12} \and
             Juan~Manuel~Espejo~Salcedo \inst{1}\and
             Rodrigo~Herrera~Camus\inst{13,14} \and
             Santi~García-Burillo\inst{15} \and
             Tadayuki~Kodama\inst{16} \and 
             Lilian~Lee \inst{1}\and
             Minju~M.~Lee \inst{17,18} \and
             Daizhong~Liu \inst{19} \and
             Dieter~Lutz \inst{1}\and
             Giovanni~Mazzolari \inst{1} \and
             Thorsten~Naab \inst{4} \and
             Amit~Nestor~Shachar \inst{20} \and
             Claudia~Pulsoni \inst{1}\and
             Alvio~Renzini \inst{21} \and
             Monica~Rubio \inst{22} \and
             Taro~T.~Shimizu \inst{1} \and
             Amiel~Sternberg \inst{1,20,23}\and
             Eckhard~Sturm \inst{1} \and
             Hannah~Übler \inst{1} \and
             Antonio~Usero\inst{12} \and
             Stijn~Wuyts\inst{24}  
         }

   \institute{Max-Planck-Institut für extraterrestrische Physik, 85748 Garching, Germany \\
              \email{jhchen@mpe.mpg.de}
              \and Departments of Physics and Astronomy, University of California,
Berkeley, CA 94720, USA
         \and
             Institut de Radioastronomie Millimétrique (IRAM), 300 Rue de la Piscine, 38400 Saint-Martin-d’Hères, France
             \and 
             Max-Planck-Institut für Astrophysik (MPA), Karl-Schwarzschild-Str. 1, D-85748 Garching, Germany 
             \and 
             Department of Physics and Astronomy “Augusto Righi”, University of Bologna, Via Piero Gobetti 93/2, 40129 Bologna, Italy 
             \and
             INAF - Astrophysics and Space Science Observatory of Bologna, Via Piero Gobetti 93/3, 40129 Bologna, Italy
             \and
             Department of Astronomy, University of Maryland, College Park, MD 20742, USA
             \and
            Joint Space-Science Institute, University of Maryland, College Park, MD 20742, USA
            \and
             University Observatory, Ludwig Maximilians University, Scheinerstr. 1, 81679 Munich, Germany
             \and
             Observatoire de Paris, LUX, Collège de France, CNRS, PSL University, Sorbonne University, 75014 Paris, France
             \and
             Sorbonne Université, UPMC Paris 6 and CNRS, UMR 7095, Institut d’Astrophysique de Paris, 98b bd. Arago, 75014 Paris, France
             \and
             Technical University of Munich, TUM School of Natural Sciences, Physics Department, 85747 Garching, Germany
             \and
             Departamento de Astronomía, Universidad de Concepción, Barrio Universitario, Concepción, Chile
             \and
             Millenium Nucleus for Galaxies (MINGAL), Concepción, Chile 
             \and
             Observatorio Astronómico Nacional (OAN-IGN)-Observatorio de Madrid, Alfonso XII, 3, 28014 Madrid, Spain
             \and
             Astronomical Institute, Tohoku University, 6-3 Aramaki, Aoba-ku, Sendai 980-8578, Japan
            \and
                Cosmic Dawn Center (DAWN), Copenhagen, Denmark
            \and
                DTU-Space, Technical University of Denmark, Elektrovej 327, DK2800 Kgs. Lyngby, Denmark
            \and Purple Mountain Observatory, Chinese Academy of Sciences, 10 Yuanhua Road, Nanjing 210023, China
            \and
                School of Physics and Astronomy, Tel Aviv University, Tel Aviv 69978, Israel 
                \and
                Osservatorio Astronomico di Padova, Vicolo dell’Osservatorio 5, Padova I-35122, Italy
            \and
            Departamento de Astronomía, Universidad de Chile, Casilla 36-D, Santiago, Chile
            \and
            Centre for Computational Astrophysics, Flatiron Institute, 162 5th Avenue, New York NY 10010, USA
            \and
            Department of Physics, University of Bath, Claverton Down, Bath, BA2 7AY, UK        
             }

   \date{Received April **, ****; accepted *** **, ****}

  \titlerunning{NOEMA-3D}
  \authorrunning{Jianhang Chen et al.}

 
  \abstract{
We present a spatially resolved study of cold molecular gas and dust in ten main-sequence galaxies at $z=1.1-1.6$, using observations of CO(4$-$3), CO(3$-$2), [C~\textsc{i}](1$-$0), and dust continuum from the \noema{} survey.
We find widespread spatially extended molecular gas and dust, with sizes comparable to those of the stellar disk, in contrast to those of centrally dominated starburst galaxies at similar redshifts.
Among the targeted molecular gas tracers, the CO line ($J=3-2$ or $J=4-3$) remains the most effective for mapping molecular gas distribution and kinematics.
The spatially resolved correlations between different molecular gas tracers exhibit about twice the scatter as their galactic-integrated correlations, indicating that interstellar medium (ISM) conditions already deviate from global averages on scales of 3$-$6~kpc. 
This likely reflects the clumpy or inhomogeneous ISM in cosmic noon star-forming galaxies.
Within our sample, both the molecular gas fraction and its depletion time are nearly constant across the galactic disks out to $2\times R_\text{e}$, supporting a global linear Kennicutt-Schmidt law.
These galaxies also have relatively small bulges, with bulge-to-total ratios between 6--24\%, and are actively forming stars.
These results provide a resolved view of how galaxies can remain on the star‑forming main sequence during their secular evolution at late cosmic noon, an evolutionary stage supported by quasi-steady gas accretion and efficient gas transport via prominent spiral arms and/or bars.
}


   \maketitle
   \nolinenumbers

\section{Introduction}

The epoch around $z\sim1-3$, often dubbed ``cosmic noon'', marks the peak of the cosmic star formation rate density (SFRD) \citep{Madau2014}.
Star-forming galaxies (SFGs) at this epoch follow a tight sequence on the stellar mass versus star formation rate (SFR) plane \citep[the main sequence, or MS;][]{Daddi2007,Elbaz2007,Rodighiero2011,Sargent2012,Speagle2014,Schreiber2015}, which supports a global paradigm that regulates the evolution of SFGs.  
To remain on the MS, galaxies require a continuous replenishment of cold gas, which can either be fresh gas accreted from the cosmic web, gas-rich mergers, or recycled gas from previous galactic feedback (see \citealt{Tacconi2020} for a review).
Consequently, the net availability of cold gas, particularly molecular gas, regulates the rise and fall of star formation within galaxies and is a key ingredient supporting galaxy growth \citep{Tacconi2010,Daddi2010,Tacchella2016,Baker2022}.
Over the past decade, one of the major achievements in understanding the cosmic star formation history has been the measurements of the cosmic evolution of molecular gas mass density \citep{Genzel2015,Scoville2016,Tacconi2018,Liu2019,Tacconi2020,Peroux2020,Walter2020,Riechers2020,Boogaard2023,Bollo2025}.  
Both the SFRD and the molecular gas density peak at \(1 < z < 3\), underscoring the vital role of molecular gas in driving the global, secular evolution of galaxies.
The next step forward is to resolve the spatial distribution, excitation conditions, and kinematics of the cold gas at cosmic noon, connecting the cold gas cycle with galactic structure formation, dynamics, and quenching mechanisms during this globally transitional period \citep[e.g.,][]{ForsterSchreiber2020,Genzel2023}.

However, spatially resolved information about the molecular gas content of SFGs remains incomplete, particularly during the peak epoch of cosmic star formation.  
Various tracers have been used to estimate the total molecular gas mass, including carbon monoxide (CO) in its many rotational transitions and isotopic lines, cold dust emission, and neutral carbon \citep[see reviews by][]{Bolatto2013,Carilli2013,Tacconi2020}.
Despite uncertainties in the individual conversion factors for different tracers, global cross-comparisons indicate that they generally yield consistent results for integrated galactic measurements \citep{Genzel2015, Scoville2016, Tacconi2018, Dunne2022}.  
However, there have been few direct, resolved comparisons of different cold-gas tracers at cosmic noon, which cautions against their use for mapping the spatial distribution of molecular gas \citep[e.g.,][]{Arriagada-Neira2025,Boogaard2026}.

In addition, our current understanding of the molecular gas distribution in SFGs is largely biased toward starburst galaxies at cosmic noon and earlier cosmic epochs. 
The majority of the observed targets are submillimeter galaxies (SMGs) or dusty star-forming galaxies (DSFGs), which mostly lie above the MS \citep[see reviews by][]{Casey2014,Hodge2020}.
Studies of these galaxies have reported compact dust morphology, with sizes often several times smaller than those traced by CO and stellar emission \citep{Hodge2015, Spilker2015, Chen2017, CalistroRivera2018, Boogaard2026}.
Observations of less extreme but still massive SFGs have shown similar trends, albeit with smaller differences between the sizes of molecular lines, dust continuum, and stars \citep{Tadaki2017,Tadaki2023,Kaasinen2020, Pantoni2021}.
The diverse results in the literature highlight global differences in star formation patterns within galaxies. Therefore, systematic, resolved studies of a representative sample of MS galaxies are necessary to better understand the internal evolution of the broader galaxy population. Additionally, differences in apparent sizes across galaxy groups caution against the indiscriminate use of various tracers for mapping molecular gas distributions and emphasize the need for systematic, resolved studies of a more representative sample.

Moreover, as it fuels star formation, the spatially resolved distribution of cold molecular gas directly reflects ongoing stellar buildup and quenching processes.
Large near-infrared integrated field unit (IFU) surveys have revealed that majority of the massive MS galaxies at cosmic noon are rotation-dominated \citep[see review from][]{ForsterSchreiber2020}, a picture further supported by the prominent spiral arms and bars observed in these systems \citep[e.g.,][]{Guo2023,LeConte2024,EspejoSalcedo2025}.  
Together, the current evidence suggests that the growth of these galaxies was likely fueled by smooth gas accretion and minor mergers, rather than disruptive major mergers.  
Importantly, kinematic studies have revealed large-scale, rapid gas inflows in such systems \citep{Genzel2023}, consistent with perturbations from early established stellar structures such as galactic bars and spiral arms \citep[][Pulsoni et al. in prep.]{Ubler2024,Huang2025,Pastras2025}.  
The resolved molecular gas in these galaxies thus completes our picture of how the molecular gas distribution contributes to the secular evolution of galaxies and to their positions relative to the MS.
However, resolving the molecular gas distribution in typical MS galaxies at cosmic noon requires very sensitive observations.
So far, this has only been done for a few cases, benefiting from fortuitous lensing magnification or rare, very deep on-source integration \citep{Liu2023,Arriagada-Neira2025,Liu2025}.

To this aim, we designed and conducted the \noema{} survey to assemble a sizable sample of massive MS SFGs at $z\sim1.1-1.6$, which enabled us to map the cold gas and dust emission on scales of a few kiloparsecs.
\noema{} uses the IRAM Northern Extended Millimeter Array (NOEMA), the successor of the Plateau de Bure Interferometer (PdBI).
NOEMA's wide bandwidth enables it to simultaneously cover multiple spectral lines and achieve ultra-deep sensitivity in the dust continuum.
\citet{Jolly2026} (hereafter Paper~1) investigated the kinematic properties of the sample and highlight the prominent radial gas flows associated with spiral arms and bars.
\citet{Pastras2025,Pastras2026} investigated the radial gas flows in two barred systems and compared them with high-resolution simulations.
In this work, we focus on the spatial distributions of molecular gas and dust and discuss how they could inform our understanding of the evolution of massive MS galaxies at cosmic noon. 

In this paper, we first briefly introduce \noema{} in Sect.\ref{sec:observation_data}, focusing on the observational setup of different cold interstellar medium (ISM) tracers.
This section also includes descriptions of the multiwavelength ancillary data and the integrated and resolved galaxy properties derived from spectral energy distribution (SED) fitting.
Section \ref{sec:analysis} covers the steps of data analysis, including size measurements, radial profiles of different galactic components, bulge-disk decomposition, and molecular gas mass measurements.
We discuss our results in Sect.\ref{sec:results}, covering the size comparison of different components, comparisons with literature results, mass-size evolution, and bulge growth.
We then compare molecular gas tracers CO, [C\,\textsc{i}], and dust continuum in Sect.\ref{sec:molecular_gas}.
Then, in Sect.\ref{sec:implications}, we discuss the implications of our results in the context of massive galaxy formation.
Finally, Sect.\ref{sec:conclusions} summarizes all our key results.
Throughout the paper, we adopt the Chabrier stellar initial mass
function \citep{Chabrier2003} and 
a flat Lambda Cold Dark Matter ($\Lambda$CDM) cosmology with H$_0$ = 70 km\,s$^{-1}$ and $\Omega_\text{m} = 0.3$.

\section{NOEMA$^\text{3D}$ and ancillary data}
\label{sec:observation_data}

\begin{table*}
  \centering
  \caption{Basic properties of the \noema{} targets}
  \label{tab:sample}
  \begin{tabular}{lccccccc}
    \hline\hline

Name & R.A. & Dec. & redshift & $\log(M_\text{mol}/M_\odot)$ & $\log(L_\text{FIR}/L_\odot)$ & $\log(M_\star/M_\odot)$ & SFR \\
 & J2000 & J2000 & & &  &  & M$_\odot$~yr$^{-1}$ \\
\hline
G4\_38065  & 14:20:05.409 & +53:01:15.545 & 1.1151 &  10.82$\pm$0.03 & 11.92 $\pm$ 0.12 & 11.43 $\pm$ 0.02 & 85 $\pm$ 4 \\
GN4\_18574 & 12:37:02.740 & +62:14:01.666 & 1.2463 & 10.86$\pm$0.02 & 12.11 $\pm$ 0.12& 10.89 $\pm$ 0.02& 129 $\pm$ 6 \\
GN4\_24517  & 12:36:21.346 & +62:15:46.001 & 1.2411 & 10.47$\pm$0.06 &  11.63 $\pm$ 0.12& 10.86 $\pm$ 0.02& 43 $\pm$ 2 \\
G4\_20371  & 14:20:25.129 & +53:00:27.576 & 1.1165 & 10.71$\pm$0.03 &  12.09 $\pm$ 0.31& 10.87 $\pm$ 0.05& 128 $\pm$ 15 \\
G4\_23011  & 14:19:11.206 & +52:48:00.354 & 1.1917 & 10.67$\pm$0.04 &  12.07 $\pm$ 0.41& 11.10 $\pm$ 0.04& 121 $\pm$ 6 \\
G4\_38232  & 14:19:48.926 & +52:58:32.027 & 1.1159 & 10.65$\pm$0.08 &  11.60 $\pm$ 0.99& 10.82 $\pm$ 0.11& 36 $\pm$ 15 \\
\hline
GN4\_32842  & 12:37:22.531 & +62:18:38.192 & 1.5233 & 10.80$\pm$0.06&  12.10 $\pm$ 0.12& 11.32 $\pm$ 0.02& 128 $\pm$ 6 \\
G4\_17555  & 14:19:19.684 & +52:48:14.485 & 1.5372 & 10.20$\pm$0.20 &  11.86 $\pm$ 0.78& 10.45 $\pm$ 0.12& 83 $\pm$ 26 \\
G4\_24078  & 14:19:15.996 & +52:49:10.543 & 1.3595 & 10.50$\pm$0.09 &  11.67 $\pm$ 0.69& 10.78 $\pm$ 0.07& 39 $\pm$ 12 \\
G4\_37375  & 14:19:06.731 & +52:50:39.684 & 1.6335 & 10.20$\pm$0.30 &  11.82 $\pm$ 0.77& 10.63 $\pm$ 0.13& 91 $\pm$ 30 \\
\hline
  \end{tabular}
  
  \raggedright\small{\textbf{Notes:} The first six galaxies are referred to as \firstgroup{}, which covers \co{} and \ci{}; the last four galaxies are referred as \secondgroup{}, covering CO(3$-$2). The quoted redshifts are derived from the integrated molecular lines discussed in this work. The molecular gas is derived from \co{} for \firstgroup{} and CO(3$-$2) for \secondgroup{}. In the last three columns, we report the main SED-based global properties, namely, far-infrared (FIR) luminosity, stellar mass, and SFR derived from UV+IR luminosities.}
\end{table*}

\begin{figure*}[htpb]
  \centering
  \includegraphics[width=0.49\textwidth]{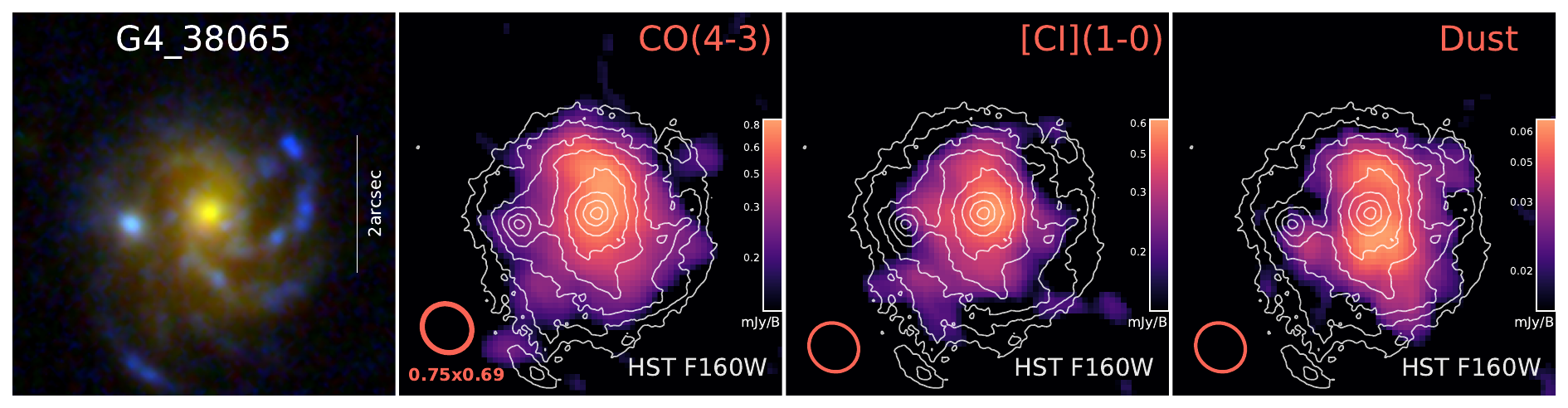}
  \includegraphics[width=0.49\textwidth]{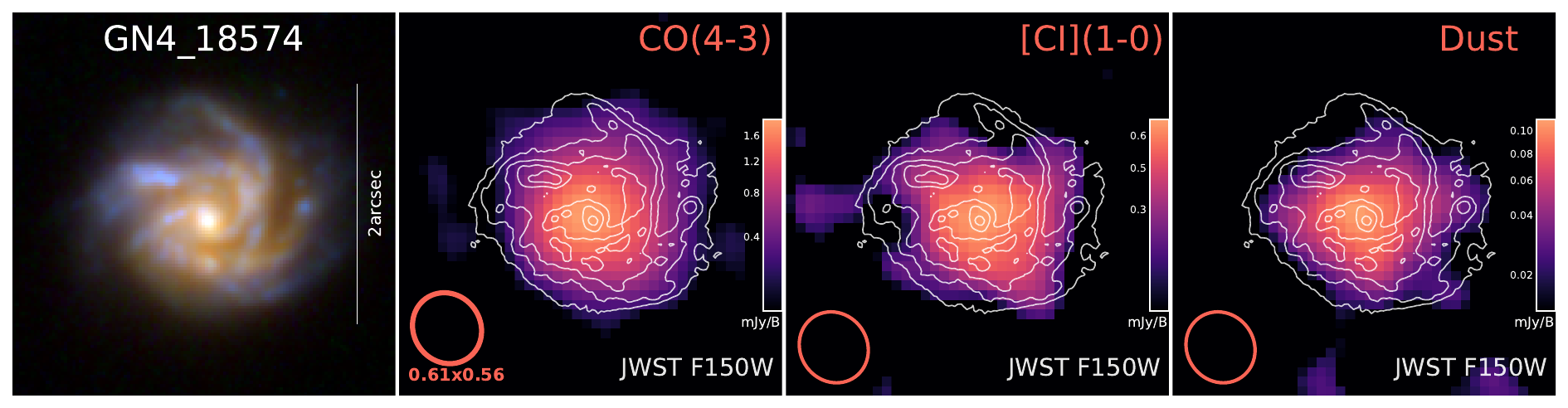}
  \includegraphics[width=0.49\textwidth]{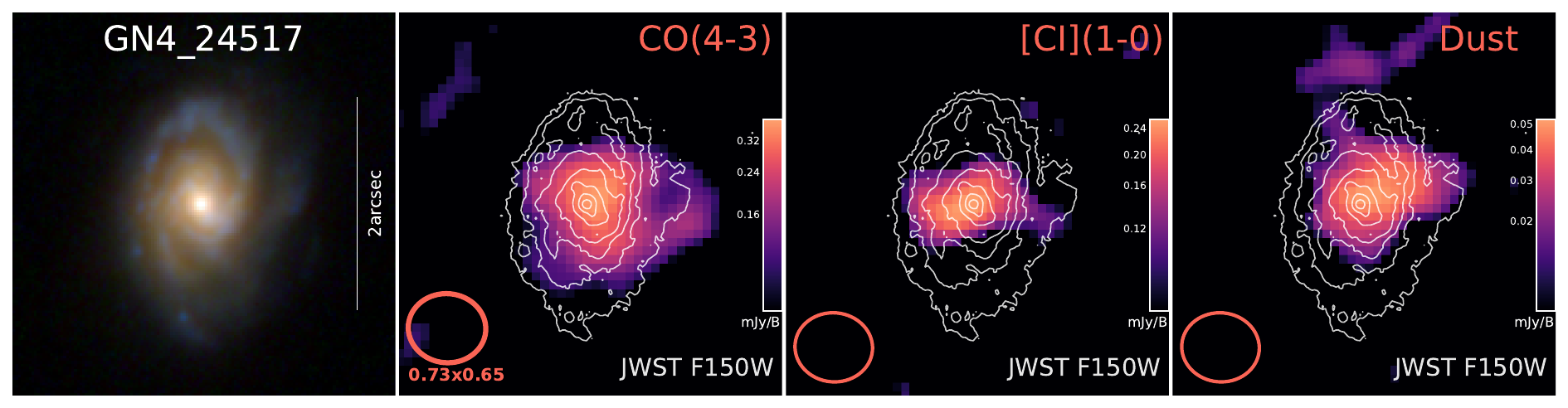}
  \includegraphics[width=0.49\textwidth]{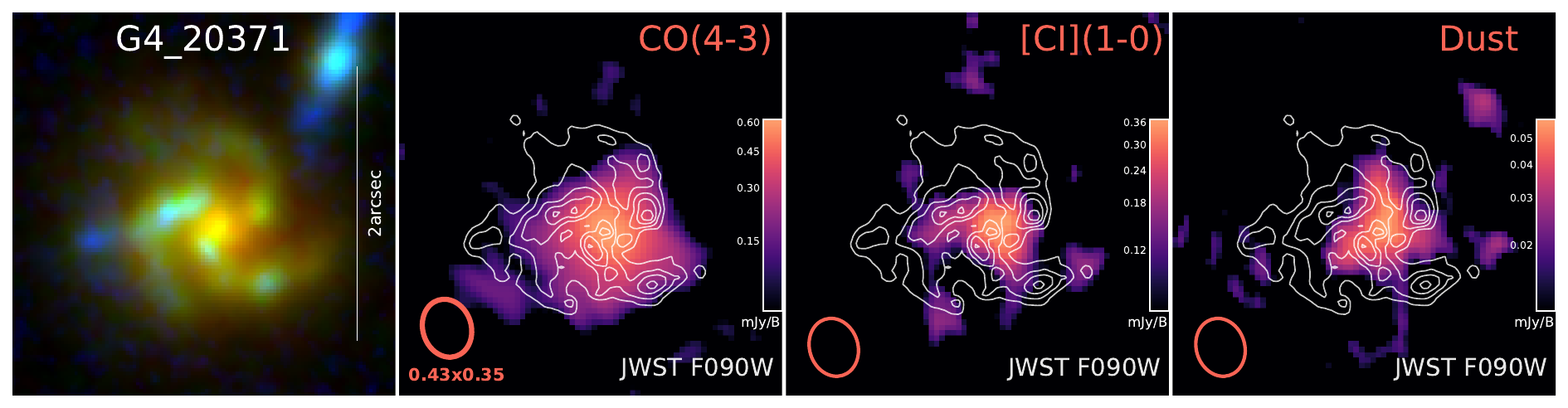}
  \includegraphics[width=0.49\textwidth]{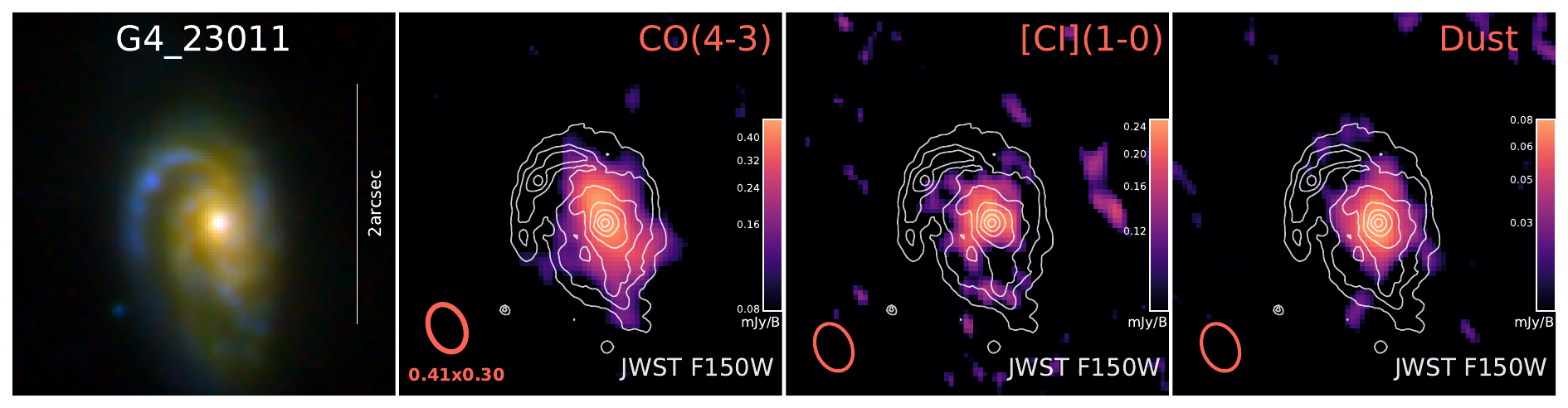}
  \includegraphics[width=0.49\textwidth]{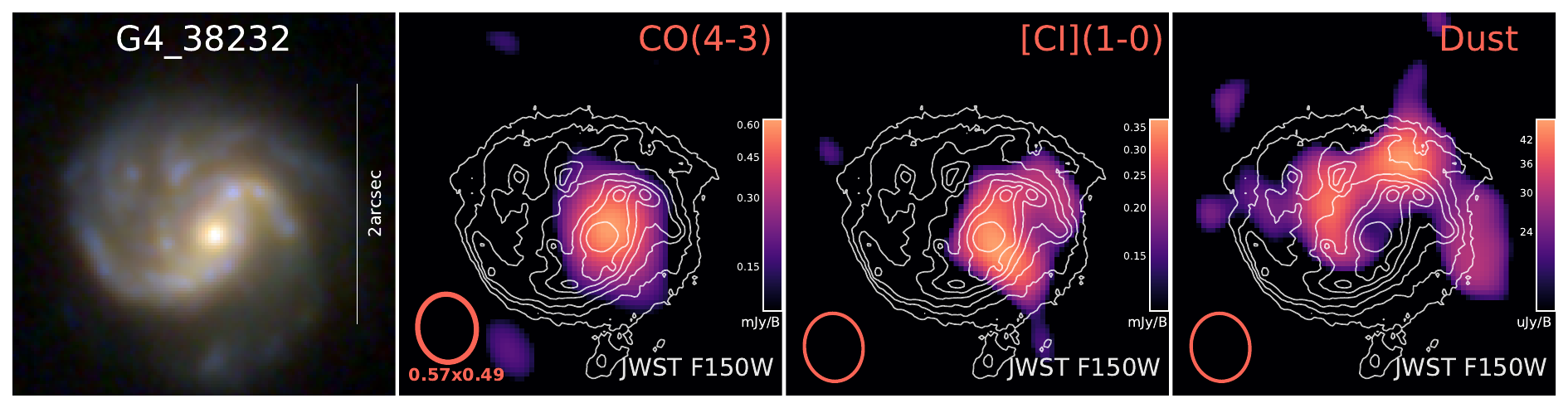}
  \caption{Spatially resolved maps of \co{}, \ci{}, and dust continuum for the six galaxies from \firstgroup{}. \textit{First column}: Color image constructed from the available broadband \HST{}/\JWST{} data for each target. \textit{Columns 2-4}: Intensity maps of \co{}, \ci{}, and dust continuum, with the color map starting at 2$\sigma$. In these maps, the white contours (starting at 5$\sigma$) represent starlight from the rest-frame optical broadband image, with the filter name indicated in the bottom-right corner. The red ellipse in the bottom-left corner illustrates the resolution beam for each image. Generally, the molecular gas tracers and the dust continuum, when well detected, exhibit similar spatial extents after accounting for S/N.}
  \label{fig:multi_tracers_co43}
\end{figure*}

\begin{figure*}[htpb]
  \centering
  \includegraphics[width=0.4\textwidth]{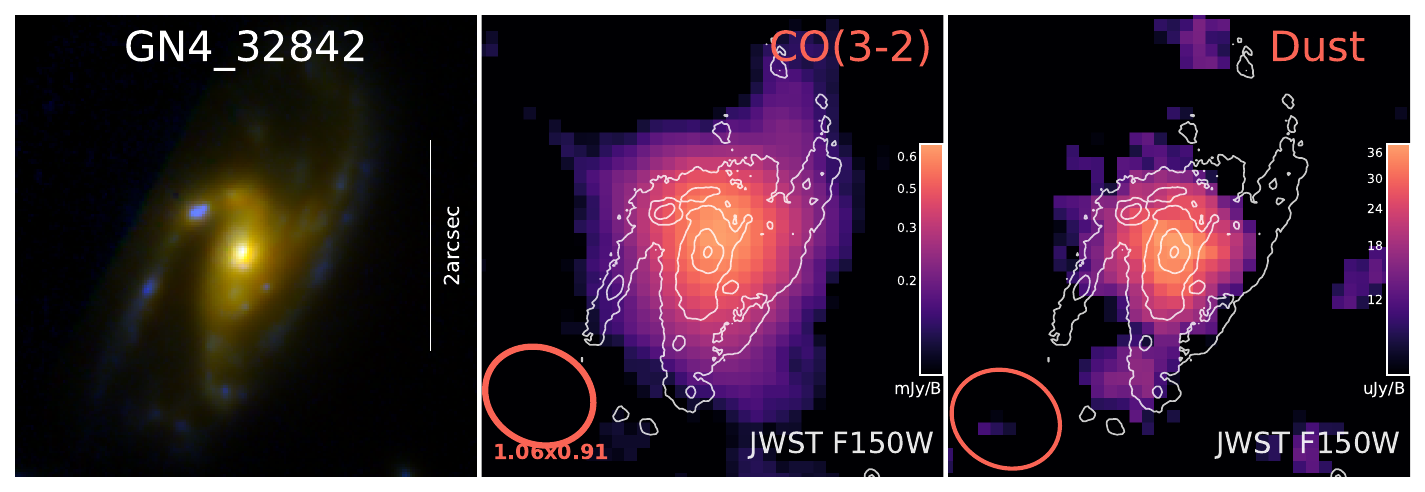}
  \includegraphics[width=0.4\textwidth]{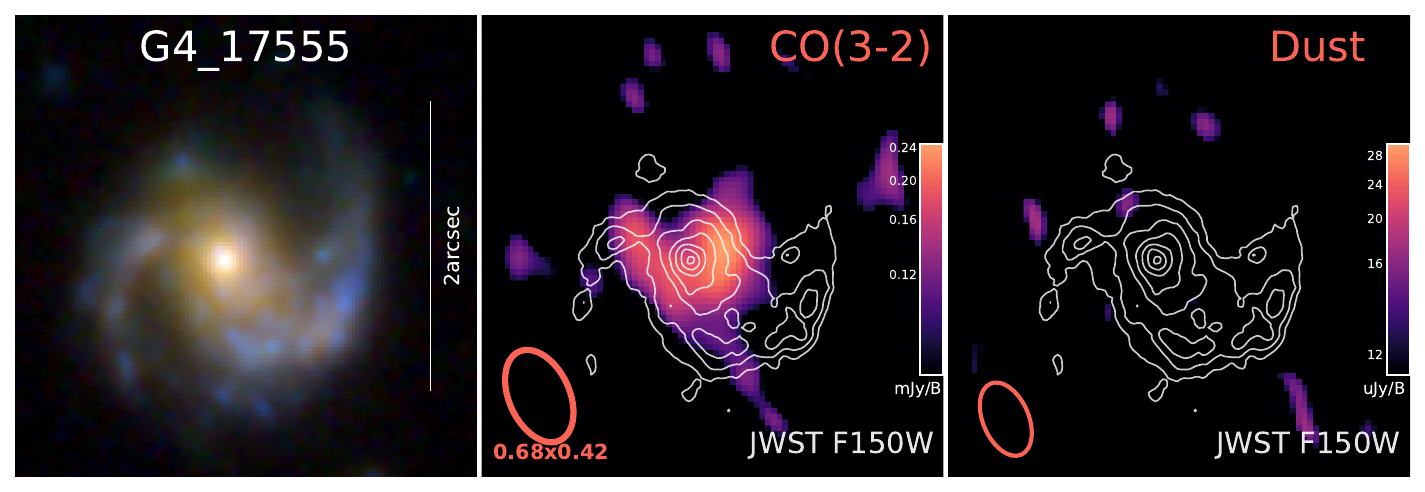}
  \includegraphics[width=0.4\textwidth]{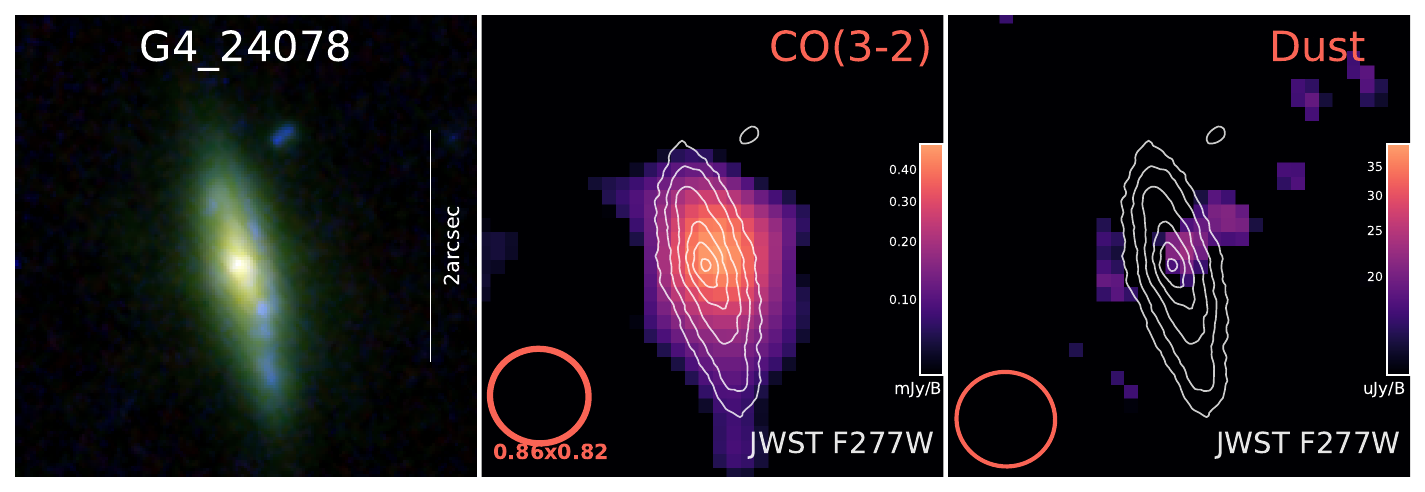}
  \includegraphics[width=0.4\textwidth]{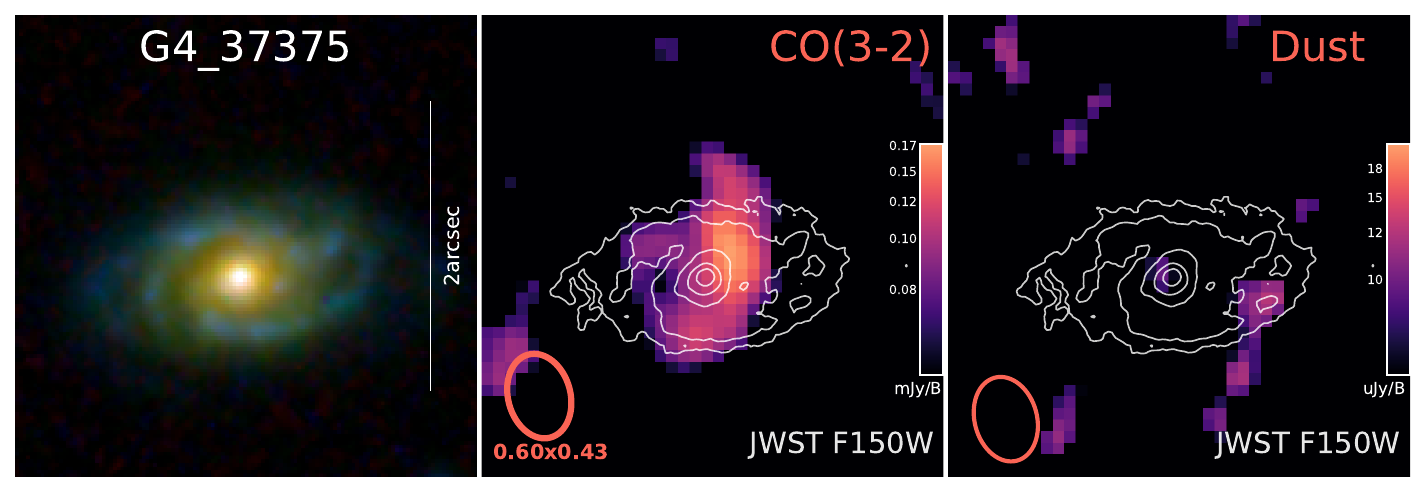}
  \caption{Spatial resolved maps of CO(3$-$2) and dust continuum for the galaxies from \secondgroup{} (same presentation as in Fig.~\ref{fig:multi_tracers_co43}). The dust emission in these targets is generally fainter than that observed in \firstgroup{}, primarily due to the lower rest-frame wavelengths involved. Among all four targets, only GN4-32842 shows a secure detection of the dust continuum.}
  \label{fig:multi_tracers_co32}
\end{figure*}

\subsection{NOEMA3D}
\label{subsec:noema}

\noema{} is the first systematic deep survey of cold molecular gas in MS galaxies around cosmic noon.
It targets ten massive MS galaxies between $z=1.1-1.6$, pushing the spatial resolution down to sub-arc-second scales (0.3$-$0.7$"$, or 3$-$6\,kpc in physical resolution).
It features ultra-deep integration and kiloparsec-scale spatial resolution, with 20$-$70 hours of on-source time, offering a unique dataset for probing the spatial distribution of cold molecular gas and its kinematics.
The global properties of these galaxies have been summarized in Table~\ref {tab:sample}.
We refer the reader to \citetalias{Jolly2026} for a full description of the observation design and data reduction; here, we summarize only the details most relevant to this paper.

The \noema{} targets were selected primarily based on their stellar mass, position on MS, and size.
Specifically, we selected massive galaxies with stellar mass log($M_\star$/M$_\odot$) $>$ 10.4, within $\Delta\text{MS} < \pm0.6$~dex, and $R_\text{e} > 3$\,kpc.
These requirements ensure that we achieve a good signal-to-noise ratio (S/N), map the extended molecular line emission, and resolve their kinematics with $20\sim70$\,hours of NOEMA on-source integration.
The \noema{} targets were observed using a combination of the A and C configurations of NOEMA.
The observations were conducted between 2019 and 2024 under optimal weather conditions in Band 2 or Band 3.
For targets around $z\sim1.2$, the receivers were configured to cover ${}^{12}$CO~$J=4-3$, hereafter \co{}, and [C\,\textsc{i}](${}^3P_1-{}^3P_0$), hereafter \ci{}, simultaneously.
For targets at $z\sim1.5$, the receivers were configured to cover ${}^{12}$CO~$J=3-2$, hereafter CO(3$-$2).
For simplicity, we refer to the $z\sim1.2$ targets covering \co{} as group-1, and the $z\sim1.5$ targets covering CO(3$-$2) as group-2.
In addition to the spectral lines, we also utilized the full available bandwidth of 15.5\,GHz to cover the continuum emission around 1.4\,mm (rest-frame 650~$\mu$m) for \firstgroup{} and 2\,mm (rest-frame 800~$\mu$m) for \secondgroup{}.

For each target, the observations were calibrated individually and then combined for imaging.  
All data were calibrated using the official \textsc{clic} pipeline, with additional flagging applied after visual inspection. The combined data were then imaged with \textsc{mapping}.  
Both the \textsc{clic} pipeline and \textsc{mapping} are available as part of the GILDAS package\footnote{https://www.iram.fr/IRAMFR/GILDAS/}.
In addition to the full data cube products as described in \citetalias{Jolly2026}, we also produced line intensity images for this work.
We first extracted the spectrum from the data cube to identify the line-emitting channels.
The image of each emission line was then created by collapsing all line-emission channels in the $uv$ plane and cleaned with natural weighting to achieve the best available sensitivity and recovery of extended emission.
The continuum image was created by collapsing all the line-free channels from both sub-bands.
In the final image products, the CO image shows a slightly larger beam size than \ci{} and the continuum, due to its lower frequency.
To facilitate spatial comparison and line ratios, we recreated the images of \ci{} and continuum using the same CLEAN beam as the CO images.
All images are shown in Fig.~\ref{fig:multi_tracers_co43} and \ref{fig:multi_tracers_co32} for \firstgroup{} and \secondgroup{} targets, respectively.

\subsection{Multiwavelength ancillary data}

\subsubsection{HST and JWST photometry}
\label{subsubsec:hst_jwst_photometry}

We collected broadband images from various public \textit{HST} and \textit{JWST} surveys (see \citetalias{Jolly2026} for a complete list of program IDs).
For \textit{JWST} images, we first retrieved high-quality reduced data from the public DAWN \textit{JWST} Archive (DJA\footnote{https://dawn-cph.github.io/dja/}) whenever available. 
For public data but not yet available on DJA, we downloaded the raw data from the MAST\footnote{https://mast.stsci.edu/search/} database and performed our own reduction using a customized \textit{JWST} pipeline \footnote{https://github.com/1054/Crab.Toolkit.JWST} (see also \citealt{EspejoSalcedo2025}).
The \textit{HST} data include both ACS and WFC3/IR observations, covering wavelengths from the F435W to F160W bands (rest-frame in $\approx 0.17$$-$0.7~$\mu$m). The \textit{JWST} data consist of multiband NIRCam images, spanning the F070W--F444W range (rest-frame in $\approx 0.3$$-$2~$\mu$m).

We created color-composite images from broadband images from \textit{JWST}/NIRCam and/or \textit{HST}/ACS-WFC3.
The RGB image is created by \textsc{trilogy} \citep{Coe2012}, assigning blue filters ($<1~\mu$m) to the blue channel, red filters ($>3~\mu$m) to the red channel, and intermediate filters to the green channel.
Since the filter coverage varies from target to target, the exact combinations differ accordingly. For the only target without \textit{JWST} coverage, G4\_38065, the color image was produced using HST filters alone, assigning the reddest filters ($>1~\mu$m) to red, the bluest filters ($<0.5~\mu$m) to blue, and the remaining ones to green. The resulting color images are shown in Figs.~\ref{fig:multi_tracers_co43} and ~\ref{fig:multi_tracers_co32}.

\subsubsection{Host galaxy properties}
\label{subsec:sed}
Table~\ref{tab:sample} lists the integrated properties of our sample. 
Here, we provide a brief summary of the photometric datasets and the main assumptions used to perform the SED fitting.

We used all publicly available photometry, with slightly different wavelength coverage depending on the data availability for each target \citepalias[see Table A.1 in][]{Jolly2026}. In particular, we extracted integrated photometry from HST/ACS and WFC3/IR, and JWST/NIRCam, complemented by mid- and far-IR data from Spitzer/IRAC (3.6$-$8~$\mu$m) and MIPS 24~$\mu$m \citep{Dickinson2003,Whitaker2014}, and from Herschel/PACS (70$-$160~$\mu$m; \citealt{Lutz2011}). In addition, we included integrated dust continuum measurements from our NOEMA observations in Band~3 (1.4~mm) or Band~2 (2~mm) (presented in Sect.~\ref{subsec:noema}).

Based on the multiwavelength data, we performed integrated SED fitting with \textsc{CIGALE} \citep{Boquien2019} to derive the main global physical properties of the galaxies, including stellar mass ($M_\star$), SFR, and FIR luminosity.
For each galaxy, we fixed the redshift to the known spectroscopic value and assumed a constant star formation history (SFH, parameterized by an exponentially declining SFH with an e-folding time $\tau = 8000$~Myr).
We adopted the stellar population synthesis models of \citet{Bruzual2003}, assumed a \citet{Chabrier2003} initial mass function, fixed the metallicity to the Solar value ($Z=0.02$), and included the nebular emission module. Dust attenuation was modeled following \citet{Calzetti2000}, while dust emission was reproduced using the templates of \citet{Dale2014}.
The stellar masses and FIR luminosity were taken directly from the CIGALE outputs, while the total SFRs were computed from the combination of UV and IR luminosity of the SED best-fit models, following the empirical relations of \citet{Kennicutt1998}.

In addition to the integrated quantities described above, we also used spatially resolved maps of stellar mass, SFR, and dust attenuation (A$_\text{V}$) obtained from a spatially resolved SED fitting of HST/ACS and WFC3/IR, and JWST/NIRCam imaging. The pixel-by-pixel SED fitting was performed under the same assumptions adopted for the integrated SED analysis, after convolving all the available HST and JWST images to the spatial resolution of the JWST/NIRCam F444W band, or to that of the HST/WFC3 F160W band when JWST data were not available. A detailed spatially resolved SED-fitting analysis will be presented in Tozzi et al.~(in prep.).

\section{Analysis}
\label{sec:analysis}

In this section, we present our methods for deriving all resolved quantities. It includes size measurements, radial profiles, bulge-disk decomposition, and molecular gas mass.

\subsection{Size measurements}

\begin{table*}
  \centering
  \caption{Size and flux measurements of \noema{} targets}
  \label{tab:size}
  \begin{tabular}{lccccccccc}
    \hline\hline

Name & ${R_{\rm e,star}}^\text{a}$ & ${R_{\rm e,CO}}^\text{b}$ & ${R_{\rm e,CI}}^\text{b}$ & ${R_{\rm e,dust}}^\text{b}$ & ${R_{\rm e,500nm}}^\text{a}$ & ${R_{\rm 80,500nm}}^\text{a}$ & ${S_{\rm CO}}\Delta \varv\ ^\text{c}$ & ${S_{\rm CI}\Delta \varv\ }^\text{c}$ & ${S_{\rm dust}}^\text{c}$ \\
     & kpc & kpc & kpc & kpc & kpc & kpc & Jy km s$^{-1}$ & Jy km s$^{-1}$ & mJy \\
\hline
G4\_38065 & $8.3\pm1.3$ & $8.7\pm0.7$ & $8.0\pm1.4$ & $9.2\pm2.4$ & $9.1\pm1.3$ & $14.5\pm1.3$& 1.44 $\pm$ 0.11& 0.86 $\pm$ 0.09& 0.41 $\pm$ 0.05\\
GN4\_18574 & $3.7\pm0.1$ & $3.5\pm0.2$ & $4.6\pm0.8$ & $2.6\pm0.5$ & $5.1\pm0.1$ & $10.3\pm0.1$& 1.27 $\pm$ 0.05& 0.58 $\pm$ 0.08& 0.30 $\pm$ 0.03\\
GN4\_24517 & $3.7\pm0.1$ & $2.6\pm0.8$ & $1.7\pm0.9$ & $2.5\pm1.4$ & $5.4\pm0.1$ & $9.6\pm0.1$& 0.52 $\pm$ 0.06& 0.22 $\pm$ 0.06& 0.11 $\pm$ 0.02\\
G4\_20371 & $4.5\pm0.1$ & $2.6\pm0.2$ & $3.5\pm0.9$ & $3.3\pm1.0$ & $7.6\pm0.1$ & $16.4\pm0.1$& 1.12 $\pm$ 0.09& 0.47 $\pm$ 0.09& 0.16 $\pm$ 0.03\\
G4\_23011 & $3.9\pm0.1$ & $5.2\pm1.5$ & $3.2\pm1.0$ & $4.1\pm1.0$ & $8.9\pm0.1$ & $20.5\pm0.1$& 1.27 $\pm$ 0.14& 0.83 $\pm$ 0.17& 0.24 $\pm$ 0.04\\
G4\_38232 & $4.4\pm0.3$ & $1.7\pm0.4$ & $3.0\pm1.0$ & - & $6.1\pm0.3$ & $6.2\pm0.3$& 0.75 $\pm$ 0.14& 0.40 $\pm$ 0.15& 0.24 $\pm$ 0.05\\
\hline
GN4\_32842 & $5.1\pm0.1$ & $6.0\pm0.8$ & - & $8.5\pm2.7$ & $7.8\pm0.1$ & $16.3\pm0.1$& 1.01 $\pm$ 0.05& - & 0.07 $\pm$ 0.01\\
G4\_17555 & $3.8\pm0.1$ & $5.8\pm1.3$ & - & - & $6.1\pm0.1$ & $9.2\pm0.1$& 0.25 $\pm$ 0.04& - & 0.05 $\pm$ 0.02\\
G4\_24078 & $4.7\pm0.2$ & $5.7\pm1.3$ & - & - & $5.7\pm0.2$ & $8.8\pm0.2$& 0.64 $\pm$ 0.10& - & 0.04 $\pm$ 0.03\\
G4\_37375 & $3.2\pm0.1$ & $3.5\pm1.2$ & - & - & $4.6\pm0.1$ & $8.3\pm0.1$& 0.22 $\pm$ 0.08& - & 0.05 $\pm$ 0.02\\

\hline
  \end{tabular}
\raggedright\small{\textbf{Notes:} \\
$^{a}$ The sizes are derived from the curve-of-growth method. \\
$^{b}$ The sizes are derived from the visibility modeling. \\
$^{c}$ The integrated flux.}
  
\end{table*}

\subsubsection{Stars}

To measure the intrinsic size of the stellar mass distribution, we must account for various observational effects, including the PSF smearing, dust attenuation, and the mass-to-light (M/L) ratio.
JWST offers deep images with spatial resolution better than 0.15\,arcsec.
It is significantly smaller than the effective radius of our targets, typically larger than 0.5\,arcsec, which minimizes the effect of PSF.
Meanwhile, \textit{JWST} also probes the rest-frame near-IR emission of these galaxies, which is less affected by dust attenuation.
To further reduce the bias introduced by the M/L ratio, we also used the stellar mass maps derived from the spatially resolved SED fitting for size measurements (see Sect.\ref{subsec:sed}).

Given the relatively rich substructures in the stellar images, we adopted the nonparametric curve-of-growth method to measure the effective radius.
We applied the curve-of-growth method directly to the broadband images and stellar mass maps to derive the stellar sizes.
We first defined the extraction aperture based on the galaxy's inclination and position angle, as derived by \citetalias{Jolly2026} (see also Table~\ref {tab:second_properties}).
Then, we gradually increased the aperture size to derive the curve of growth. 
The radius that includes half the total mass was later defined as the effective radius along the major axis ($R_\text{e}$).
One example of this method is shown in Fig.~\ref{appendixfig:size_measurements}. 
We find that the measured size depends on the observing wavelength, with a smaller size observed at longer wavelengths.  
For \noema{} galaxies, the mass-based size is closest to the reddest available filter, F444W (F160W for G4\_38065).
To compare with the literature results, we also derived the effective radius $R_\text{e,500nm}$ and the size including 80\% stellar emission $R_\text{80,500nm}$ at the rest-frame optical wavelength ($\sim5000$\,\AA). Across the paper, we quote the effective radius from the stellar mass as the reference size $R_\text{e,star}$.
All the derived effective sizes are summarized in Table~\ref{tab:size}.

\subsubsection{Molecular lines and dust continuum}

We used a different strategy to measure the sizes of the molecular gas and dust continuum with \noema{}.
First, the spatial resolution of NOEMA is worse than that of \textit{HST} and \textit{JWST}, making the PSF a dominant factor in size measurements.
Because of this, we used a parametric forward model to correct for PSF smearing.
To approximate the intrinsic light distribution, we used a general 2D S\'ersic profile, which is sufficiently flexible to model our data.
Second, interferometric observations sample the Fourier space of projected sky images and use nonlinear CLEAN algorithms to recover sky objects from the dirty images. The imaging process leaves correlated noise in the final image, and its gridding and weighting schemes result in the loss of part of the information contained in the $uv$ data.
There have been many discussions about the advantages of modeling in the $uv$ space, which benefits from improved noise behavior and maximal information recovery \citep[e.g.,][]{Marti-Vidal2014,Spilker2016,Tazzari2018,Kurtovic2024}.
Therefore, we performed the 2D S\'ersic fitting directly in the $uv$ plane to better recover the intrinsic spatial extent of the cold gas and dust.
We first modeled the data from the CLEANed image to obtain the initial model parameters.
Then, we constructed the model and transformed it into visibility, from where we compare with observed visibility to perform $\chi^2$ minimization.
Depending on the integrated S/N of the data, we occasionally had to fix some parameters, such as disk ellipticity and S\'ersic index, to achieve convergence in the fitting.
The best-fit models and final residuals are shown in Figs.~\ref{appendixfig:uvfitting_co43}--\ref{appendixfig:uvfitting_co32}.
All best-fit parameters are summarized in Table~\ref{appendixtab:uvmodelling}, while the derived major-axis effective radius in kiloparsecs is summarized in Table~\ref{tab:size}.

For the final size comparison, we adopted the CO size measurement as the reference for the molecular gas size in \firstgroup{}.
This is because 1) we generally find comparable size measurements between \co{} and \ci{} under the assumption of an intrinsic S\'ersic profile, with their differences lying within the 1~$\sigma$ error bars (see also Table~\ref{tab:size});
moreover, 2) \co{} is brighter than [C\,\textsc{i}] (see their integrated values and errors in Table~\ref{tab:sample}), providing more robust measurements without requiring additional fixed parameters.
However, future follow-up observations covering additional CO and [C\,\textsc{i}] transitions are required to correct their excitation conditions and optical depth.

Dust size measurements are known to be influenced by optical depth and dust temperature distributions.
The unprecedented continuum sensitivity of \noema{} allows us to probe the fully resolved dust emission at the $\sim$1.4\,mm (\firstgroup{}, rest-frame $\sim650\,\mu$m) and 2\,mm (\secondgroup, rest-frame $\sim800\,\mu$m).
At these wavelengths, dust emission is largely optically thin and proportional to the distribution of the total dust mass \citep{Scoville2016}, which is also less sensitive to the radial temperature variations.
Due to the unknown dust temperature gradients, we still used the constant-temperature assumption, which likely underestimates the extent of the intrinsic dust size \citep[see also][]{Cochrane2019,Mitsuhashi2024}.
However, the typical error introduced is small, within 0.2~dex (see more detailed discussion in Appendix~\ref{appendix:dust_temperature}).

\begin{table*}
  \centering
  \caption{Derived properties of the \noema{} galaxies}
  \label{tab:second_properties}
  \begin{tabular}{lcccccccc}
    \hline\hline
Name & inc$^\text{a}$ & PA$^\text{a}$ & $n_\text{bulge}$$^\text{b}$ & $R_\text{e,bulge}$$^\text{b}$ & $n_\text{disk}$$^\text{b}$ & $R_\text{e,disk}$$^\text{b}$ & $\log(\Sigma_\text{1kpc}/(\text{M}_\odot\,\text{kpc}^{-2}))$ $^\text{c}$ & B/T$^\text{d}$ \\ 
  & deg & deg & & kpc & & kpc & & \\ 
\hline
G4\_38065 & 18 & 12 & 1.4$\pm$0.2 & 1.0$\pm$0.1 & 1.01$\pm$0.10 & 11.0$\pm$0.2 & 9.76 & 0.09 \\
GN4\_18574 & 12& -40 & 1 & 0.3$\pm$0.2& 0.59$\pm$0.05 & 3.66$\pm$0.09 & 9.85 & 0.06 \\
GN4\_24517 & 41& 154 & 2 & 0.69$\pm$0.06 & 1 & 4.23$\pm$0.05 & 9.86 & 0.23 \\
G4\_20371 & 32 & -148 & 1 & 0.8$\pm$0.2 & 0.60$\pm$0.10 & 4.46$\pm$0.24 & 9.74 & 0.20 \\
G4\_23011 & 51 & 190 & 1 & 0.35$\pm$0.04 & 1.12$\pm$0.04 & 4.00$\pm$0.05 & 10.30 & 0.10 \\
G4\_38232 & 23 & -63 & 2 & 0.31$\pm$0.01& 1 & 5.01$\pm$0.06 & 9.77 & 0.15 \\
\hline
GN4\_32842 & 20& -66& 2 & 0.60$\pm$0.08 & 0.9$\pm$0.1 & 6.24$\pm$0.09 & 10.10 & 0.16 \\
G4\_17555 & 25 & 4 & 2 & 0.73$\pm$0.08 & 1 & 4.76$\pm$0.13 & 9.66 & 0.24 \\
G4\_24078 & 76 & 14 & 1 & 0.30$\pm$0.03& 1.49$\pm$0.02 & 3.08$\pm$0.02 & 9.94 & 0.07 \\
G4\_37375 & 50 & 100& 2.8$\pm$0.4 & 0.49$\pm$0.03& 1 & 3.77$\pm$0.02 & 9.91 & 0.20 \\

\hline
  \end{tabular}
  
\raggedright\small{\textbf{Notes:}\\
$^\text{a}$ The inclination (inc) and position angle (PA) of the disk based on kinematic and stellar morphology \citepalias{Jolly2026}\\
$^\text{b}$ S\'ersic index and effective radius of the bulge and disk\\
$^\text{c}$ The stellar density in the central
1\,kpc region\\
$^\text{d}$ The stellar bulge to total ratio from bulge-disk decomposition of the stellar image
}
  
\end{table*}

\begin{figure*}[htpb]
  \includegraphics[width=0.24\textwidth]{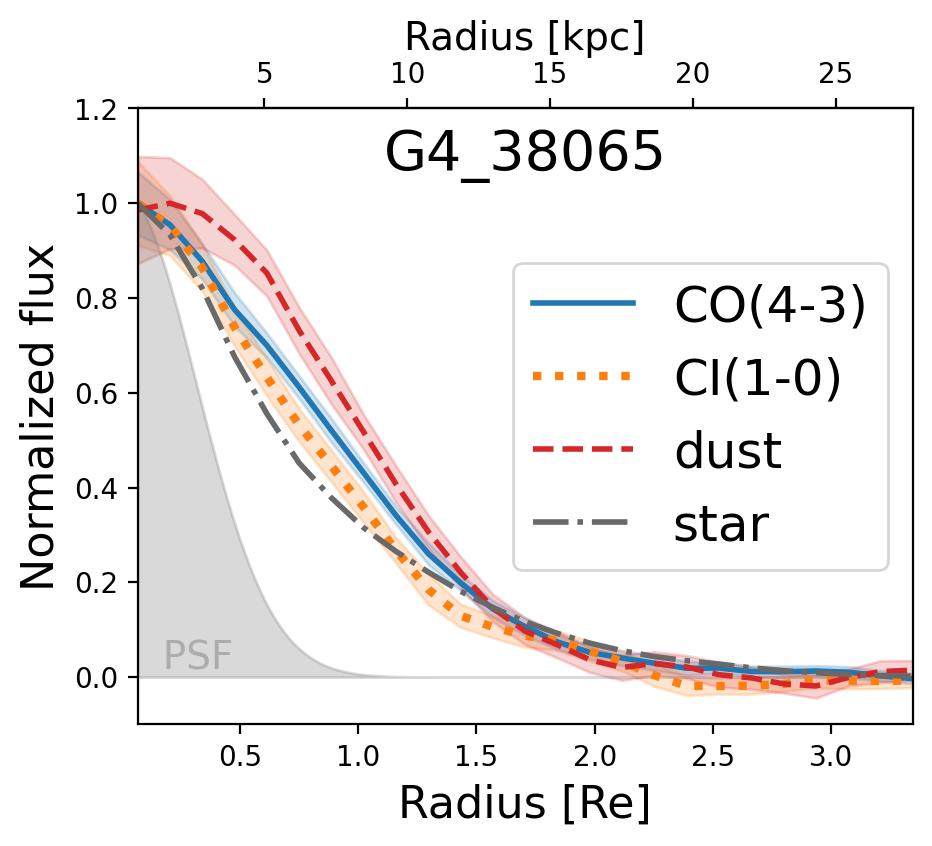}
  \includegraphics[width=0.24\textwidth]{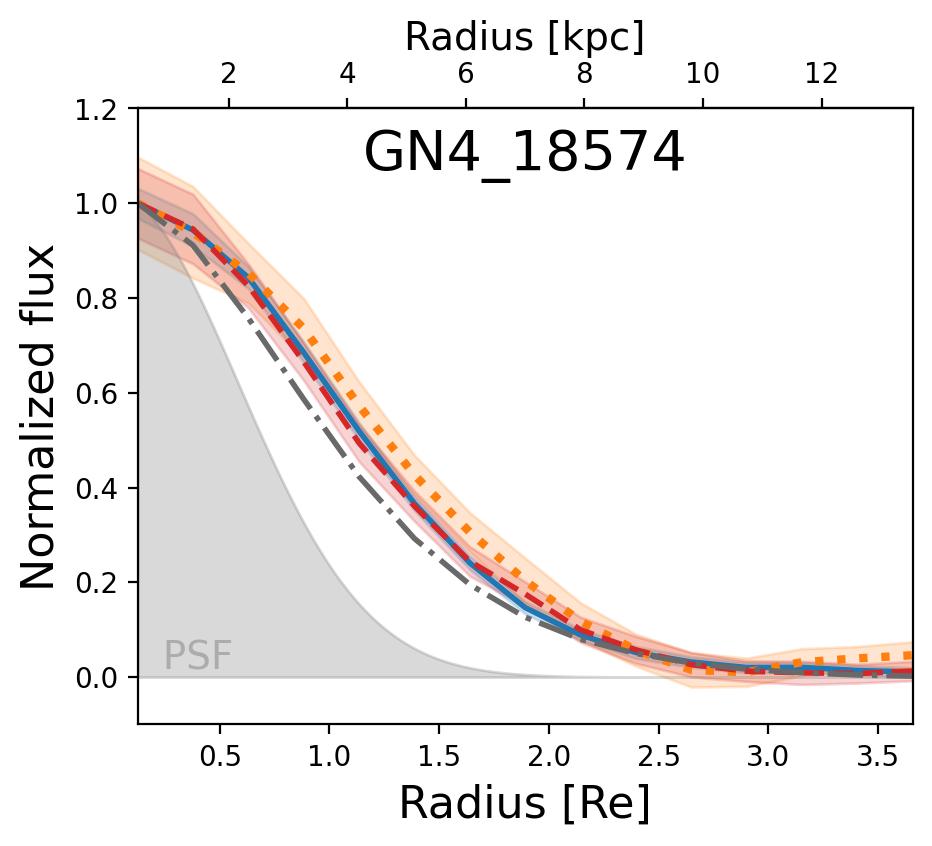}
  \includegraphics[width=0.24\textwidth]{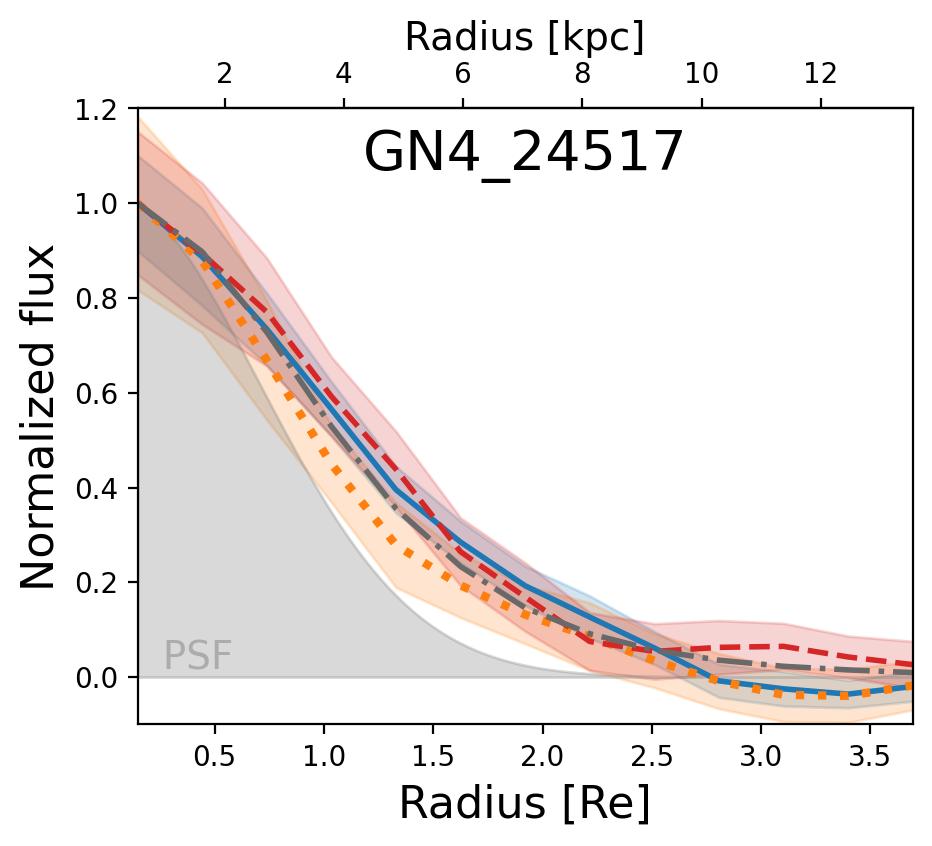}
  \includegraphics[width=0.24\textwidth]{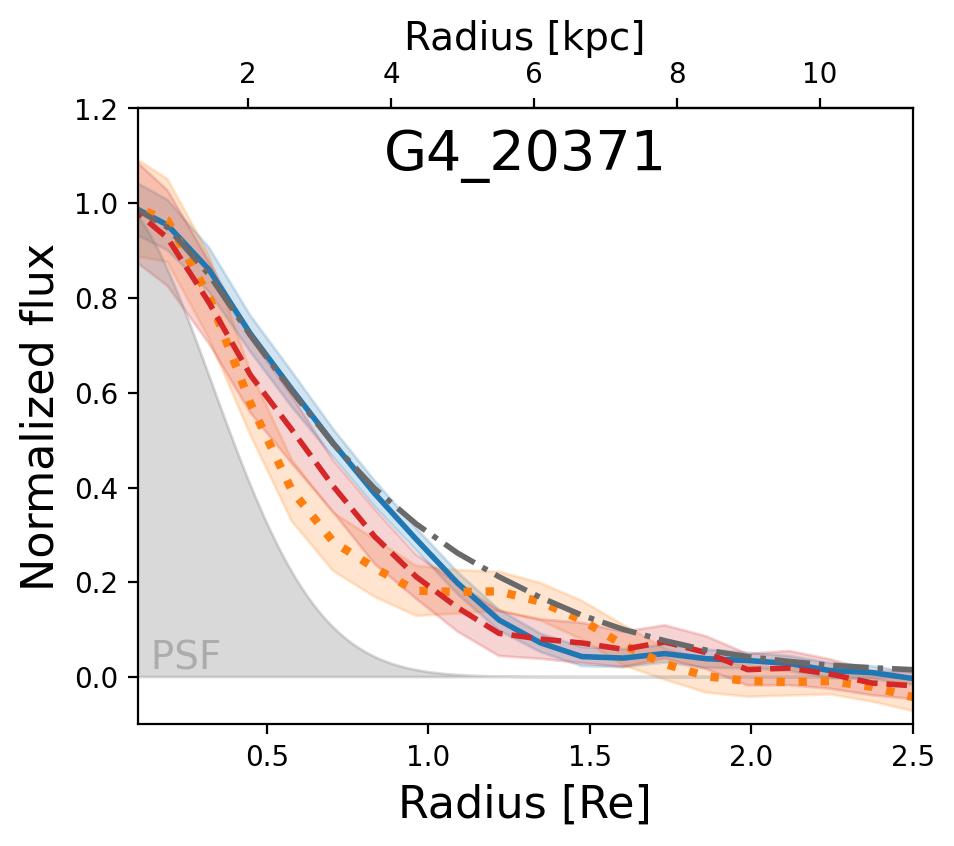}
  \includegraphics[width=0.24\textwidth]{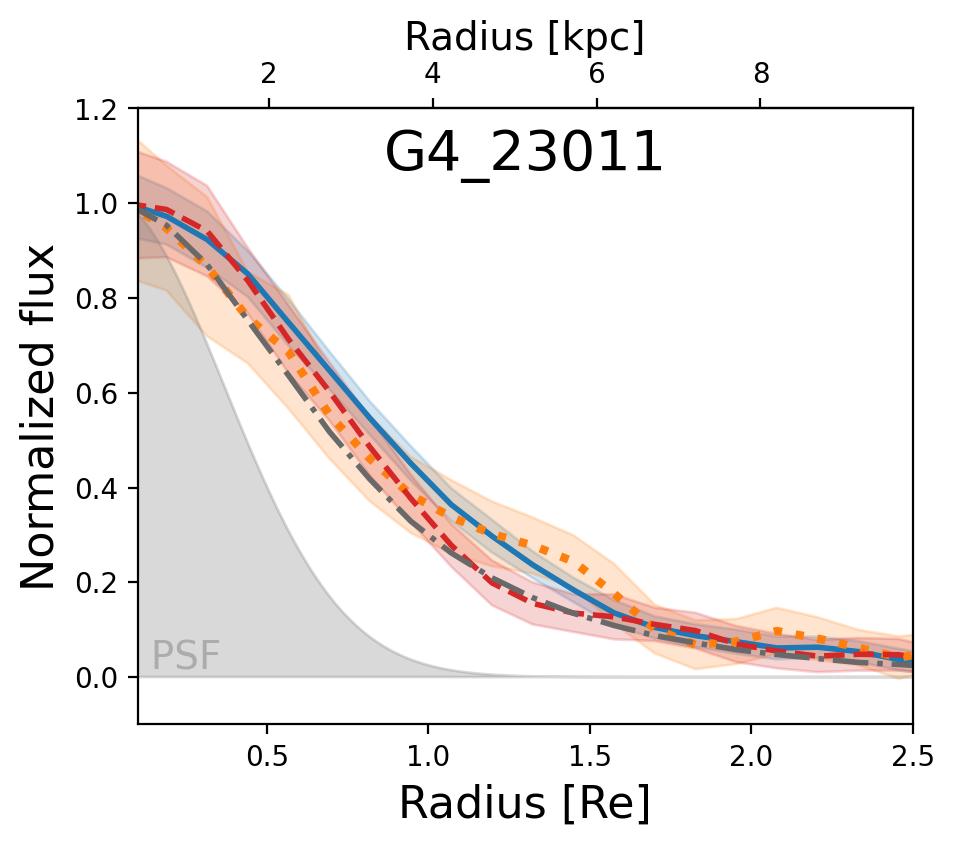}
  \includegraphics[width=0.24\textwidth]{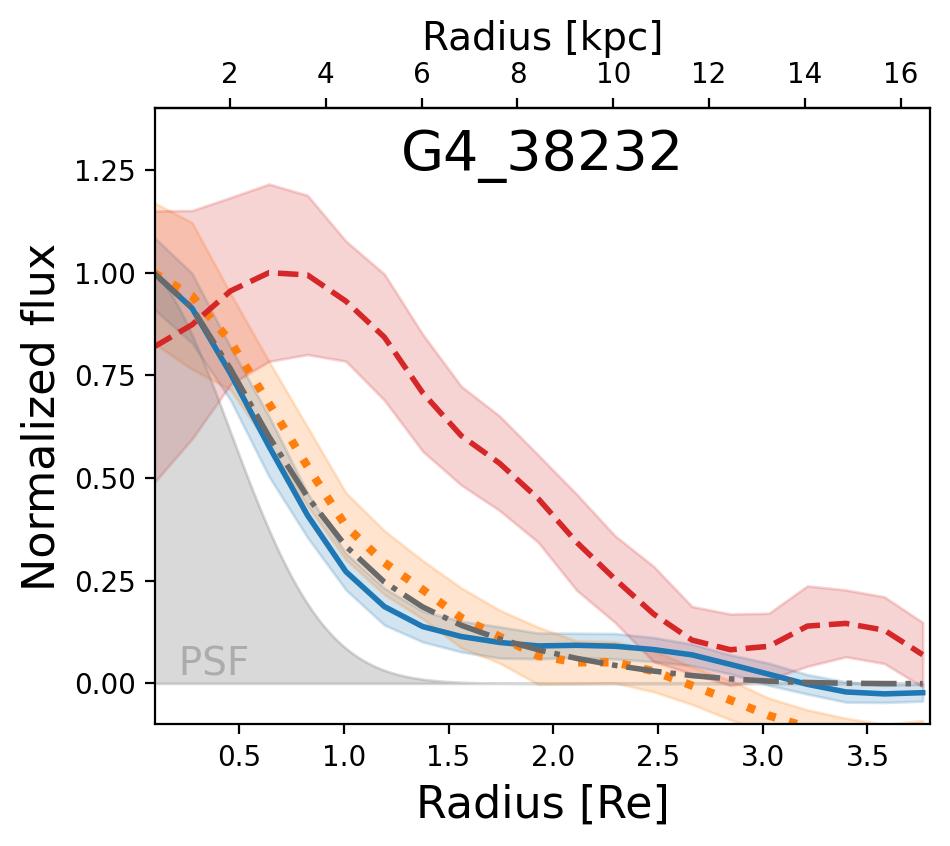}
  \newline
  \includegraphics[width=0.24\textwidth]{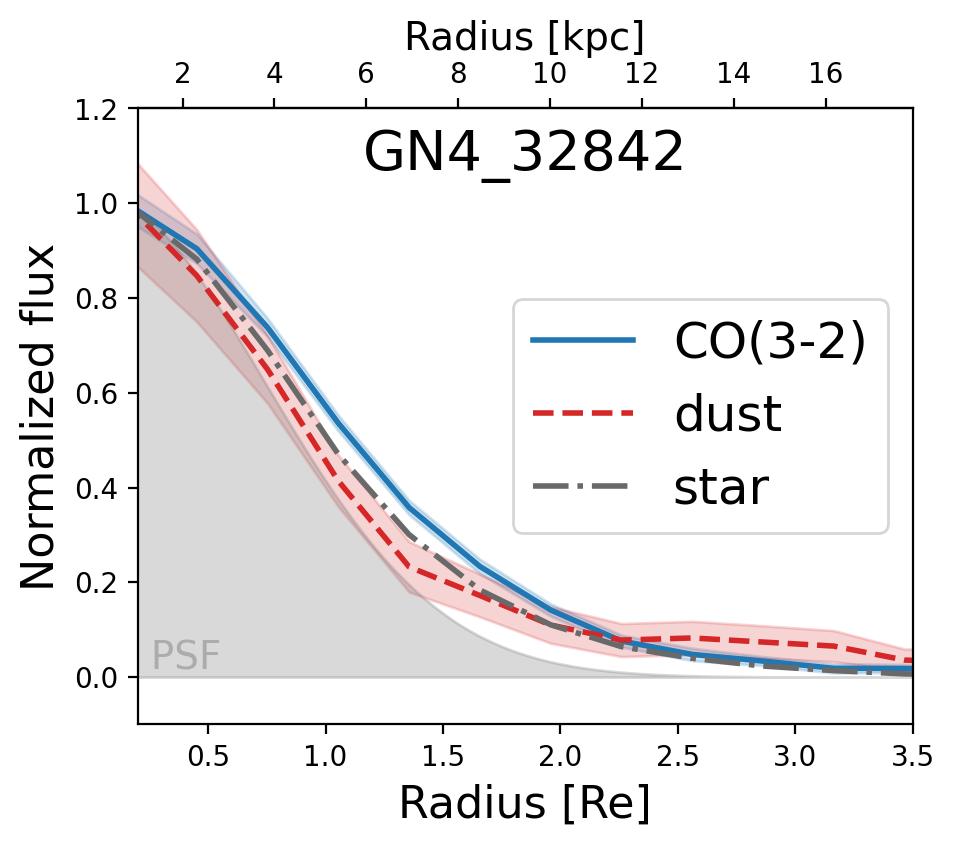}
  \includegraphics[width=0.24\textwidth]{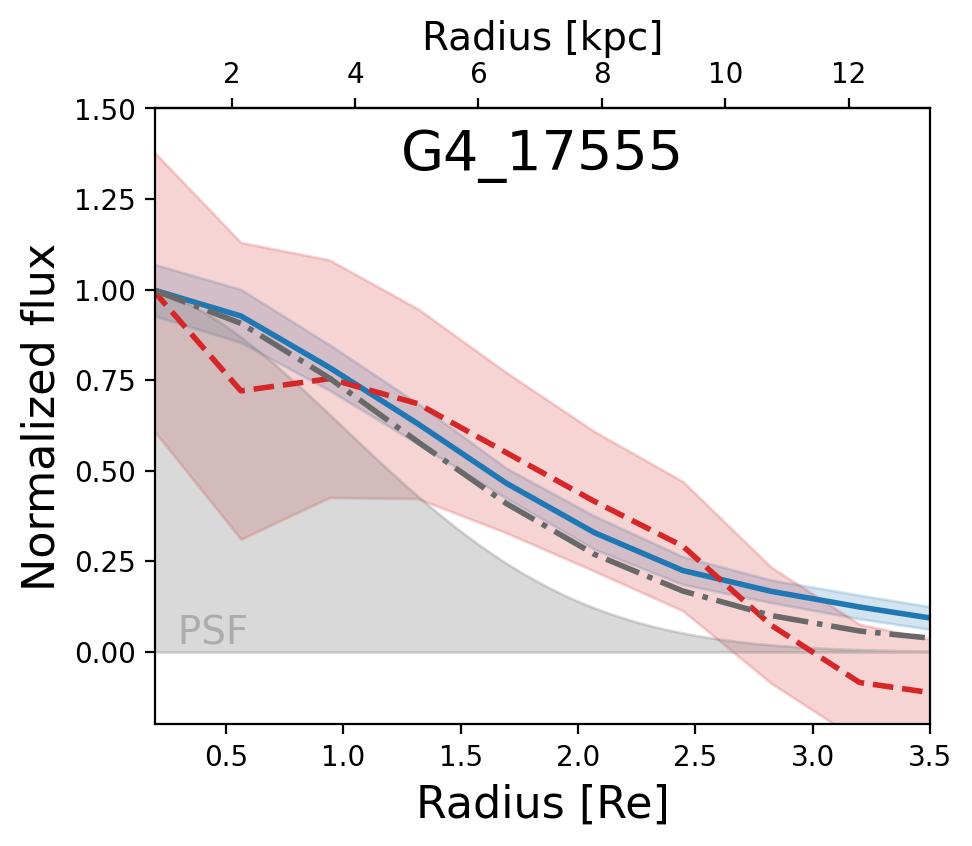}
  \includegraphics[width=0.24\textwidth]{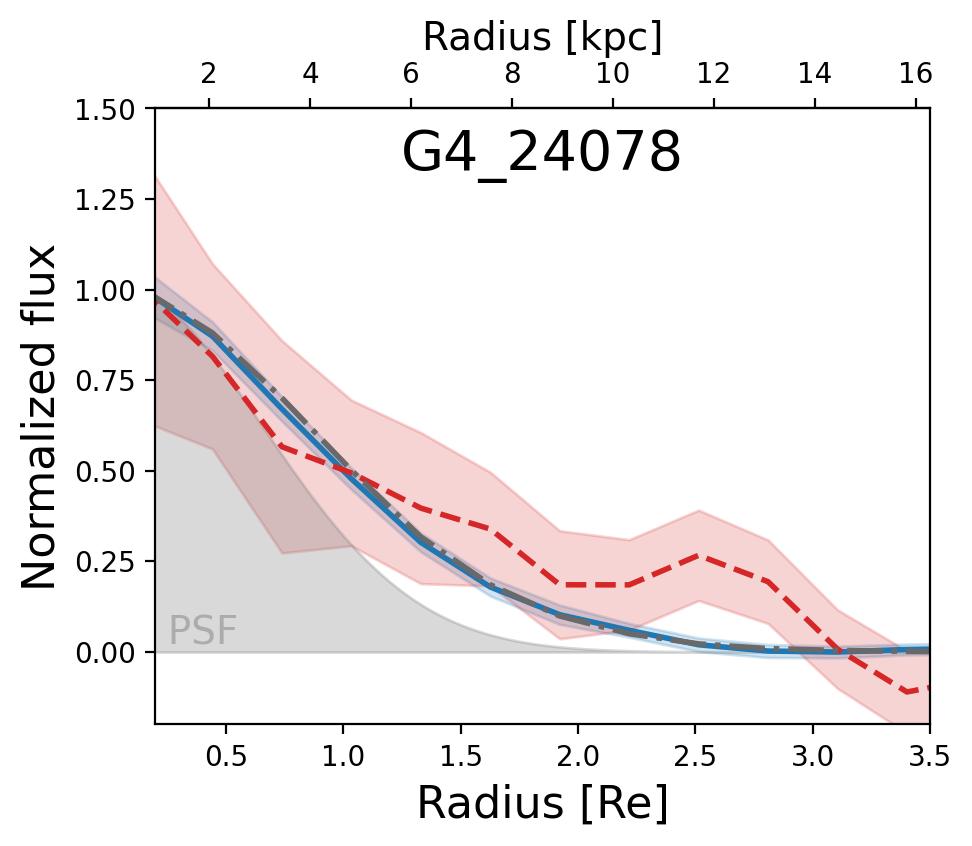}
  \includegraphics[width=0.24\textwidth]{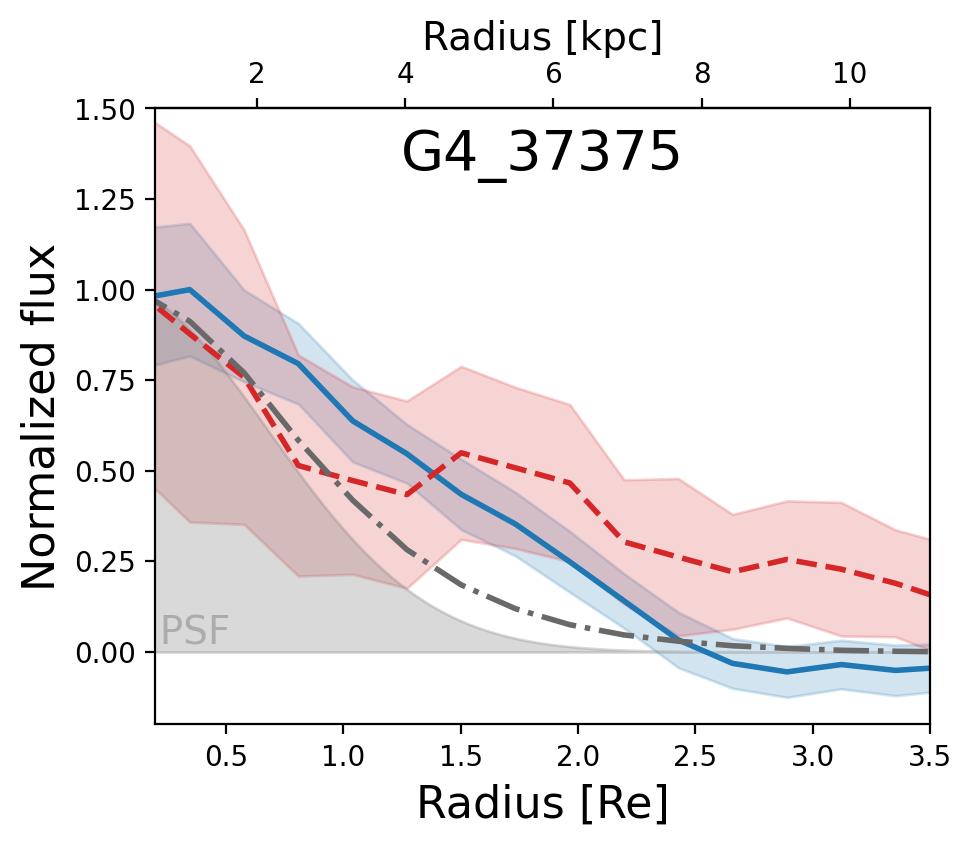}
  \caption{Radial flux profiles of various observables for \noema{} targets. In each panel, the radial profile of each component was normalized by its peak value. The gray-shaded area shows the PSF profile. The radius on the bottom x-axis is scaled to the stellar effective radius, while the physical sizes in kiloparsecs are displayed along the top x-axis. The radial profiles of the different components are largely comparable across the sample (except for G4-38232), suggesting that gas, dust, and stars are generally well mixed.}
  \label{fig:radial_profiles}
\end{figure*}

\subsection{Flux density profiles}

We averaged the data within radial elliptical annuli to construct the radial profiles of different cold gas tracers and stars.
To ensure consistent comparison between the observables, we first convolved the stellar emission down to the same resolution as the CO images.
Here, we mainly compared with the stellar profile from \textit{JWST}/NIRCam F444W (\textit{HST}/WFC3 F160W for G4-38065), which provides the closest approximation to the stellar mass profile (see the example in Fig.~\ref{appendixfig:size_measurements}).
We did not use the resolved SED map here to keep the comparison to direct observables.
To achieve uniform sampling across the different components, we applied the same aperture, following the disk geometry described in Table~\ref{tab:second_properties}, to all components. We gradually increased the aperture radius, in steps of half the beam size, to extract the radial profiles.
The derived radial profiles of CO(3$-$2), \co{}, \ci{}, dust, and stars are shown in Fig.~\ref{fig:radial_profiles}.

In general, we find that most galaxies show comparable radial profiles across different components, with differences largely within their $1~\sigma$ variations.
However, not all the cold ISM tracers follow the same radial profiles.
In G4-38065 and G4-38232, we observed more extended dust emission, mainly driven by off-center dust emission.
Overall, such global similarity in radial profiles is significantly different from what has been observed in more extreme DSFGs, where the dust is found to be more compact than the gas and stars \citep{Hodge2015,Chen2017,Tadaki2017a,CalistroRivera2018,Boogaard2026}.

\begin{figure*}
\begin{center}
  \includegraphics[scale=0.35]{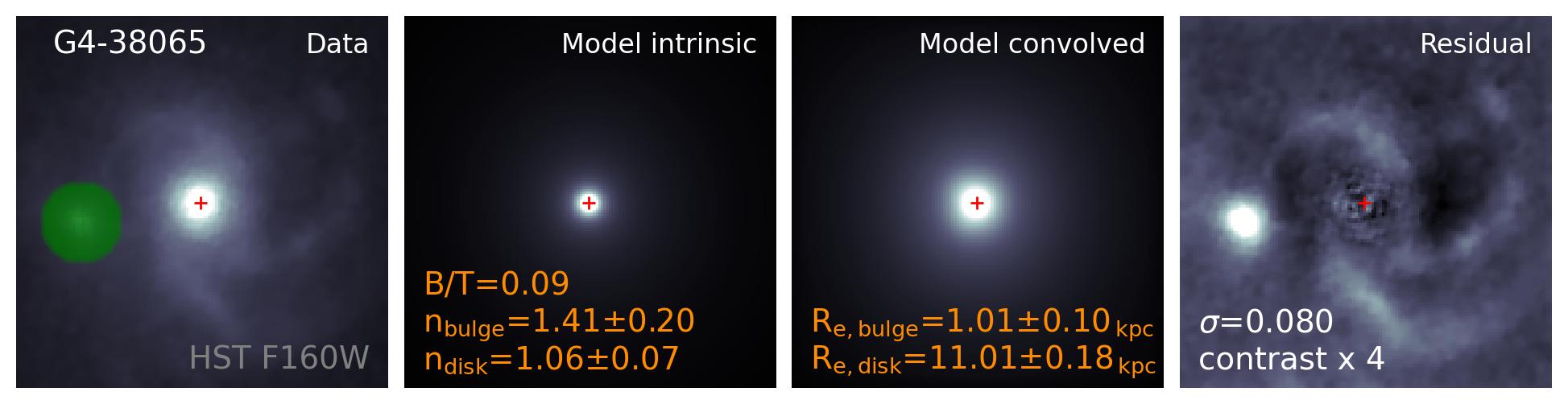}
  \includegraphics[scale=0.35]{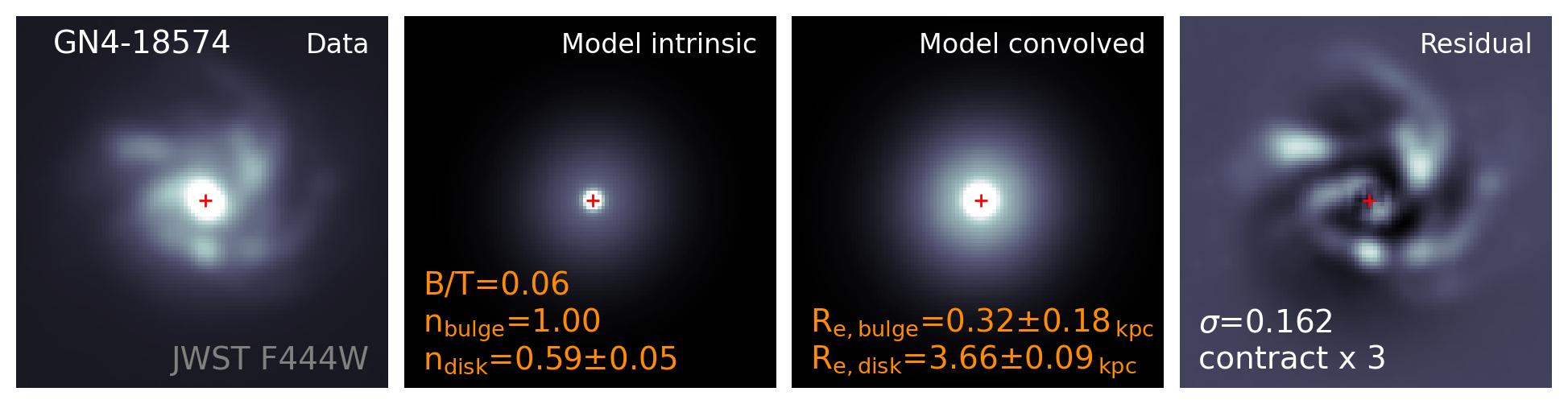}
  \includegraphics[scale=0.35]{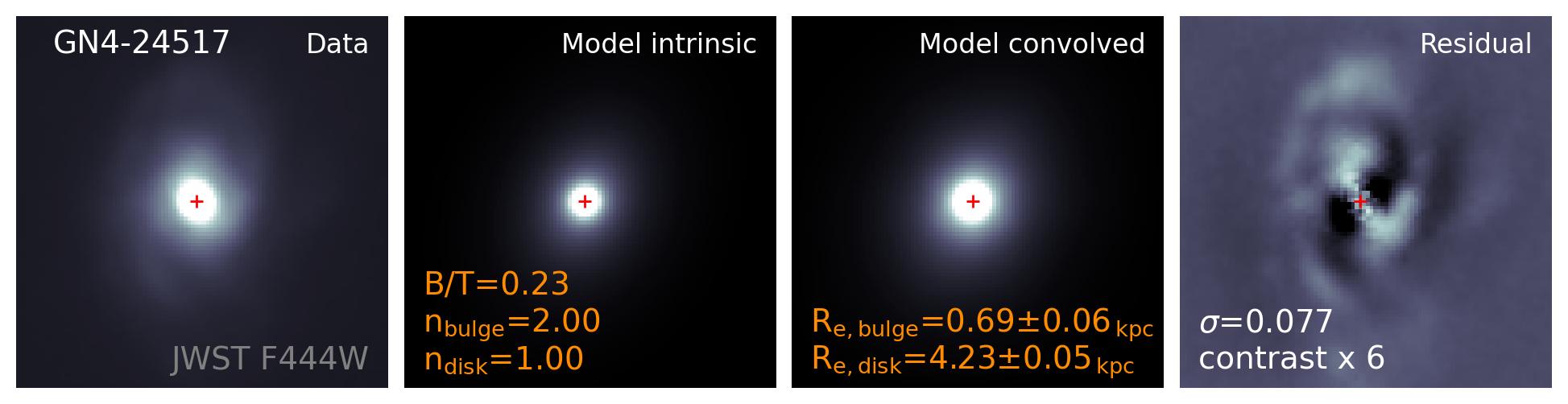}
  \includegraphics[scale=0.35]{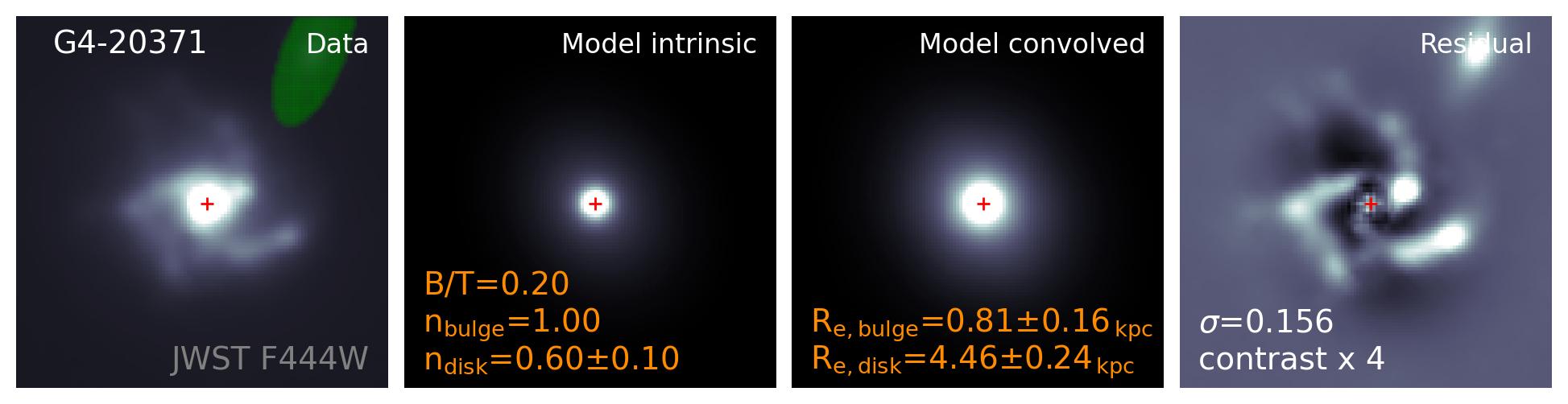}
  \includegraphics[scale=0.35]{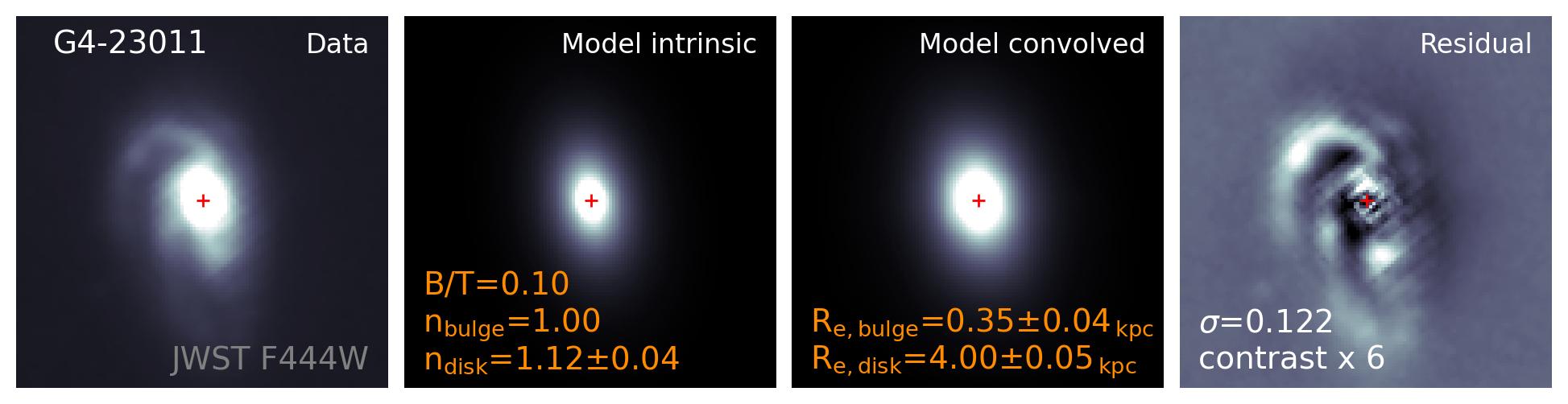}
  \includegraphics[scale=0.35]{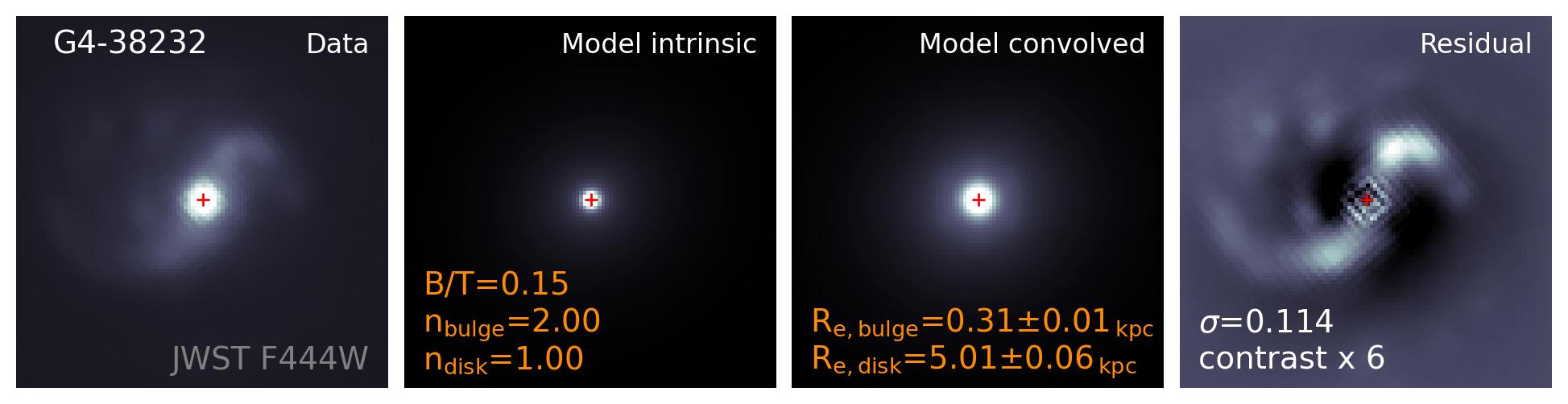}
  \includegraphics[scale=0.35]{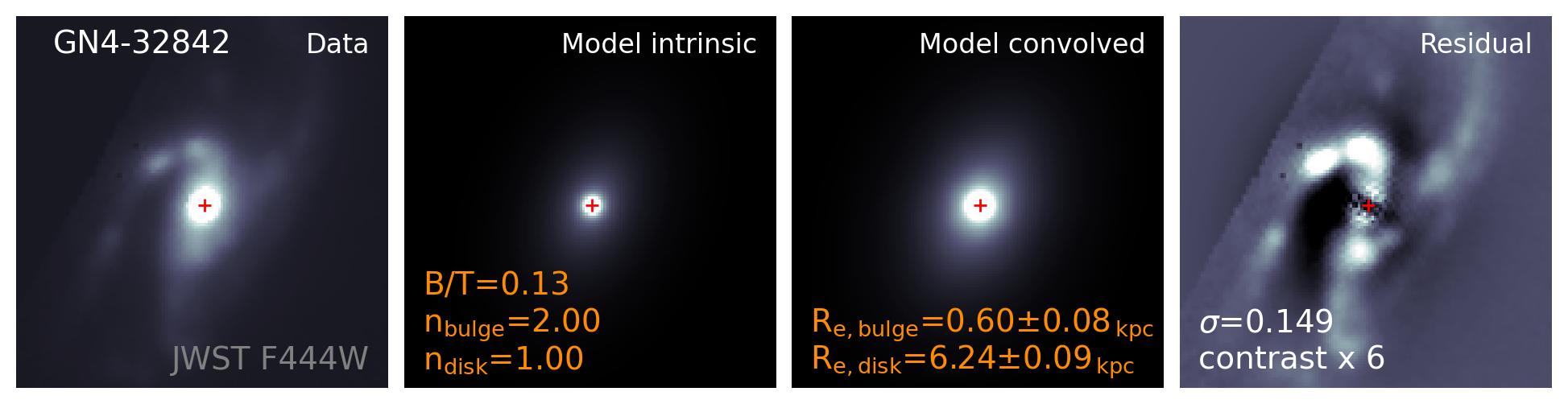}
  \includegraphics[scale=0.35]{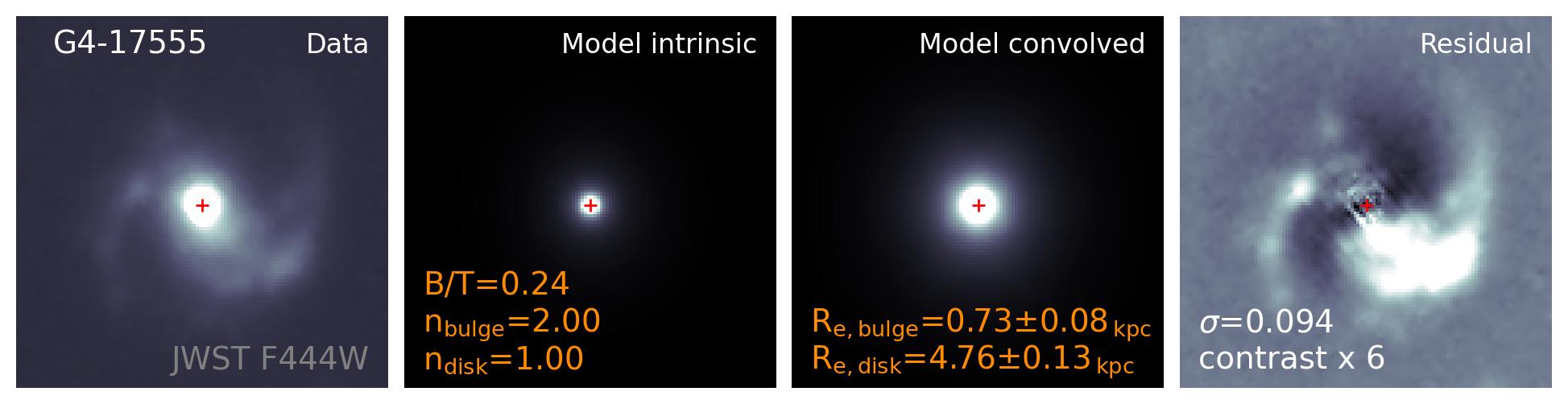}
  \includegraphics[scale=0.35]{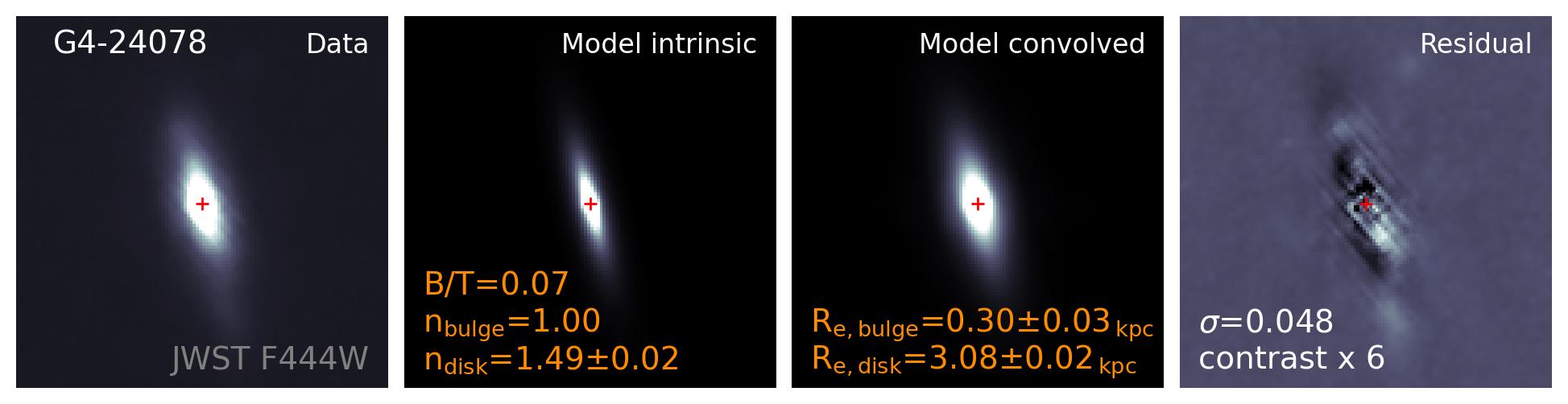}
  \includegraphics[scale=0.35]{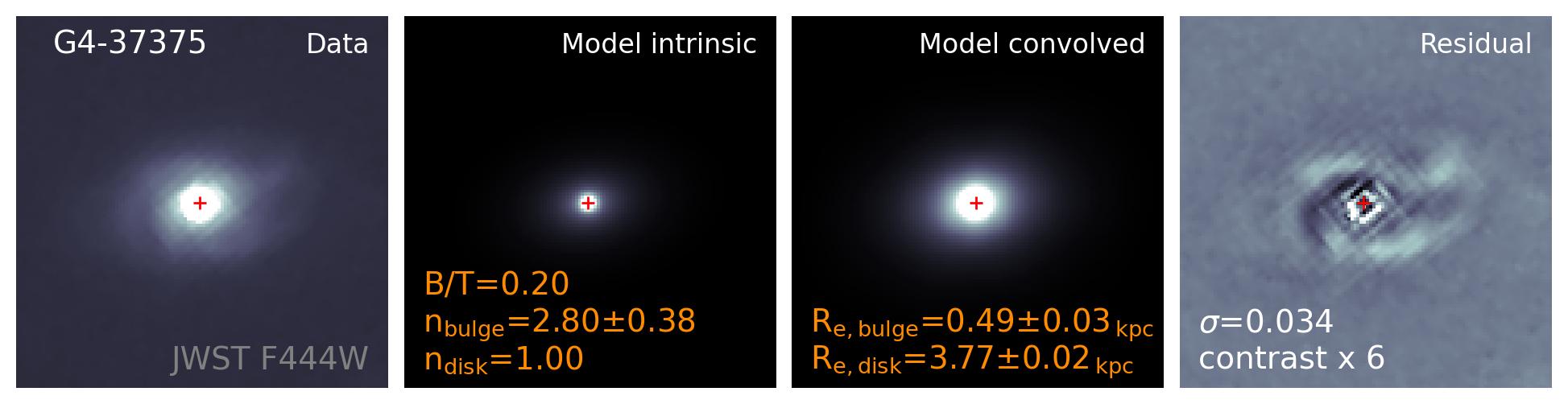}
\end{center}
\caption{Combined bulge and disk modeling of the stellar emission, based on the reddest available broadband filter (mostly F444W, except F160W for G4-38065). For each galaxy, the four columns display the observed stellar emission, the intrinsic bulge and disk model, the model convolved with the PSF, and the residuals after subtracting the best-fit model. We also masked close companions and foreground or background galaxies during modeling, as indicated by the transparent green mask. The derived B/T, S\'ersic index, and effective radius for the bulge and disk components are highlighted in orange in the model columns and listed in Table~\ref{tab:second_properties}. The Sérsic parameters without errors were fixed during fitting. We observed a globally small bulge contribution in these galaxies, which may indicate either ongoing bulge formation or a nuclear disk. The bulge component is largely absent in bluer filters, likely due to stronger dust attenuation in the central region (for an indepth discussion, see Sect.~\ref{subsec:bulge}).}
\label{fig:bulge-fit}
\end{figure*}

\subsection{Bulge-disk decomposition}

Most of our galaxies exhibit prominent central stellar light concentration in the rest-frame near-IR, resembling the bulge component widely seen in nearby galaxies.
For simplicity, we continue to refer to these features as ``bulges'', with the caveat that they may not fully correspond to the classical spherical bulge component observed in more nearby systems \citep[see discussions in][]{Gadotti2026, LeConte2026}.

We performed bulge--disk decomposition using two 2D Sérsic profiles to characterize the bulge profile and quantify the bulge-to-total ratio (B/T).
While the extended stellar disk is not strongly affected by smearing, the compact central bulge lies largely within the PSF and thus requires forward modeling that accounts for PSF smearing.
We modeled the bulge as a spheroid with its axis ratio fixed to 1.
To reduce degeneracy between the two Sérsic components and better separate the bulge contribution, we applied the following constraints to our model before the global fitting.
We first tied the centers of the bulge and the disk together.
We then derived the bulge center by fitting only the central 2\,kpc region, which was later fixed during the global fitting.
For the disk component, we adopted its position angle and inclination from a joint analysis of isophote fitting and the kinematics \citepalias[see more details in][]{Jolly2026}.
We also restricted our modeling to the reddest available filters (F444W and F160W for G4-38065) to minimize dust attenuation.
During fitting, the combined disk and bulge model was convolved with the model PSF, constructed using the Space Telescope PSF (STPSF) toolkit\footnote{https://stpsf.readthedocs.io} and stacked in the same manner as the science data from different exposures \citep{Pastras2026}.
Fitting was performed in two iterations. 
The first fitting used a differential evolution algorithm to search a wide parameter space and calculate the initial guess for the second least-squares minimization. 
The best-fit parameters are summarized in Table~\ref{tab:second_properties}.
The images, along with the best-fit models and residuals, are shown in Fig.~\ref{fig:bulge-fit}.

Even with all these constraints, the fitting still fails to converge on most of the bulge components.
In such cases, we fixed the Sérsic index \(n\) of the bulge component to 1.  
Two main reasons explain why we selected a lower Sérsic index rather than the widely adopted value of \(n = 4\) \citep[e.g.,][]{Tacchella2015}.  
First, based on two successful fits, the Sérsic indices are typically much lower than 4.  
Second, we tested a grid of Sérsic indices for the bulge and found that \(n = 1\) consistently provides a better fit than larger Sérsic indices.
We note that the present near-IR spatial resolution is insufficient to robustly constrain the bulge’s shape; nevertheless, the adopted $n_{\text{bulge}}$ has a relatively small effect on the derived B/T and does not change our conclusions.
Four galaxies in our sample also show a presence of galactic bar.
In these cases, we fixed both the S\'ersic index of the bulge ($n=2$) and the disk ($n=1$) to avoid bias from the bar component.
For these barred galaxies, fixing the Sérsic index to 1 typically yields a slightly lower B/T and a poorer fit, but it does not affect the conclusions.

\subsection{Molecular gas mass}

In this subsection, we discuss various scaling relations used to estimate the molecular gas mass for our observations. 
The \firstgroup{} targets provide us with the opportunity to compare three tracers simultaneously, namely \co{}, \ci{}, and the dust continuum, and assess the potential impact of variations in excitation/dust temperature on the global molecular gas estimates.
All \secondgroup{} galaxies were detected in CO(3--2), a reliable tracer of molecular gas in massive main‑sequence galaxies \citep{Tacconi2020}. Only one of the four \secondgroup{} galaxies was detected in dust continuum, so we estimated their molecular gas masses solely from the CO(3--2) line.
We converted these observables into molecular gas mass using the empirical calibration derived directly from CO(1-0) whenever possible to avoid any cross-calibration issues.
The final molecular gas mass, $M_\text{mol}$, represents the total molecular gas mass, including the 36 percent mass contribution from helium.

\subsubsection{CO}

Studies of the CO spectral line energy distribution (SLED) have been conducted in numerous SFGs around cosmic noon. However, due to the faintness of CO(1-0), direct calibration compared to CO(1-0) is very limited \citep[e.g.,][]{Aravena2010,Aravena2014,Bolatto2015,Henriquez-Brocal2022,FriasCastillo2023,Prajapati2026}. In this study, we adopted the calibration proposed by \citet{Tacconi2020}, who carefully compiled various literature data, homogenized the analysis, and conducted convergence tests \citep[see also][]{Genzel2015,Tacconi2018}. We then estimated the molecular gas mass with the following equations:

\begin{equation}
\begin{split}
  & L^\prime_\text{CO} = 3.25\times10^{7}\,S_\text{CO}\,\Delta \varv\,v_\text{obs}^{-2} D_\text{L}^2 (1+z)^{-3} \,\,\text{K\,km\,s}^{-1}\text{pc}^2\\
  & M_\text{mol} = \alpha_\text{CO}\times R_\text{1J} \times L^\prime_{\text{CO}_\text{J-(J-1)}}\,\,\text{M}_\odot
\end{split}
\end{equation}

Here, $S_\text{CO}\Delta \varv$ is measured in Jy km s$^{-1}$; the observed frequency $v_\text{obs}$ is in GHz; and the luminosity distance $D_\text{L}$ is in Mpc. Following \citet{Tacconi2020}, we adopted the recommended $\alpha_\text{CO}=4.36\pm0.9$ $\text{M}_\odot$\,K\,km\,s$^{-1}$\,pc$^{2}$, and a CO ladder correlation of $R_{14}=2.4$ and $R_{13}=1.8$ for \firstgroup{} and \secondgroup{} targets, respectively.

\subsubsection{Dust continuum}
\label{subsec:dust}

Long-wavelength dust emission has gained popularity as an alternative method for measuring molecular gas mass \citep[e.g.,][]{Eales2012,Magdis2012,Scoville2014}. 
At millimeter wavelengths, dust emission is primarily optically thin, with the observed flux being proportional to its mass \citep{Scoville2014, Scoville2016, Scoville2017}.
Our observations were mostly conducted at rest-frame wavelengths of 650~$\mu$m and 800~$\mu$m; thus, we can safely use these measurements to trace the distribution of molecular gas mass. 
We adopted the conversion from \citet{Scoville2016}, which has been empirically calibrated using a sample of SFGs with CO(1-0) observations at various redshifts, SFRs, stellar masses, and spanning three orders of magnitude in dust emission:

\begin{equation}
\begin{split}
  & L_{v_{\rm 850\mu m}} = 4\pi\,S_{v_\text{obs}} \times K\,\left(\frac{D^2_\text{L}}{1+z}\right) \,\,\text{W Hz}^{-1}\\
  & K = \left(\frac{\rm 353 GHz}{v_\text{rest}}\right)^{3+\beta}\left(\frac{e^{\text{h}v_\text{rest}/k/T_\text{d}}-1}{e^{16.956/T_\text{d}}-1}\right)\\
  & M_\text{mol} = \frac{L_{v_{\rm 850\mu m}}}{\alpha_{v_{\rm 850\mu m}}} \,\,\text{M}_\odot
\end{split}
\end{equation}

Here, $K$ is the K-correction between the observed frequency and 850$\mu$m. We adopted $T_\text{d}=25$~K, $\beta=1.8$, and $\alpha_{v_{\rm 850\mu m}} = 6.7\pm1.7\times10^{12} \text{W\,Hz}^{-1}\text{M}_\odot^{-1}$, as recommended by \citet{Scoville2016}. It is noteworthy that the 25\,K dust temperature is a mass-weighted average, since most dust in the ISM is supposed to be cold. \citet{Scoville2016} cautioned against using a variable dust temperature in the scaling relation for converting dust continuum to total molecular gas.
This is because the observed dust SED is biased toward the luminosity-weighted dust temperature, which is influenced by the high temperatures of luminous star-forming regions. In contrast, mass-weighted dust is dominated by the more extended cloud envelopes, which remain less variable across galaxies.

\begin{figure*}[htpb]
  \centering
  \includegraphics[width=0.32\textwidth]{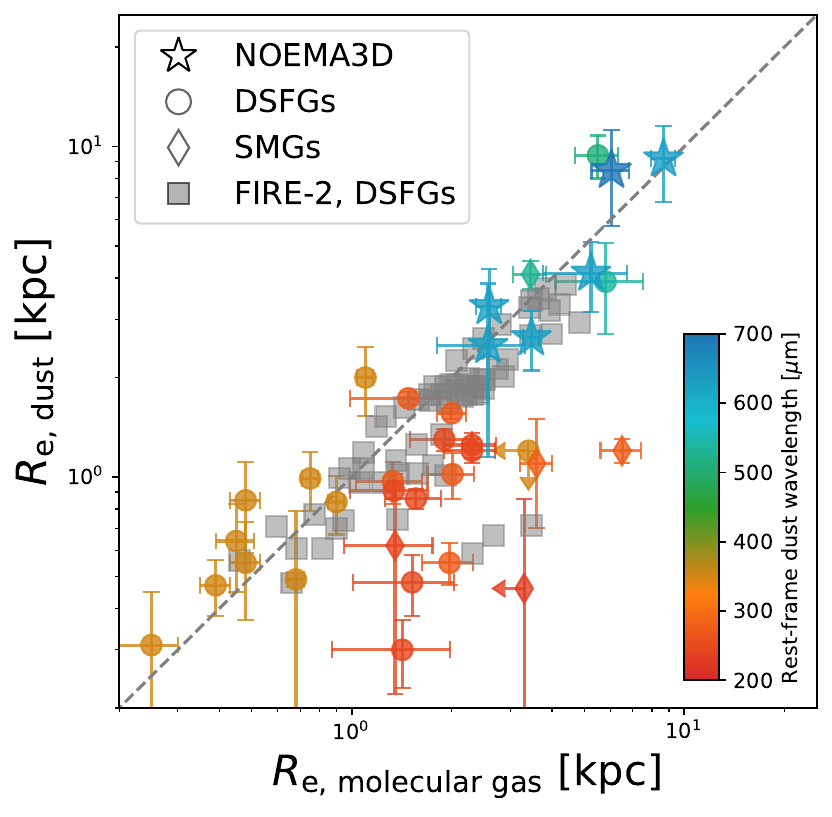}
  \includegraphics[width=0.32\textwidth]{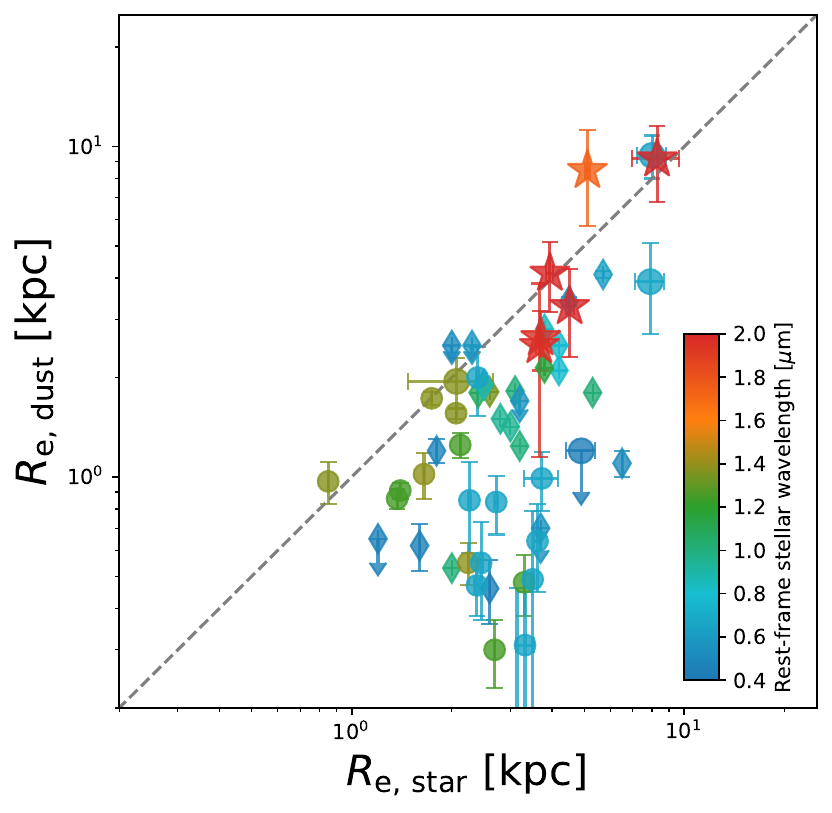}
  \includegraphics[width=0.32\textwidth]{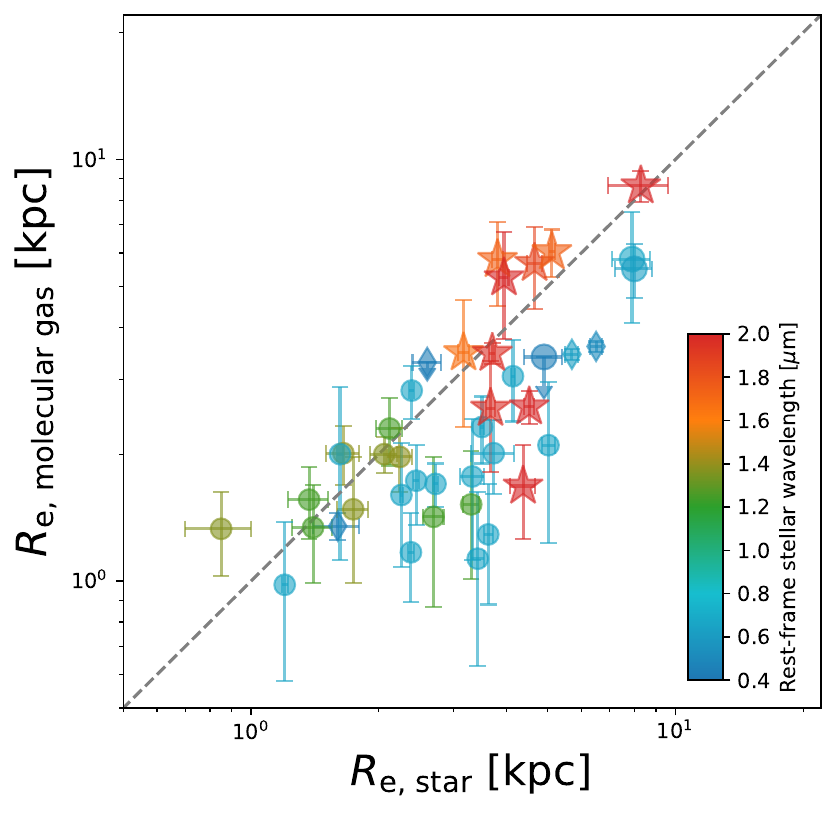}
  \caption{Global size comparison among different components across various galaxy types. \textit{Left:} Dust size vs. molecular gas size, as traced by CO. \textit{Middle:} Dust size versus stellar size. \textit{Right:} Molecular gas size versus stellar size. We include measurements from the literature concerning size estimates in DSFGs \citep[circles;][]{Tadaki2017, Kaasinen2020,Ikeda2022,Tadaki2023}, as well as from more extreme starburst galaxies \citep[diamonds;][]{Chen2017, CalistroRivera2018, Pantoni2021, Smail2023, Hodge2025}.
  In each panel, the color bar shows the rest-frame wavelength of the dust emission or stellar emission. For comparison, we also present simulation predictions for the far-infrared (FIR) bright galaxies from the FIRE-2 simulation \citep[grey squars;][]{Cochrane2019}. When combining all these measurements, our findings indicate that the sizes of stars, molecular gas, and cold dust are largely comparable in MS galaxies. However, more extreme starburst galaxies typically exhibit a more compact distribution of stars and dust. These differences likely reflect the distinct star formation patterns observed in MS galaxies compared to those in compact starburst galaxies.} 
  \label{fig:size_comparison}
\end{figure*}

\subsubsection{CI}

\ci{} is sometimes considered as an alternative molecular gas mass tracer to CO(1-0) \citep{Walter2011,Alaghband-Zadeh2013,Bothwell2017,Valentino2018,Valentino2020,Bourne2019,Lee2021,Arriagada-Neira2025}. It has a low excitation temperature of about 24~K and a critical density similar to that of CO(1-0). Additionally, it has been found to correlate well with CO(1-0) in resolved nearby galaxies \citep{Jiao2019,Liu2023}. Its notable advantage lies in its better availability for $z>1$ SFGs, where CO(1-0) becomes significantly harder to detect and resolve \citep{Papadopoulos2004}. Among its two commonly observed lines, \ci{} is regarded as more reliable than [C\,\textsc{i}](2-1), as the latter can be strongly sub-thermally excited \citep{Harrington2021,Papadopoulos2022}. Based on an empirical analysis of the three most commonly used molecular gas tracers, \citet{Dunne2022} demonstrates that \ci{} exhibits the least intrinsic scatter in the parameter space we probe.

Following \citet{Dunne2022}, we estimated the molecular mass from \ci{}, using

\begin{equation}
\begin{split}
  & L^\prime_\text{CI} = 3.25\times10^{7}\,S_\text{CI}\,\Delta \varv\,v_\text{obs}^{-2} D_\text{L}^2 (1+z)^{-3} \,\,\text{K\,km\,s}^{-1}\text{pc}^2\\
  & M_\text{mol} = \alpha_\text{CI} L^\prime_\text{CI} \,\,\text{M}_\odot
\end{split}
\end{equation}

Here, $D_L$ is the luminosity distance in Mpc, and $S_\text{CI}\Delta \varv$ is the flux density of \ci{} in Jy\,km\,s$^{-1}$. We adopted the best-fit $\alpha_\text{CI}=18.7\pm0.6$, derived from direct minimization with a CO(1-0) sample of MS galaxies reported by \citet{Dunne2022}.

In summary, at an integrated level, all three molecular gas tracers return gas masses in reasonable agreement. However, as suggested by \citet{Tacconi2020}, a fair comparison can only be made after properly correcting for zero-point offsets between methods. In this work, we adopted separate sources of scaling relations that were calibrated largely independently to better reveal offsets or differences from each individual tracer.

\section{Results}
\label{sec:results}

In this section, we discuss the evolutionary patterns of \noema{} galaxies, based on the physical properties derived and presented in  Sect.\ref{sec:analysis}. 
We also compare the results with literatures.

\begin{figure}
  \centering
  \includegraphics[width=0.48\textwidth]{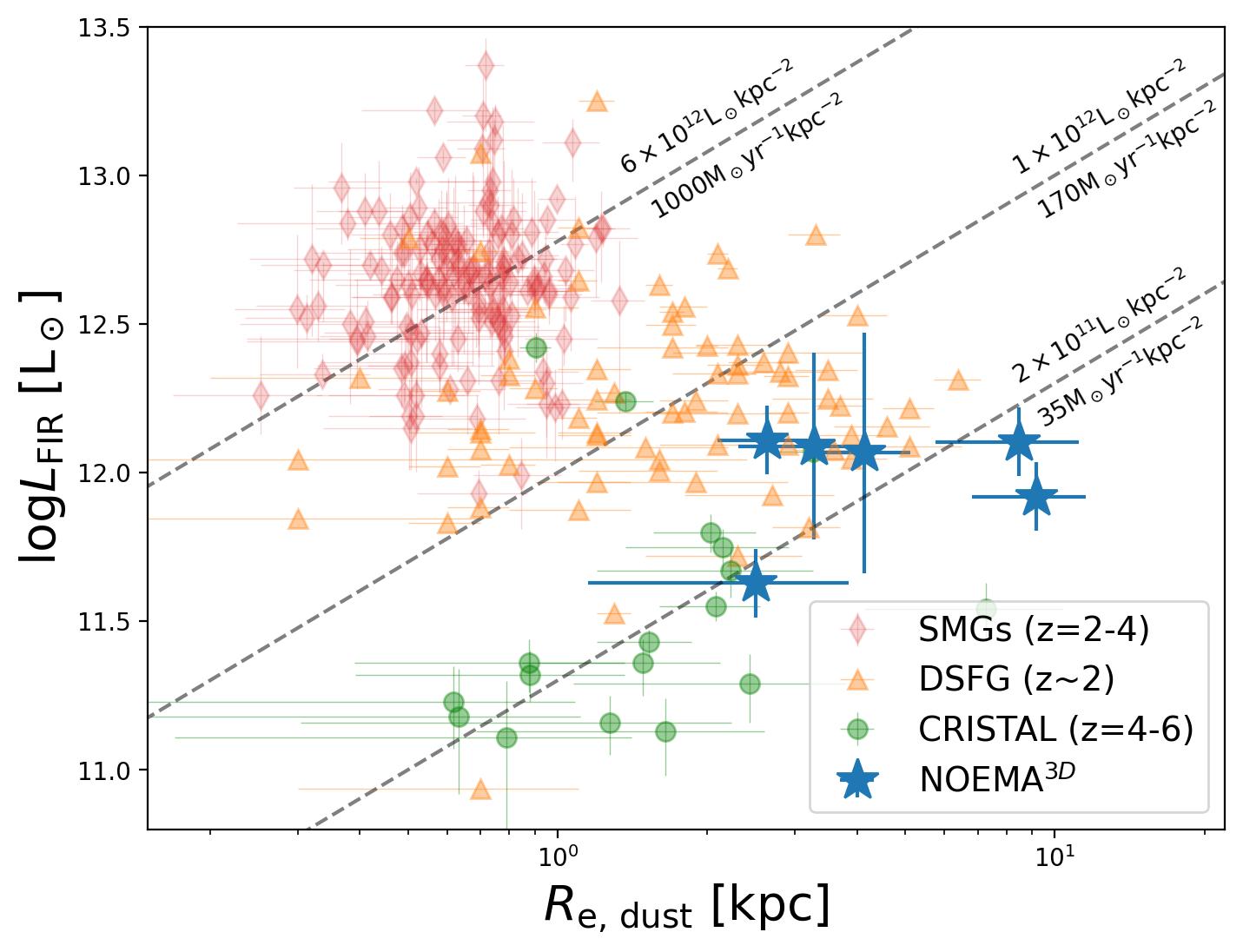}
  \caption{$L_\text{FIR}$-size relation for dust emission with respect to different galaxy types and redshifts. We include the SMGs from \citet{Gullberg2019}, the MS galaxies from the CRISTAL survey \citep{Mitsuhashi2024}, and a sample of DSFGs mixed of both MS and SMGs from \citet{Tadaki2020}. All galaxies collected in this study were observed using modern submillimeter interferometers, specifically NOEMA and ALMA. The dashed lines represent constant FIR luminosity densities, with their corresponding values labeled. These dashed lines also approximate constant SFR densities if a constant scaling factor is assumed \citep{Kennicutt2012}. The apparent separation between SMGs and MS galaxies stems from their different SFR densities.}
  \label{fig:dust_size_LIR}
\end{figure}

\begin{figure*}[htpb]
  \centering
  \includegraphics[width=0.32\linewidth]{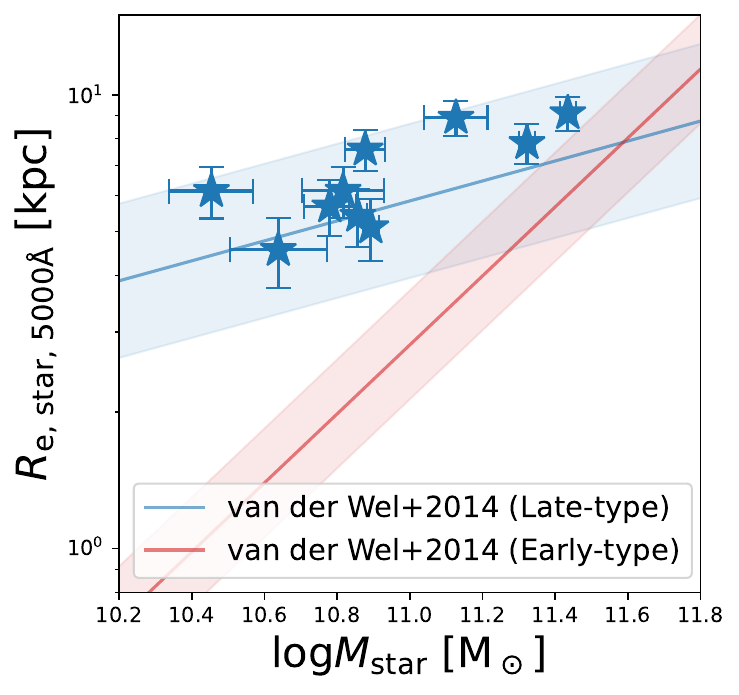}
  \includegraphics[width=0.33\linewidth]{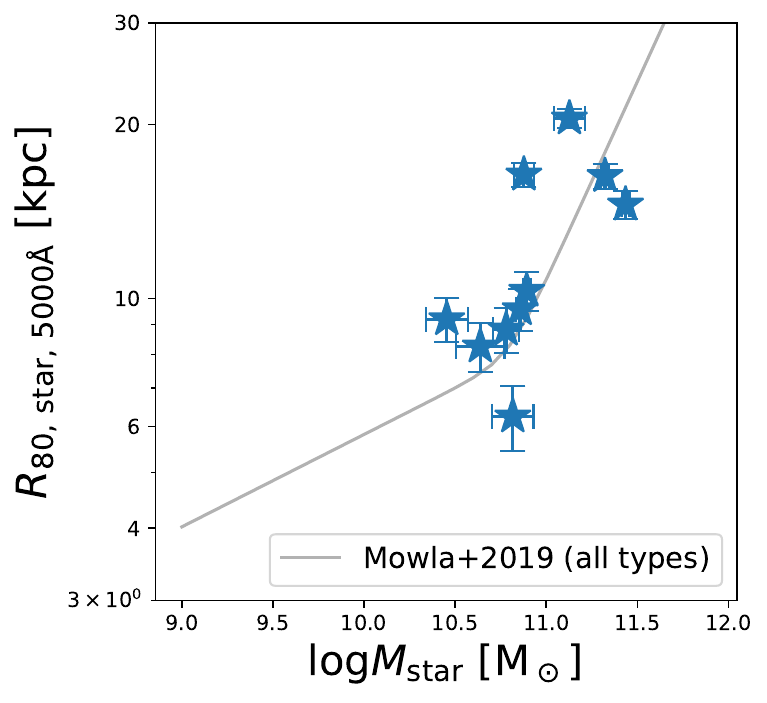}
  \includegraphics[width=0.32\linewidth]{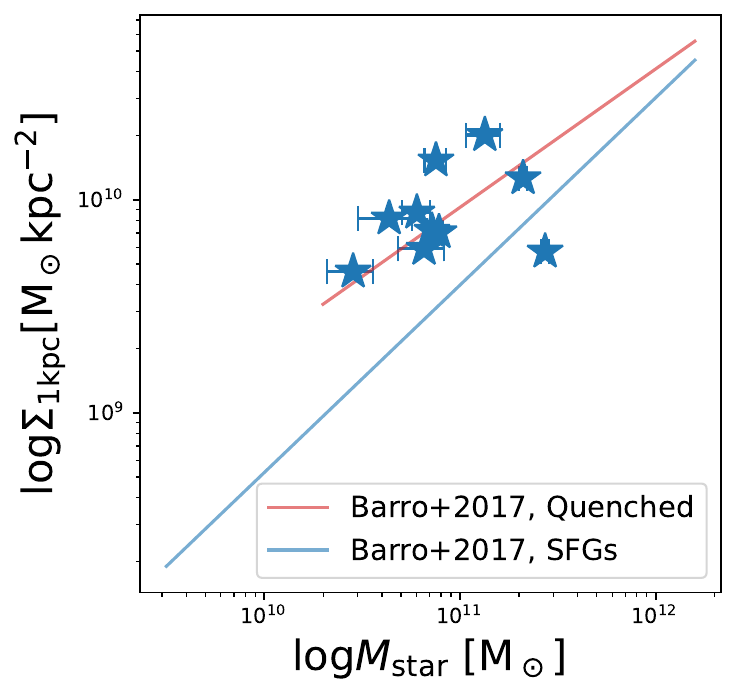}
  \caption{Mass and size related scaling relations for \noema{} galaxies. The size measurement is based on the effective radii $R_e$ and $R_{80}$, which represent the radii that encompass 50 percent and 80 percent of the total stellar emission at a rest frame of 5000 \AA, respectively. Regarding the mass--size relation based on the effective radius, our sample probes SFGs with extended disks at $z\sim1.25$ \citep{vanderWel2014}. In the $R_{80}$ mass-size relation, our galaxies are positioned just above the transitional stellar mass corresponding to the turnover point in the mass--size slope \citep{Mowla2019}. Regarding the central stellar density, $\Sigma_\text{1kpc}$, our galaxies have a density similar to that of compact, quenched galaxies at similar redshifts \citep{Barro2017}. Together, these comparisons indicate that our galaxies represent the typical massive SFGs at cosmic noon, likely in an evolutionary stage before the gradual transition to quenched early-type galaxies.}
  \label{fig:mass-size-relation}
\end{figure*}

\subsection{Comparable size between gas, dust, and stars}

In Fig.~\ref{fig:size_comparison}, we compare all available size measurements from different components for \noema{} targets.
The stellar, molecular gas, and dust sizes are scattered around the one-to-one correlations, with their deviations generally smaller than the $1\sigma$ measurement uncertainties.  
The most prominent outlier is G4-38232, where the molecular size is significantly smaller than the stellar size.
This galaxy also exhibits a noncentral-peaked dust emission, and its low S/N ($<3\sigma$) does not enable a robust measurement on its size.
Among most of the \secondgroup{} targets (3 of 4), we did not reliably detect dust emission, nor can we measure their sizes.
We also compared the \noema{} targets with other galaxy types and simulations in Fig.~\ref{fig:size_comparison}.
We included DSFGs \citep[e.g.,][]{Tadaki2017,Tadaki2023,Kaasinen2020,Pantoni2021} and more starburst SMGs \citep[e.g.,][]{Chen2017,CalistroRivera2018,Pantoni2021,Smail2023,Hodge2025}.
All these studies have a spatial resolution better than 0.7$"$ and have their size measured from spatially resolved observations.
We also included the predictions from \citet{Cochrane2019}, which conducted detailed radiative transfer simulations within the Feedback In Realistic Environments 2 (FIRE-2) framework.

Although the \noema{} galaxies exhibit comparable sizes in dust, gas, and stellar components, literature results show substantial scatter. In the comparison between dust and molecular gas, the dust size become smaller at the shorter rest-frame dust wavelengths. At the rest-frame wavelengths smaller than \(400\,\mu\)m, the dust continuum primarily traces ongoing star formation rather than the molecular gas distribution. This effect is most pronounced for the subset of SMGs. 
The simulated FIRE-2 galaxies nicely reproduced the relationship between dust and molecular gas sizes, confirming that the spatial extent of dust emission is sensitive to both dust temperature and mass \citep{Cochrane2019}. 
When comparing stellar sizes to those of molecular gas and dust, stellar sizes measured from rest-frame optical to ultraviolet wavelengths systematically show larger values. Given that most of these ALMA-selected galaxies are dust-rich, the concentrated dust could significantly attenuate the central stellar emission, thereby biasing the stellar effective size toward larger values. 
In simulations, FIRE-2 galaxies systematically predict larger dust sizes than stellar sizes, contrary to observations. The systematically larger dust size is likely due to the strong feedback models implemented in the simulations.
Similarly, the scatter in the comparison between molecular gas and stellar sizes is also likely due to dust attenuation.
Overall, we emphasize the need to fully account for observational biases in order to understand the size differences among components.

With the large availability of dust measurements for high-$z$ SFGs, we also show in Fig.~\ref{fig:dust_size_LIR} the correlation between FIR luminosity and dust size for different galaxy types across a broader cosmic history.
We include the most extreme starburst galaxies, SMGs, from a recent survey by \citet{Gullberg2019} at redshifts $z=2$$-$4.
We also incorporate a large sample of DSFGs from \citet{Tadaki2020}, which includes SFGs mostly located between MS galaxies and SMGs.
To expand the sample of MS galaxies, we also include size measurements from the CRISTAL survey at z$\sim$5 \citep{Mitsuhashi2024}.
Although there are fewer available dust-size and FIR-luminosity measurements for MS galaxies by more than an order of magnitude than for SMGs, the intrinsic number density of MS galaxies is more than a magnitude higher \citep[e.g.,][]{Chen2023}.
Thus, the current samples are significantly biased toward brighter starburst galaxies.

It is clear that normal star-forming MS galaxies and compact starbursts, such as SMGs, occupy distinct regions in the luminosity--size diagram.
Starburst SMGs are brighter in FIR emission due to their elevated SFR, which are concentrated in the central starburst core ($<$1~kpc in size), resulting in an extreme average SFR surface density, $\sim$1000\,$\text{M}_\odot\,\text{kpc}^{-2}$.
MS galaxies exhibit SFR about 1$-$2 orders of magnitude lower and are more extended, leading to a much lower average star-formation rate surface density.
Even at much higher redshifts, MS galaxies from CRISTAL show a similar star-formation rate surface density.
The CRISTAL galaxies show slightly smaller disks but also lower stellar masses and SFRs, keeping their surface densities comparable to those of MS galaxies at cosmic noon.
It should be noted that the extended size is not a characteristic feature of all MS galaxies. Many compact starburst DSFGs and even SMGs are also found to lie on the MS \citep[e.g.,][]{Elbaz2018, Gomez-Guijarro2022, Dudzeviciute2020}. The MS is a statistical description of galaxy evolution rather than a strict evolutionary pathway. As we will discuss in Sect.\ref{sec:implications}, the evolution of individual galaxies can wander around the MS, mediated by internal gas transport and redistribution.

We also compared 2D morphology of the different molecular gas tracers in Appendix \ref{appendix:2d_morphology}.
The CO generally exhibits larger asymmetries than \ci{} and dust, likely due to its higher signal-to-noise ratio (S/N) at large radii.
Because of the limited S/N in the \ci{} and dust continuum data, we focus primarily on the radial profiles in the following discussion.

\subsection{Size evolution}
\label{subsec:size_evolution}

The mass-size evolution is one of the most fundamental tests for galaxy formation theories. 
Early studies based on HST have revealed two distinct populations: red quenched galaxies and blue SFGs, each following distinct mass-size relations \citep{Toft2007, vanderWel2014}. 
When we place our galaxies on the mass-size relation of cosmic noon galaxies (see Fig.~\ref{fig:mass-size-relation}), they align with the blue sequence.
This supports that the size and stellar mass growth of the \noema{} galaxies is primarily driven by gas accretion and in situ star formation, instead of galaxy major mergers \citep{vanDokkum2010}.

In the $R_{80}-M_\star$ relation, the \noema{} galaxies predominantly occupy the region around the transitional mass where the change in size becomes steeper (see Fig.~\ref{fig:mass-size-relation}).  
The stellar mass at this turning point corresponds to the stellar mass at the peak of the stellar-to-halo mass relation (SHMR) \citep{Mowla2019}, indicating that their halo masses are generally $\sim25$ times their total stellar mass \citep{Leauthaud2012}.
This provides an estimate for the total halo mass of our targets, in the range of $0.8-6\times10^{12}\text{M}_\odot$. 
At z$\sim$1-1.5, these halos are approximately at the boundary where cold gas accretion remains feasible, a topic we explore further in Sect.\ref{subsec:gas_origin}.

The environment can play a key role in shaping galaxy sizes. 
The \noema{} galaxies were selected specifically to avoid any close neighbors. 
From the most recent deep \textit{JWST} images, none experiences an appreciable gravitational interaction with neighboring galaxies.
Their overall smooth morphology and regular kinematics \citepalias{Jolly2026} indicate that they are typically isolated disk galaxies. 
Moreover, environmental analysis of the CANDELS fields \citep{Chartab2020} allows us to define the density contrast of each region based on the environmental overdensity $\delta=\sigma/\bar{\sigma}-1$. 
Here, $\bar{\sigma}$ represents the background number density, while $\sigma$ is the number density of target galaxies within a volume of $1h^{-3}$Mpc$^3$; thus, $\delta=0$ indicates no overdensity.
According to this definition, the \noema{} galaxies are predominantly (8 out of 10) located in regions with density contrasts $\delta\sim[-0.03,0.6]$, close to the mean density of MS galaxies at cosmic noon. 
Galaxies G4-17555 and G4-24078 are situated in slightly denser regions with $\sigma=1.2$ and $\sigma=1.7$, respectively, but these values remain lower than the typical density contrasts found in (proto-)clusters at cosmic noon, where $\delta>2$ \citep[e.g.,][]{Overzier2016}. 
Therefore, we do not expect significant environmental effects on the most recent evolution of \noema{} galaxies. 

\subsection{Subdominant but ongoing formation of galactic bulge}
\label{subsec:bulge}

The formation of bulges has been recognized as a key factor in modulating global star formation, particularly in explaining the bimodality observed in galaxy populations \citep{Drory2007,Bluck2014}.
The bulge components appear ubiquitous in the \noema{} sample, but they are less dominant than the host galaxy.
Our findings indicate that the central bulge typically constitutes 6 to 24 percent of the total light of the galaxy at rest-frame near-IR.
Although the bulge Sérsic index cannot be robustly constrained from the current JWST images, our tests do favor a less peaked (S\'ersic index $n<3$) central bulge compared to classical bulges, which is closer to the argument for a compact nuclear disk in the centers of galaxies \citep{Gadotti2026, LeConte2026}.
We also calculated the central stellar density within the inner 1 kpc region, denoted as $\Sigma_{\rm 1kpc}$, which ranges between $2-5 \times 10^9 \,\text{M}_\odot \text{kpc}^{-2}$. 
Given the stellar masses of \noema{} galaxies, the $\Sigma_{\rm 1kpc}$ values approach the densities of typical compact quenched galaxies at similar redshifts \citep{Barro2017}, as illustrated in Fig.~\ref{fig:mass-size-relation}. 
Such a high stellar density may result from a past central starburst event.
However, unlike compact quenched systems, \noema{} galaxies still exhibit active star formation and large molecular gas reservoirs as revealed by their dust and molecular line emissions.
The presence of the central concentrated dust also explains why the bulge stands out mainly in the rest-frame near-infrared, as illustrated in Figs.~\ref{fig:multi_tracers_co43} and \ref{fig:multi_tracers_co32} by their reddish color in the galactic center.
Therefore, our observations suggest these massive disk galaxies at cosmic noon have not yet developed a full-fledged bulge component, and it is very likely that we are witnessing the ongoing formation of central bulges alongside the secular growth of galaxies \citep{Genzel2008}.

\section{Comparison of molecular gas tracers}
\label{sec:molecular_gas}

\begin{figure*}[htpb]
  \centering
  \includegraphics[width=0.95\textwidth]{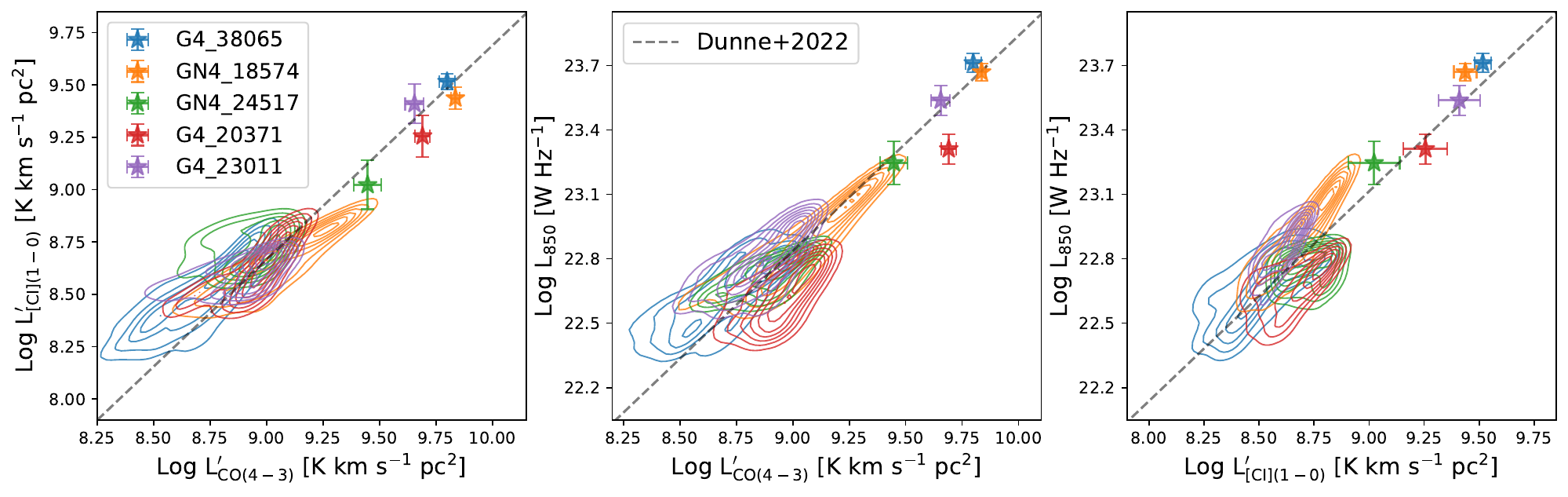}
  \caption{Spatially resolved luminosity analyzed for three molecular gas tracers: \co{}, \ci{}, and dust continuum (scaled to 850$\mu$m). In each panel, the colored stars represent the integrated ratios, while the colored contours show the number density of the resolved luminosities of each tracer from each target. Each resolved measurement is extracted from the resolved map with an aperture that is half the beam size and has an S/N larger than 3 for both lines. The dashed line illustrates the integrated correlation derived from a large compilation of different samples \citep{Dunne2022}. The best-fit correlation for \co{} is scaled from the CO(1-0) line, assuming a global CO ladder correction of $R_{41}=2.4$ \citep{Tacconi2020}. In general, the resolved correlations follow the integrated correlation, but they exhibit nearly twice the scatter, likely due to clumpy, inhomogeneous ISM conditions.}
  \label{fig:integrate_vs_resolved}
\end{figure*}

\begin{figure*}[htpb]
  \centering
  \includegraphics[width=0.95\textwidth]{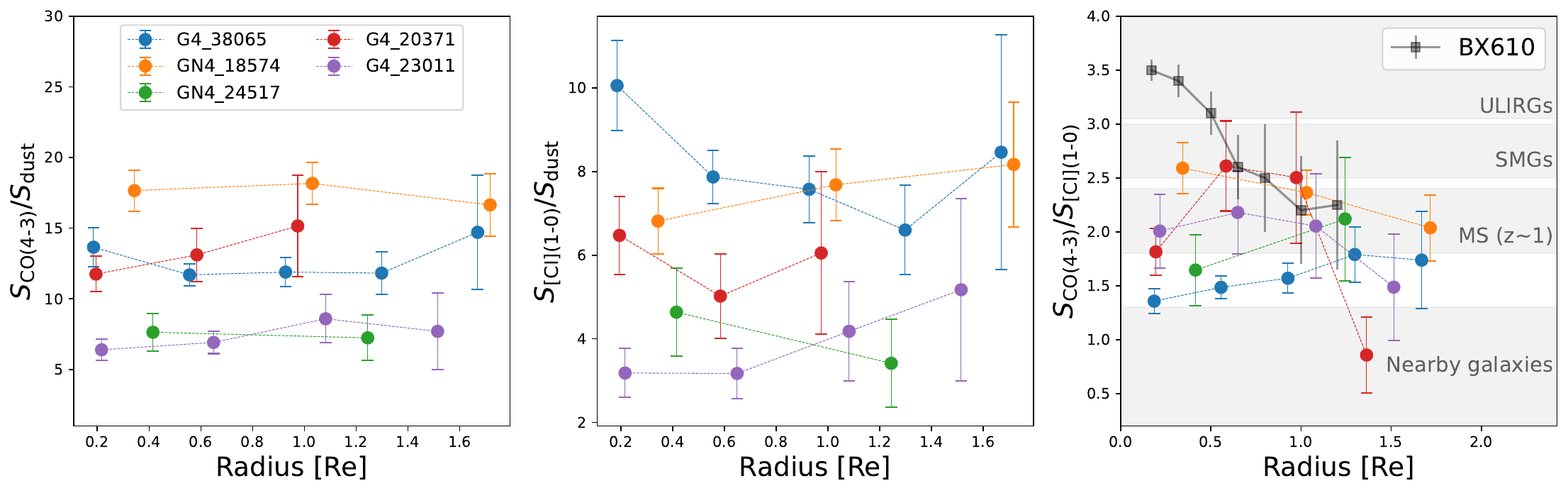}
  \caption{Flux ratios of different cold ISM gas tracers. For the \co{}/\ci{} ratio, we also present measurements from BX610, another MS galaxy observed with ALMA \citep{Arriagada-Neira2025}. The integrated line ratios for different galaxy types are shown as gray-shaded regions. Most \noema{} galaxies show only mild radial variations at the physical resolution of 3$-$6\,kpc, though their line ratios can vary significantly from galaxy to galaxy.}
  \label{fig:flux_ratio}
\end{figure*}

\subsection{The spatially resolved luminosity correlations}

We begin our comparison by breaking down the integrated luminosity correlation into resolved correlations with our \firstgroup{} targets.
The integrated line luminosity is derived from the maximum flux of the curve-of-growth measurements. For the resolved measurements, we used an aperture with a radius equal to half the beam size to sample the entire map and retained only measurements in which both lines were detected above 3$\sigma$ level. 
The results of the mutual comparison among dust, \co{}, and \ci{} are shown in Fig.~\ref{fig:integrate_vs_resolved}.
The colored contours show the number density of the resolved measurements.
To facilitate comparisons with the literature, we converted our measurements to the more commonly used reference. 
For instance, following the convention of \citet{Scoville2016}, we converted the dust luminosity to 850\,\(\mu\)m using the K-correction computed from a modified blackbody with a dust temperature of 25\,K and a dust emissivity of 1.8. 
The overall correction factor for the \firstgroup{} galaxies is approximately \(0.5\pm0.1\), which does not significantly affect the measurements discussed here. 
In accordance with the recommendations of \citet{Tacconi2020}, we adopted a global correction factor of 2.4 to convert \co{} to CO(1-0). 
We also compared our results with the global correlations derived by \citet{Dunne2022}, who compiled the largest current sample of galaxies with CO, \ci{}, and dust emissions.

Unsurprisingly, the integrated measurements align well with the correlations derived from the literature. 
Of particular interest is the consistency observed between \co{} and \ci{}, which follow with the relationships derived from CO(1-0) and CO(2-1).
This suggests that the global CO SLED correction of 2.4 performs reasonably well for these massive MS galaxies \citep[see also][]{Tacconi2020}. 
The correlations with the dust continuum also follow the galactic integrated correlation from \citet{Dunne2022}.
By comparison, the resolved correlations on 3$-$6\,kpc scales exhibit nearly two times larger scatter.
The increased scatter likely reflects the inhomogeneous ISM conditions on these scales, which could be rooted in the generally clumpy structure observed on similar or smaller scales \citep{ForsterSchreiber2011, Guo2012, Claeyssens2023, Kalita2025}, or due to the widespread presence of radial gas flows in these gas-rich SFGs \citep{Genzel2023, Pastras2025, Ubler2024}.

\begin{figure*}[t]
  \centering
  \includegraphics[width=1.0\textwidth]{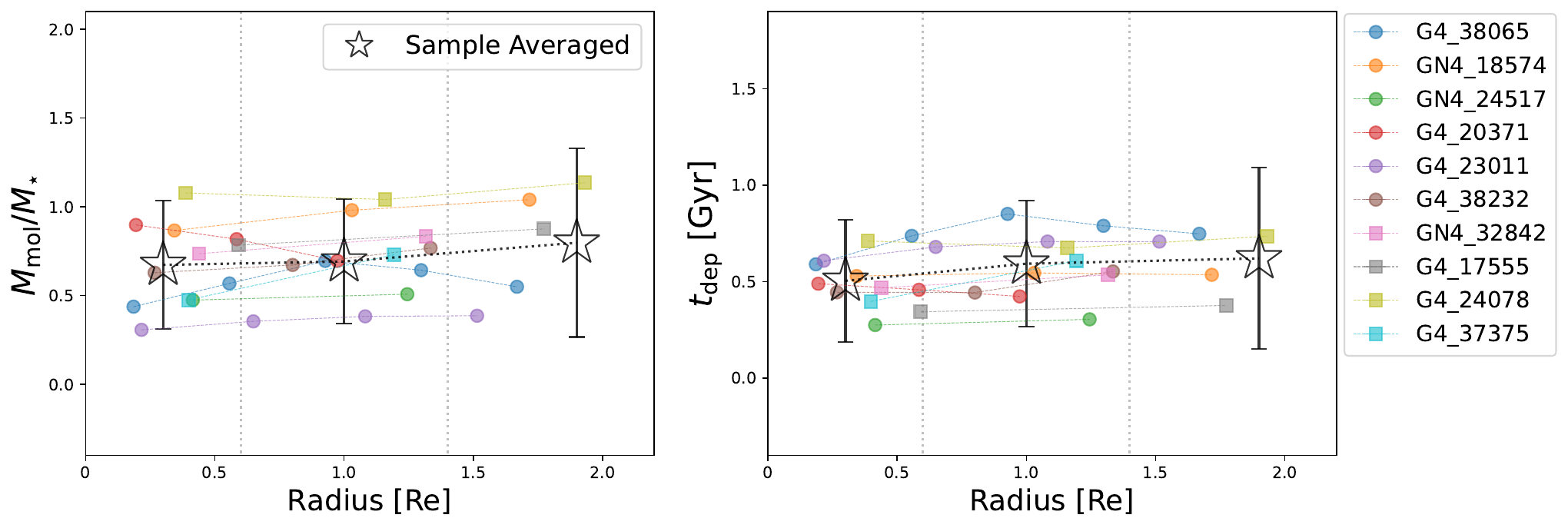}
  \caption{Radial distribution of molecular gas fraction and its depletion timescale. The colored symbols show the measurements for each galaxy. The open stars show the sample-averaged values. For the \firstgroup{} targets, both the molecular gas fraction and the depletion timescale are weighted averages combining all three available tracers. Appendix \ref{appendixfig:fgas_tdep_radial} shows the results for each individual tracer. Different galaxies exhibit diverse molecular gas fractions and depletion timescales, but the overall sample shows a rather flat gas fraction and depletion timescale across the disk, out to 2$\times R_\text{e}$.}
  \label{fig:fgas_tdep_radial}
\end{figure*}

\subsection{The line flux ratios}
To further understand the behavior of different gas tracers as a function of galactic radius, we extracted radial profiles of their flux ratios, as illustrated in Fig.~\ref{fig:flux_ratio}.  
These ratios were derived from the radial profiles of each tracer, sampled at intervals equal to half the beam size.  
A flux ratio was computed only when both tracers were detected with a significance greater than \(3\sigma\) after integrating the signal within each radial annulus.
For our comparison, we used the flux densities directly since all three tracers were observed at similar frequencies. 
The conversion from flux ratio to luminosity ratio requires only a correction for their frequency differences, which is on the order of \(\sim1.1\) and is generally smaller than their error bars.

Among the flux ratios, \co{}/dust shows the flattest distribution. 
In theory, both \co{}/dust and \ci{}/dust can be considered first approximations of the gas-to-dust ratio. 
However, in star-forming regions, \co{} ($T_\text{ext}\sim55$\,K, $n_\text{crit}\sim2\times10^4\,\text{cm}^{-3}$) has nearly two times higher excitation temperature and a order of magnitude higher critical density than \ci{} ($T_\text{ext}\sim24$\,K, $n_\text{crit}\sim5\times10^2\,\text{cm}^{-2}$). 
Thus, \co{}/dust is more representative of the gas-to-dust ratio in denser gas than \ci{}/dust. 
The nearly flat \co{}/dust ratio in our sample indicates that \co{} and the dust continuum around rest-frame $650\mu$m trace similar gas clouds within each galaxy, consistent with simulations by \citet{Cochrane2019}. 
However, the absolute differences between individual galaxies are on the order of \(\sim0.6\)~dex. 
The \ci{}/dust ratio exhibits larger error bars in several galaxies compared to \co{}/dust due to their lower S/N. 
The physical drivers behind the radial variations remain unclear. 
While metallicity has been implicated in the substantial variations of gas-to-dust ratios in nearby resolved galaxies \citep[e.g.,][]{Leroy2011}, it is unlikely to hold for the \noema{} galaxies given their high stellar mass, especially considering that most resolved studies favored rather flat metallicity gradient among massive SFGs \citep[e.g.,][]{Wang2017,Wuyts2016,ForsterSchreiber2018,Ju2025,Fujimoto2025,Lee2026}.
Another more likely reason could be the different excitation mechanisms of \co{} and \ci{} at various radii, as both the CO SLED and [C\,\textsc{i}](2-1)/\ci{} ratios have been found to vary significantly with radiation fields and gas densities \citep{Papadopoulos2012a, Narayanan2014, Valentino2020, Boogaard2020}. 
Such spatially resolved excitation variations have already been observed in MS galaxies at cosmic noon \citep[e.g.,][]{Daddi2015}.

In Fig.~\ref{fig:flux_ratio}, \co{}/\ci{} ratio shows quite some scatter with no clear variation trend across the radii. 
Due to their different critical densities, \co{}/\ci{} serves as a first-order approximation for the dense-to-total molecular gas ratio \citep{Geach2012, Papadopoulos2012}.
The radial variations in the \noema{} targets can thus be attributed to changes in the relative abundance of dense versus diffuse molecular gas, which explains well the decreasing trends found in galaxies such as GN4\_18574 and G4\_23011.  
Other galaxies, such as G4\_38065 and GN4\_24517, show a rather flat profile, suggesting that the dense star-forming clouds also extend into the outer disk.
Integrated \co{}/\ci{} ratios have been widely observed to vary among different SFGs \citep{Valentino2020}, with relatively high values ($>$3) in extreme ULIRGs and lower values of $\sim$1 in nearby star-forming disk galaxies. 
The ratio in the central ($<\!R_\text{e}$) region of all galaxies is largely consistent with the integrated ratios of MS galaxies at cosmic noon, although exhibiting a variation of about 0.3~dex.
We also present the radial profile derived from BX610 \citep[][see Fig.~\ref{fig:flux_ratio}]{Arriagada-Neira2025}, another massive MS galaxy at \(z\sim2.2\).
BX610's central region shows a line ratio similar to that of ULIRGs, likely driven by an enhanced SFR fueled by efficient gas inflow \citep{Genzel2023}.
Theoretically, an enhancement of \ci{} could also suggest a deficit of CO due to cosmic-ray-induced CO destruction in molecular clouds \citep{Papadopoulos2010, Bisbas2015}.
We do not observe any significant decrease in the line ratios at the centers of our \noema{} targets, possibly canceled out by the brighter \co{} emission due to the higher excitation temperature.

\subsection{Spatially resolved gas fraction and depletion timescale}
\label{subsec:radial_fgas_tdep}

With the availability of spatially resolved stellar mass, SFR, and molecular gas mass, we can derive the spatially resolved molecular gas fraction ($\mu_\text{mol}=M_\text{mol}/M_\star$) and gas depletion timescale ($t_\text{dep} = M_\text{mol}/\text{SFR}$). 
The integrated molecular gas fraction and depletion timescale of galaxies have been the subject of extensive discussion over the past decades \citep[e.g.,][]{Genzel2015, Scoville2017, Tacconi2018}. 
Both quantities are found to show a strong dependence on redshift and on their relative distance from the spine of MS. 
The spatially resolved depletion timescales and molecular gas fractions provide a new perspective for quantifying their significance in a galaxy's internal evolution.

We first examined the radial profile of molecular gas fraction in our galaxies.
A high gas fraction is a key factor to support efficient star formation.
The radial profiles of the molecular gas fraction derived from different tracers are shown in Fig.~\ref{appendixfig:fgas_tdep_radial}.
In Fig.~\ref{fig:fgas_tdep_radial}, we show the S/N weighted radial profile considering all the available molecular gas tracers.
Most galaxies exhibit a generally flat molecular gas fraction with radius, with the radial variation well within the 1$\sigma$ measurement uncertainties.
Meanwhile, variations between galaxies are significant, with molecular gas fractions ranging from 30\% to 100\%.
Overall, the observed galaxy-to-galaxy variations are likely due to their different evolutionary stages relative to the MS.
The sample-averaged molecular gas fraction in different $R_\text{e}$ bins, shown as gray stars in Fig.~\ref{fig:fgas_tdep_radial}, also favors a nearly flat trend out to $2\times R_\text{e}$ across our sample.

We also derived the radial profile of the molecular gas depletion timescale, which reflects how long the currently available molecular gas can sustain the current SFR.
Its inverse is also widely referred to as star formation efficiency.
In Fig.~\ref{fig:fgas_tdep_radial}, the \noema{} galaxies exhibit a relatively flat radial profile in molecular gas depletion timescale, with their median timescales range between $0.3-0.8$~Gyr.
Such flat radial depletion timescales indicate that star formation at different radii generally follows the same Kennicutt--Schmidt law, similar to that found in the nearby Universe \citep{Bigiel2011,Muraoka2019}.
It also gives a natural explanation that the galactic integrated molecular gas depletion timescale shows little or no dependence on galaxy size \citep{Tacconi2020}.
The sample-averaged depletion timescale also favors a general flat trend, similar to that of the molecular gas fraction.

\subsection{The caveats of different molecular gas tracers}

CO, including the \co{} and CO(3$-$2) transitions, remains the brightest tracer compared to \ci{} and dust continuum within our observing setups, and it has been consistently detected for all of the \noema{} targets.  
It is well known that typical MS galaxies exhibit a diverse CO SLED \citep{Daddi2015,Valentino2020a,Boogaard2020}, which advises caution when using mid-J CO alone as a tracer of total molecular gas mass.
Meanwhile, $\alpha_\text{CO}$ could also vary at different galactic environments, making it tricky to reveal the true molecular gas mass distribution using a single global conversion.
Nevertheless, the extended spatial distribution and brightness of CO(3-2) or \co{} showcases its advantage to be a powerful tracer for molecular gas kinematics \citepalias{Jolly2026}.

\ci{} has been widely proposed as an alternative to CO(1-0) \cite[e.g.,][]{Papadopoulos2004,Prajapati2026}. 
Our results generally support that \ci{} traces \co{} fairly well on galactic scales of a few kiloparsecs, but their ratio shows a modest radial and galaxy-to-galaxy variation.
Based on our observations alone, we cannot provide a definitive verdict on the reliability of \ci{} in tracing the total molecular gas mass. 
However, one drawback to \ci{} is that it is typically two times weaker than \co{}, requiring significantly longer telescope time to spatially resolve \ci{}, especially for studies focusing on the kinematics or dynamics of molecular gas.

Dust is generally faint in our observations.  
Within the \firstgroup{} targets, the dust continuum surface brightness is approximately 7$-$18 times fainter than that of \co{} around rest-frame 650\,$\mu$m.  
The dust continuum is even fainter compared to CO(3$-$2) at rest-frame 800\,$\mu$m; among the \secondgroup{} targets, only GN4\_32842 shows a clear detection of dust emission.  
The successful detections of dust continuum in \noema{} galaxies clearly demonstrate the powerful wideband capability of NOEMA, where the 15.5~GHz bandwidth compensates for a sensitivity about 6 times better than that for the spectral line.  
The intrinsic brightness of dust emission scales approximately with $\nu^2$ in the Rayleigh--Jeans tail, so observations at higher frequencies could also benefit from brighter dust emission; however, the comparison between literature measurements also shows a wavelength dependence. Measurements at short rest-frame wavelengths are likely biased toward hot dust emission driven by concentrated starburst activity.

Although there is some radial variation among molecular gas tracers from our observations, their overall spatial extent is similar, highlighting their utility for tracing molecular gas distribution. 
More importantly, the ability to use multiple independent molecular gas tracers could minimize the radial variations that arise from applying a uniform conversion factor to any single tracer, thereby making our derived molecular gas fraction and depletion timescale more robust.

\section{The formation of massive SFGs}
\label{sec:implications}

Our observations provide a new, exquisite, deep view of the molecular gas distribution in ten massive MS galaxies at cosmic noon.  
We analyzed the available multiwavelength data, constraining their mass--size relation, bulge growth, and the molecular gas distribution.  
The remaining key question is how this evidence informs our understanding of the general evolution of massive SFGs around cosmic noon.
In this section, we compare our new observations with existing studies of similarly gas-rich but compact starburst galaxies to investigate how different gas accretion modes drive the evolution of massive galaxies.

\subsection{The origin of the cold gas}
\label{subsec:gas_origin}

Direct gas accretion is argued to be the most common channel supporting the growth of SFGs at cosmic noon \citep{Genel2010,Somerville2015,Naab2017,Tacconi2020}.
Within this framework, the available sources of cold gas include gas-rich minor mergers, gas streams from the cosmic web, and cooling flows from recycled gas.
Simulations indicate that such gas accretion tends to be corotating and coplanar with the disk itself \citep[e.g.,][]{Stewart2017,Buck2020}.
Depending on the smoothness of gas accretion, this scenario predicts different consequences for global star formation patterns and quenching \citep{Dekel2009a,Dekel2014}.  
With steadier gas accretion, the accreted gas first settles in the galactic disk and reaches a global equilibrium with the galactic integrated SFR, matching the ``bathtub model" closely \citep[e.g.,][]{Bouche2010,Lilly2013}.
In contrast, more stochastic gas accretion--including gas-rich minor mergers--can be more chaotic. It could significantly increase the gas fraction of the galactic disk over a short timescale and enhance the disk instability \citep{Genzel2008,Zolotov2015,Tacchella2016}.
In \noema{} galaxies, extended gas-rich disks are observed, consistent with feeding by steady gas accretion.  

The halo properties of \noema{} galaxies also support steady cold gas accretion.
Based on the halo mass estimation (see Sect.\ref{subsec:size_evolution}), most \noema{} galaxies have halo mass around $0.8-6\times10^{12}~\text{M}_\odot$. 
According to \citet{Dekel2006}, halos of these masses lie close to the boundary of where cold accretion through the cosmic web is still possible.
Even without the direct cold accretion, the cooling flows from the hot halo can still support their current SFR and mass growth \citep[e.g.,][]{vandeVoort2011,Sultan2026}.

\begin{figure*}[htpb]
  \centering
  \includegraphics[width=\textwidth]{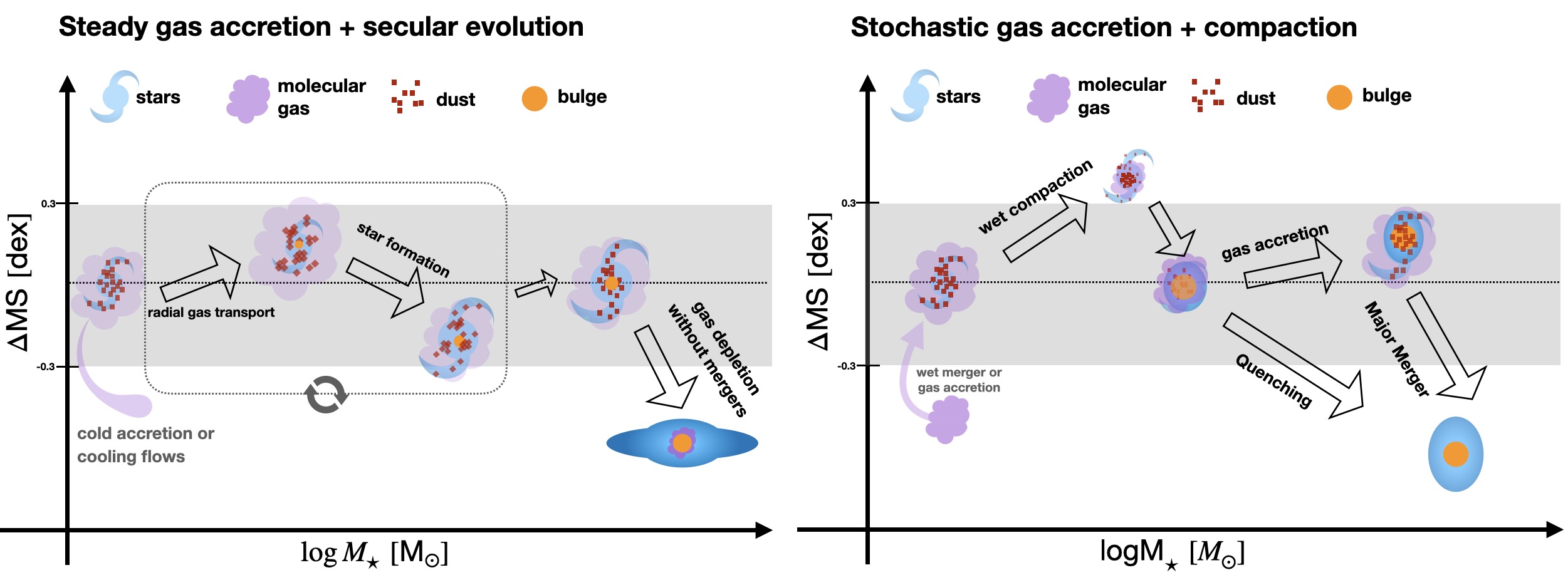}
  \caption{Sketch of the different galaxy evolutionary pathways with two distinct accretion modes. SFGs supported by semi-steady gas accretion typically remain on the MS, with stellar, molecular gas, and dust components of comparable sizes. These galaxies are usually quenched when cold gas accretion terminates, driven by the growth of halo mass. Another subset of SFGs, primarily starburst systems as well as some compact MS galaxies, experience star formation triggered by gas compaction due to stochastic gas accretion or gas-rich mergers. The turbulent gas-rich disk then triggers wet compaction, leading to a concentrated molecular gas and a central starburst, which rapidly depletes the available cold gas and builds a dominant central bulge. The prominence of the central bulge results in rapid quenching and a generally compact morphology. The massive disk galaxies observed by \noema{} align well with the MS galaxies sustained by semi-steady gas accretion.}
  \label{fig:toymodels}
\end{figure*}

\subsection{The transport of cold gas}

Under the assumption of semi-steady gas accretion, accreted gas first settles in the outer galactic disk, which can extend up to $\sim$10~kpc \citep{Keres2005,Dekel2009a,Danovich2015}, comparable to the extent of the observed molecular gas in \noema{} galaxies.
To sustain global star formation, particularly the growth of the central bulge, gas must be transported inward efficiently.

In systems with prominent galactic structures such as spiral arms and stellar bars, gravitational torques represent a widely discussed channel for radial gas flows \citep{Pastras2026}.
Observational evidence of net gas inflow has been extensively reported in the nearby Universe \citep{Wong2004,Davies2009,Sormani2023}. 
The inferred net gas inflow rates range from 0.1$-$10~M$_\odot\text{yr}^{-1}$, with velocities on the order of a few kilometers per second \citep{Schmidt2016,DiTeodoro2021,Lopez-Coba2025}. 
The inflow rates, though often calculated with large uncertainties, are generally comparable to the ongoing SFR in these systems. Similar net inflows have also been reported in cosmological zoom-in simulations \citep[e.g.,][]{Trapp2022}.

At higher redshifts, galactic disks are generally more gas-rich and turbulent, potentially supporting stronger gas inflow rates. 
Prominent disk and bar structures were difficult to confirm in the \textit{HST} era, where only rest-frame optical images were available for galaxies at cosmic noon. 
With the advent of \textit{JWST}, we can now clearly identify and resolve these stellar structures.
All the \noema{} galaxies show prominent spiral arms (see Figs.~\ref{fig:multi_tracers_co43},\ref{fig:multi_tracers_co32}.
A stellar bar is also visible in GN4\_32842, G4\_38232, and possibly in GN4\_24517 and G4\_17555.
Similar spiral arms and bars have also been identified in galaxies up to $z\sim5$ \citep{Costantin2023,Guo2023,LeConte2024,Tsukui2025,EspejoSalcedo2025,Huang2025}, highlighting that their role in driving gas flows may have begun much earlier in cosmic history.
Observationally, \citet{Genzel2023} has shown evidence of gas radial flows in cosmic noon galaxies, in correspondence with bars and spiral arms. 
They reported radial flow velocities up to $\sim$100~km~s$^{-1}$, almost an order of magnitude higher than in local galaxies. 
\citet{Pastras2025} also provide the first evidence of gas inflow in the barred galaxy GN4-32842, indicating rapid molecular gas inflow at a rate comparable to the ongoing SFR \citep[see also][]{Pastras2026}.
\citetalias{Jolly2026} systematically analyzed radial gas flows in all \noema{} galaxies, finding that the ubiquitous presence of gas inflow and outflow, with a net gas inflow rate largely comparable to the integrated galactic SFR.
The generally flat radial profiles of gas fraction and depletion time further support the view that net gas inflow and star formation have reached a local equilibrium within the disk, with no significant local gas accumulation or deficit.

\subsection{Star formation quenching}

As the direct fuel of star formation, the spatial distribution of molecular gas is closely connected to star formation and its quenching processes.
In \noema{} galaxies, the comparable size between cold gas and stellar disk, as well as the flat molecular gas fraction and depletion timescale across the disks, support a balance between the local SFR and the available molecular gas.
The resolved gas-depletion timescales in the center and outer disks are similar to the integrated gas-depletion timescale of typical SFGs, which is around 0.6~Gyr \citep{Tacconi2020}, indicating that these galaxies do not quench their star-forming activity on a timescale shorter than their dynamical timescale.

By comparison, the observed molecular gas in compact starburst galaxies at similar redshifts is mostly concentrated in the center with a much shorter depletion timescale of around 100~Myr \citep{Tacconi2008,Dudzeviciute2020}.
Wet compaction provides a compelling framework for explaining their formation, in which dissipative compaction first leads to a highly concentrated molecular gas reservoir, triggering boosted star formation activity.
This, in turn, quickly depletes the available cold gas, forming a dominant bulge component, and naturally explains the presence of compact, quenched galaxies at similar cosmic epochs \citep{vanDokkum2008,Barro2013,Dekel2014,Zolotov2015,Tacchella2016}.
Galaxies caught in the process of quenching their starburst core could have exhibited a ring-like molecular gas distribution around the central quenched core \citep[e.g.,][]{Liu2023}. 
 
For the quenching of the \noema{} galaxies, we therefore believe their halo plays a more important role during their secular evolution. 
The formation of the bulge component with steady gas transport could be delayed relative to that in compact starburst galaxies, weakening the bulge's role in earlier quenching processes.
Therefore, steady gas accretion may lead to the coexistence of a bulge component and ongoing star formation in the central region.
As the entire galaxy grows, the halo mass will eventually play a leading role in quenching global star formation by preventing or slowing down cold gas accretion and recycling. 
With declining cold gas accretion, the galaxy will ultimately quench due to the depletion of available gas, but the disk morphology is more likely to be preserved.
Without future disruptive merger processes, the \noema{} galaxies could be progenitors of quenched disk galaxies with less dominant bulge components, such as lenticular galaxies with extended stellar disks \citep{Laurikainen2010,Emsellem2011,Saha2018,Moster2020}, and possibly massive red spirals \citep[e.g.,][]{Masters2010,DiTeodoro2021a}.

Active galactic nuclei (AGNs) could also play a distinct role in quenching, depending on the gas feeding conditions.  
During dissipative gas inflow, not only does the bulge grow rapidly, but so does the central supermassive black hole (SMBH), which may trigger strong AGN feedback shortly after bulge formation \citep{Bluck2014,Genzel2014}.  
This powerful AGN feedback can drive outflows that quickly remove available molecular gas or heat and ionize it \citep{Fabian2012,King2015}.  
In cases of steady growth of the central bulge, as we found for \noema{} galaxies, SMBH growth may lag behind that of the host galaxy, resulting in later and weaker AGN feedback.  
Future X-ray follow-up observations and spectral line diagnostics will provide more decisive tests of the importance of AGNs in quenching star formation in these galaxies.

In Fig.~\ref{fig:toymodels}, we illustrate the two different gas accretion scenarios by summarizing all their possible observational signatures.
One notable difference between the two scenarios is the spatial distribution of molecular gas, dust, and stars.
In the steady gas accretion scenario, these three components typically exhibit similar extents, and the galaxy will only wander gently around the canonical MS. 
In contrast, in stochastic gas accretion, galaxies exhibit a more compact star formation pattern traced by their compact dust emission.
During this stage, the galaxy can deviate significantly from the MS.
It should be noted that there is likely no clear distinction between steady and stochastic gas accretion in the dynamic, complex galactic environment.
Throughout the lifetime of a typical MS galaxy, both accretion modes could occur at different evolutionary stages and both modes contribute to the scatter of the MS observed from statistical studies.

Another line of observational evidence for secular evolution driven by steady gas accretion is the prevalence of massive, yet extended, disk galaxies in the early Universe.
\textit{JWST} has uncovered multiple giant disk galaxies ($R_{\rm e}$ > 3\,kpc) at $z>1$ \citep{Fudamoto2022, Wang2025a, Umehata2025,Xiao2025,Ubler2024,Boogaard2026}.
In these systems, disks and bar-like structures are commonly observed, but exhibit a less dominant central bulge, features that align with steady gas accretion being the primary driver of their evolution.
\citet{Jiang2025} systematically analyzed these ``bulge-less'' disk galaxies in the TNG simulation, supporting the idea that these galaxies reside in mildly overdense environments where gas is supplied steadily along cosmic filaments while avoiding disruptive mergers during their evolution.
Overall, depending on the dominance of these two different accretion modes, galaxies can evolve into different morphologies by the end of their star formation.

\section{Conclusions}
\label{sec:conclusions}

In this work, we studied the spatial distribution of molecular gas, traced by CO(3$-$2)/\co{}, [C~\textsc{i}], and dust continuum from the \noema{} survey.
We analyzed the size, radial profiles, and spatial distributions of different galactic components by combining \noema{} data with existing multiband images.
Our main conclusions are as follows:

\begin{enumerate}
  \item The deep, wideband NOEMA observations reveal a widespread presence of extended cold molecular gas and dust in MS galaxies at cosmic noon. Their sizes are largely comparable to those of the stellar disks, in contrast to the size differences observed in more compact DSFGs, such as traditional SMGs.
  \item All \noema{} galaxies exhibit a prominent central stellar concentration in the rest-frame near-IR broadband images. Although they resemble the bulge component of more nearby galaxies, our bulge-disk decomposition indicates a less dominant bulge, contributing only 6--24\% of the host galaxy's light in the near-IR. The bulge regions still show active star formation and a large molecular gas reservoir, supporting an ongoing bulge formation.
  \item Different molecular gas tracers are generally consistent with one another when observed at comparable sensitivities. In massive SFGs, the mid-J CO transition, such as CO(3$-$2) and \co{}, is generally brighter than other tracers at similar frequencies, such as [C\,\textsc{i}] and dust continuum, and thus remains the most practical tracer for mapping the spatial distribution and kinematics of molecular gas. With the implementation of ultra-wide bandwidth in modern interferometers, dust continuum proves to be another promising tool.
  \item The spatially resolved \co{}, \ci{}, and dust continuum generally follow the galactic integrated correlations, but show roughly two times larger scatter. The increased scatter indicates that ISM conditions begin to deviate from the globally averaged conditions already on scales of 3-6~kpc. This is consistent with the idea that the ISM in these star-forming regions has substructures on kiloparsec scales, possibly reflecting ongoing gas accretion and clumpy star formation. 
  \item The resolved molecular gas fraction and molecular gas depletion timescale generally show a mild radial variation in most of these extended MS galaxies at cosmic noon, and it is largely consistent with their galactic integrated ones. It indicates that steady gas accretion and radial gas transport support the buildup of extended disk galaxies, in contrast to the stochastic gas accretion and wet compaction inferred from the compact starburst galaxies at a similar cosmic time.
\end{enumerate}

We highlight the crucial role of increased bandwidth and long baselines provided by NOEMA for such a large survey.
The ongoing dual-band implementation will enable simultaneous sampling of multiple molecular lines to constrain their excitation temperatures and of multiband dust emission to model the dust SED, making it even more efficient to fully map the ISM properties of galaxies across cosmic time.
The more sensitive ALMA is also allocating significant time to these cosmic noon MS galaxies, and it is already moving toward implementing a bandwidth upgrade \citep{Carpenter2023}.
These efforts will provide a significantly larger sample to fully understand the formation and evolution of typical massive disk galaxies since cosmic noon.

\begin{acknowledgements}
We are grateful to the anonymous referee's useful feedback.
JC is very grateful for the inspiring discussions with Rachel Cochrane, Sandro Tacchella, Padelis Papadopoulos, Zhi-Yu Zhang, and Ryota Ikeda.
This work is based on observations carried out under project number L19MD with the IRAM NOEMA Interferometer.
We thank the IRAM staff and the IRAM Partnership (MPG in Germany, CNRS in France, and IGN in Spain) for their hard work and support of the NOEMA upgrade at IRAM, which made the \noema{} project possible.
JC, NMFS, CB, GT, JMES, and LL   acknowledge funding by the European Union (ERC Advanced Grant GALPHYS, 101055023).  
GM and HÜ acknowledge funding by the European Union (ERC APEX, 101164796).
Views and opinions expressed are, however, those of the author(s) only and do not necessarily reflect those of the European Union or the European Research Council. Neither the European Union nor the granting authority can be held responsible for them.
HÜ thanks the Max Planck Society for support through the Lise Meitner Excellence Program.
TN acknowledges the support of the Deutsche Forschungsgemeinschaft (DFG, German Research Foundation) under Germany’s Excellence Strategy - EXC-2094 - 390783311 of the DFG Cluster of Excellence ‘ORIGINS’. 
SGB and AU acknowledge support from the Spanish grant PID2022-138560NB-I00, funded by MCIN/AEI/10.13039/501100011033/FEDER, EU. 
This work is based in part on observations made with the NASA/ESA/CSA James Webb Space Telescope. 
The data were obtained from the Mikulski Archive for Space Telescopes at the Space Telescope Science Institute, which is operated by the Association of Universities for Research in Astronomy, Inc., under NASA contract NAS 5-03127 for JWST. 
These observations are associated with programs \#1181, \#1264, \#1345, \#1895, \#2234, \#2514, \#2926, \#3577, \#3794, \#5398, \#5407, and \#6434 and can be accessed via DOI:10.17909/8r87-nv19. 
This research is also based on observations made with the NASA/ESA Hubble Space Telescope obtained from the Space Telescope Science Institute, which is operated by the Association of Universities for Research in Astronomy, Inc., under NASA contract NAS 5–26555. These observations are associated with the CANDELS Multi-Cycle Treasury Program \citep{Grogin2011,Koekemoer2011}.
Some of the data products presented herein were retrieved from the Dawn \textit{JWST} Archive (DJA). DJA is an initiative of the Cosmic Dawn Center (DAWN), which is funded by the Danish National Research Foundation under grant DNRF140.

\end{acknowledgements}

%
%

 \bibliographystyle{aa} 
 \bibliography{aa60515-26} 

@article{Dekel2006,
  ids = {Ouchi2006},
  title = {Galaxy Bimodality Due to Cold Flows and Shock Heating},
  author = {Dekel, Avishai and Birnboim, Yuval},
  year = 2006,
  month = may,
  journal = {MNRAS},
  volume = {368},
  pages = {2--20},
  issn = {0035-8711},
  doi = {10.1111/j.1365-2966.2006.10145.x},
  urldate = {2023-02-13}
}

@article{Overzier2016,
  title = {The Realm of the Galaxy Protoclusters. {{A}} Review},
  author = {Overzier, Roderik A.},
  year = 2016,
  month = nov,
  journal = {A\&ARv},
  volume = {24},
  number = {1},
  pages = {14},
  issn = {0935-4956},
  doi = {10.1007/s00159-016-0100-3},
  urldate = {2023-02-13},
  langid = {english}
}

@article{Chen2023,
       author = {{Chen}, Jianhang and {Ivison}, R.~J. and {Zwaan}, Martin A. and {Smail}, Ian and {Klitsch}, Anne and {P{\'e}roux}, C{\'e}line and {Popping}, Gerg{\"o} and {Biggs}, Andrew D. and {Szakacs}, Roland and {Hamanowicz}, Aleksandra and {Lagos}, Claudia},
        title = "{ALMACAL IX: Multiband ALMA survey for dusty star-forming galaxies and the resolved fractions of the cosmic infrared background}",
      journal = {MNRAS},
         year = 2023,
        month = jan,
       volume = {518},
       number = {1},
        pages = {1378-1397},
          doi = {10.1093/mnras/stac2989},
archivePrefix = {arXiv},
       eprint = {2210.09329},
 primaryClass = {astro-ph.GA},
       adsurl = {https://ui.adsabs.harvard.edu/abs/2023MNRAS.518.1378C}
}

@article{Tacconi2008,
  title = {Submillimeter {{Galaxies}} at {\emph{z}} {$\sim$} 2: {{Evidence}} for {{Major Mergers}} and {{Constraints}} on {{Lifetimes}}, {{IMF}}, and {{CO}}-{{H}} {\textsubscript{2}} {{Conversion Factor}}},
  shorttitle = {Submillimeter {{Galaxies}} at {\emph{z}} {$\sim$} 2},
  author = {Tacconi, L. J. and Genzel, R. and Smail, I. and Neri, R. and Chapman, S. C. and Ivison, R. J. and Blain, A. and Cox, P. and Omont, A. and Bertoldi, F. and Greve, T. and F{\"o}rster Schreiber, N. M. and Genel, S. and Lutz, D. and Swinbank, A. M. and Shapley, A. E. and Erb, D. K. and Cimatti, A. and Daddi, E. and Baker, A. J.},
  year = 2008,
  month = jun,
  journal = {ApJ},
  volume = {680},
  number = {1},
  pages = {246--262},
  issn = {0004-637X, 1538-4357},
  doi = {10.1086/587168},
  url = {https://iopscience.iop.org/article/10.1086/587168},
  urldate = {2023-08-18},
  langid = {english}
}

@article{Alaghband-Zadeh2013,
  title = {Using [{{C I}}] to Probe the Interstellar Medium in z {$\sim$} 2.5 Sub-Millimeter Galaxies},
  author = {{Alaghband-Zadeh}, S. and Chapman, S. C. and Swinbank, A. M. and Smail, Ian and Danielson, A. L. R. and Decarli, R. and Ivison, R. J. and Meijerink, R. and Weiss, A. and {van der Werf}, {\relax Paul}. P.},
  year = 2013,
  month = oct,
  journal = {MNRAS},
  volume = {435},
  pages = {1493--1510},
  publisher = {OUP},
  issn = {0035-8711},
  doi = {10.1093/mnras/stt1390},
  urldate = {2025-02-12}
}

@article{Aravena2010,
  title = {Cold {{Molecular Gas}} in {{Massive}}, {{Star-forming Disk Galaxies}} at z = 1.5},
  author = {Aravena, M. and Carilli, C. and Daddi, E. and Wagg, J. and Walter, F. and Riechers, D. and Dannerbauer, H. and Morrison, G. E. and Stern, D. and Krips, M.},
  year = 2010,
  month = jul,
  journal = {ApJ},
  volume = {718},
  pages = {177--183},
  issn = {0004-637X},
  doi = {10.1088/0004-637X/718/1/177},
  urldate = {2025-08-21}
}

@article{Aravena2014,
  title = {{{CO}}(1-0) Line Imaging of Massive Star-Forming Disc Galaxies at {{Z}}=1.5-2.2},
  author = {Aravena, M. and Hodge, J. A. and Wagg, J. and Carilli, C. L. and Daddi, E. and Dannerbauer, H. and Lentati, L. and Riechers, D. A. and Sargent, M. and Walter, F.},
  year = 2014,
  month = jul,
  journal = {MNRAS},
  volume = {442},
  pages = {558--564},
  issn = {0035-8711},
  doi = {10.1093/mnras/stu838},
  urldate = {2025-08-21}
}

@article{Baker2022,
  title = {The {{ALMaQUEST}} Survey {{IX}}: {{The}} Nature of the Resolved Star Forming Main Sequence},
  shorttitle = {The {{ALMaQUEST}} Survey {{IX}}},
  author = {Baker, William M. and Maiolino, Roberto and Bluck, Asa F. L. and Lin, Lihwai and Ellison, Sara L. and Belfiore, Francesco and Pan, Hsi-An and Thorp, Mallory},
  year = 2022,
  month = mar,
  journal = {MNRAS},
  volume = {510},
  pages = {3622--3628},
  publisher = {OUP},
  issn = {0035-8711},
  doi = {10.1093/mnras/stab3672},
  urldate = {2025-03-15}
}

@article{Barro2013,
  title = {{{CANDELS}}: {{The Progenitors}} of {{Compact Quiescent Galaxies}} at z \textasciitilde{} 2},
  shorttitle = {{{CANDELS}}},
  author = {Barro, Guillermo and Faber, S. M. and {P{\'e}rez-Gonz{\'a}lez}, Pablo G. and Koo, David C. and Williams, Christina C. and Kocevski, Dale D. and Trump, Jonathan R. and Mozena, Mark and McGrath, Elizabeth and {van der Wel}, Arjen and Wuyts, Stijn and Bell, Eric F. and Croton, Darren J. and Ceverino, Daniel and Dekel, Avishai and Ashby, M. L. N. and Cheung, Edmond and Ferguson, Henry C. and Fontana, Adriano and Fang, Jerome and Giavalisco, Mauro and Grogin, Norman A. and Guo, Yicheng and Hathi, Nimish P. and Hopkins, Philip F. and Huang, Kuang-Han and Koekemoer, Anton M. and Kartaltepe, Jeyhan S. and Lee, Kyoung-Soo and Newman, Jeffrey A. and Porter, Lauren A. and Primack, Joel R. and Ryan, Russell E. and Rosario, David and Somerville, Rachel S. and Salvato, Mara and Hsu, Li-Ting},
  year = 2013,
  month = mar,
  journal = {ApJ},
  volume = {765},
  pages = {104},
  issn = {0004-637X},
  doi = {10.1088/0004-637X/765/2/104},
  urldate = {2025-08-26}
}

@article{Barro2017,
  title = {Structural and {{Star-forming Relations}} since z {$\sim$} 3: {{Connecting Compact Star-forming}} and {{Quiescent Galaxies}}},
  shorttitle = {Structural and {{Star-forming Relations}} since z {$\sim$} 3},
  author = {Barro, Guillermo and Faber, S. M. and Koo, David C. and Dekel, Avishai and Fang, Jerome J. and Trump, Jonathan R. and {P{\'e}rez-Gonz{\'a}lez}, Pablo G. and Pacifici, Camilla and Primack, Joel R. and Somerville, Rachel S. and Yan, Haojing and Guo, Yicheng and Liu, Fengshan and Ceverino, Daniel and Kocevski, Dale D. and McGrath, Elizabeth},
  year = 2017,
  month = may,
  journal = {ApJ},
  volume = {840},
  pages = {47},
  issn = {0004-637X},
  doi = {10.3847/1538-4357/aa6b05},
  urldate = {2025-09-11}
}

@article{Bisbas2015,
  title = {Effective {{Destruction}} of {{CO}} by {{Cosmic Rays}}: {{Implications}} for {{Tracing H2 Gas}} in the {{Universe}}},
  shorttitle = {Effective {{Destruction}} of {{CO}} by {{Cosmic Rays}}},
  author = {Bisbas, Thomas G. and Papadopoulos, Padelis P. and Viti, Serena},
  year = 2015,
  month = apr,
  journal = {ApJ},
  volume = {803},
  pages = {37},
  issn = {0004-637X},
  doi = {10.1088/0004-637X/803/1/37},
  urldate = {2025-08-20}
}

@article{Bluck2014,
  title = {Bulge Mass Is King: {{The}} Dominant Role of the Bulge in Determining the Fraction of Passive Galaxies in the {{Sloan Digital Sky Survey}}},
  shorttitle = {Bulge Mass Is King},
  author = {Bluck, Asa F. L. and Mendel, J. Trevor and Ellison, Sara L. and Moreno, Jorge and Simard, Luc and Patton, David R. and Starkenburg, Else},
  year = 2014,
  month = jun,
  journal = {MNRAS},
  volume = {441},
  pages = {599--629},
  publisher = {OUP},
  issn = {0035-8711},
  doi = {10.1093/mnras/stu594},
  urldate = {2025-08-11}
}

@article{Bolatto2013,
  title = {The {{CO-to-H}} {$_2$} {{Conversion Factor}}},
  author = {Bolatto, Alberto D. and Wolfire, Mark and Leroy, Adam K.},
  year = 2013,
  month = aug,
  journal = {ARA\&A},
  volume = {51},
  number = {1},
  pages = {207--268},
  issn = {0066-4146, 1545-4282},
  doi = {10.1146/annurev-astro-082812-140944},
  urldate = {2022-06-01},
  langid = {english}
}

@article{Bolatto2015,
  title = {{{HIGH-RESOLUTION IMAGING OF PHIBSS}} {\emph{z}} {$\sim$} 2 {{MAIN-SEQUENCE GALAXIES IN CO}} {{{\emph{J}}}} = 1 {$\rightarrow$} 0},
  author = {Bolatto, A. D. and Warren, S. R. and Leroy, A. K. and Tacconi, L. J. and Bouch{\'e}, N. and Schreiber, N. M. F{\"o}rster and Genzel, R. and Cooper, M. C. and Fisher, D. B. and Combes, F. and {Garc{\'i}a-Burillo}, S. and Burkert, A. and Bournaud, F. and Weiss, A. and Saintonge, A. and Wuyts, S. and Sternberg, A.},
  year = 2015,
  month = aug,
  journal = {ApJ},
  volume = {809},
  number = {2},
  pages = {175},
  issn = {1538-4357},
  doi = {10.1088/0004-637X/809/2/175},
  urldate = {2024-03-13},
  langid = {english}
}

@article{Boogaard2020,
  title = {The {{ALMA Spectroscopic Survey}} in the {{Hubble Ultra Deep Field}}: {{CO Excitation}} and {{Atomic Carbon}} in {{Star-forming Galaxies}} at z = 1-3},
  shorttitle = {The {{ALMA Spectroscopic Survey}} in the {{Hubble Ultra Deep Field}}},
  author = {Boogaard, Leindert A. and {van der Werf}, Paul and Weiss, Axel and Popping, Gerg{\"o} and Decarli, Roberto and Walter, Fabian and Aravena, Manuel and Bouwens, Rychard and Riechers, Dominik and {Gonz{\'a}lez-L{\'o}pez}, Jorge and Smail, Ian and Carilli, Chris and Kaasinen, Melanie and Daddi, Emanuele and Cox, Pierre and {D{\'i}az-Santos}, Tanio and Inami, Hanae and Cortes, Paulo C. and Wagg, Jeff},
  year = 2020,
  month = oct,
  journal = {ApJ},
  volume = {902},
  pages = {109},
  issn = {0004-637X},
  doi = {10.3847/1538-4357/abb82f},
  urldate = {2025-08-21}
}

@article{Boquien2019,
  title = {{{CIGALE}}: A Python {{Code Investigating GALaxy Emission}}},
  shorttitle = {{{CIGALE}}},
  author = {Boquien, M. and Burgarella, D. and Roehlly, Y. and Buat, V. and Ciesla, L. and Corre, D. and Inoue, A. K. and Salas, H.},
  year = 2019,
  month = feb,
  journal = {A\&A},
  volume = {622},
  pages = {A103},
  issn = {0004-6361},
  doi = {10.1051/0004-6361/201834156},
  urldate = {2025-06-16}
}

@article{Bothwell2017,
  title = {{{ALMA}} Observations of Atomic Carbon in z {$\sim$} 4 Dusty Star-Forming Galaxies},
  author = {Bothwell, M. S. and Aguirre, J. E. and Aravena, M. and Bethermin, M. and Bisbas, T. G. and Chapman, S. C. and De Breuck, C. and Gonzalez, A. H. and Greve, T. R. and Hezaveh, Y. and Ma, J. and Malkan, M. and Marrone, D. P. and Murphy, E. J. and Spilker, J. S. and Strandet, M. and Vieira, J. D. and Wei{\ss}, A.},
  year = 2017,
  month = apr,
  journal = {MNRAS},
  volume = {466},
  pages = {2825--2841},
  issn = {0035-8711},
  doi = {10.1093/mnras/stw3270},
  urldate = {2025-08-21}
}

@article{Bourne2019,
  title = {The Relationship between Dust and [{{C I}}] at z = 1 and Beyond},
  author = {Bourne, N. and Dunlop, J. S. and Simpson, J. M. and Rowlands, K. E. and Geach, J. E. and McLeod, D. J.},
  year = 2019,
  month = jan,
  journal = {MNRAS},
  volume = {482},
  pages = {3135--3161},
  issn = {0035-8711},
  doi = {10.1093/mnras/sty2773},
  urldate = {2025-08-20}
}

@article{CalistroRivera2018,
  title = {Resolving the {{ISM}} at the {{Peak}} of {{Cosmic Star Formation}} with {{ALMA}}: {{The Distribution}} of {{CO}} and {{Dust Continuum}} in z {$\sim$} 2.5 {{Submillimeter Galaxies}}},
  shorttitle = {Resolving the {{ISM}} at the {{Peak}} of {{Cosmic Star Formation}} with {{ALMA}}},
  author = {Calistro Rivera, Gabriela and Hodge, J. A. and Smail, Ian and Swinbank, A. M. and Weiss, A. and Wardlow, J. L. and Walter, F. and Rybak, M. and Chen, Chian-Chou and Brandt, W. N. and Coppin, K. and Cunha, E. Da and Dannerbauer, H. and Greve, T. R. and Karim, A. and Knudsen, K. K. and Schinnerer, E. and Simpson, J. M. and Venemans, B. and Werf, P. P. Van Der},
  year = 2018,
  month = aug,
  journal = {ApJ},
  volume = {863},
  number = {1},
  pages = {56},
  issn = {0004-637X, 1538-4357},
  doi = {10.3847/1538-4357/aacffa},
  urldate = {2025-02-06},
  langid = {english}
}

@misc{Carpenter2023,
  title = {The {{ALMA Wideband Sensitivity Upgrade}}},
  author = {Carpenter, John and Brogan, Crystal and Iono, Daisuke and Mroczkowski, Tony},
  year = 2023,
  month = feb,
  publisher = {arXiv},
  address = {eprint: arXiv:2211.00195},
  doi = {10.48550/arXiv.2211.00195},
  urldate = {2025-08-18}
}

@article{Chabrier2003,
  title = {Galactic {{Stellar}} and {{Substellar Initial Mass Function}}},
  author = {Chabrier, Gilles},
  year = 2003,
  month = jul,
  journal = {PASP},
  volume = {115},
  pages = {763--795},
  issn = {0004-6280},
  doi = {10.1086/376392},
  urldate = {2025-08-19}
}

@article{Chartab2020,
  title = {Large-Scale {{Structures}} in the {{CANDELS Fields}}: {{The Role}} of the {{Environment}} in {{Star Formation Activity}}},
  shorttitle = {Large-Scale {{Structures}} in the {{CANDELS Fields}}},
  author = {Chartab, Nima and Mobasher, Bahram and Darvish, Behnam and Finkelstein, Steve and Guo, Yicheng and Kodra, Dritan and Lee, Kyoung-Soo and Newman, Jeffrey A. and Pacifici, Camilla and Papovich, Casey and Sattari, Zahra and Shahidi, Abtin and Dickinson, Mark E. and Faber, Sandra M. and Ferguson, Henry C. and Giavalisco, Mauro and Jafariyazani, Marziye},
  year = 2020,
  month = feb,
  journal = {ApJ},
  volume = {890},
  number = {1},
  pages = {7},
  publisher = {The American Astronomical Society},
  issn = {0004-637X},
  doi = {10.3847/1538-4357/ab61fd},
  urldate = {2025-08-20},
  langid = {english}
}

@article{Chen2017,
  title = {A {{Spatially Resolved Study}} of {{Cold Dust}}, {{Molecular Gas}}, {{H}} Ii {{Regions}}, and {{Stars}} in the z = 2.12 {{Submillimeter Galaxy ALESS67}}.1},
  author = {Chen, Chian-Chou and Hodge, J. A. and Smail, Ian and Swinbank, A. M. and Walter, Fabian and Simpson, J. M. and Rivera, Gabriela Calistro and Bertoldi, F. and Brandt, W. N. and Chapman, S. C. and Cunha, Elisabete Da and Dannerbauer, H. and Breuck, C. De and Harrison, C. M. and Ivison, R. J. and Karim, A. and Knudsen, K. K. and Wardlow, J. L. and Wei{\ss}, A. and Werf, P. P. Van Der},
  year = 2017,
  month = sep,
  journal = {ApJ},
  volume = {846},
  number = {2},
  pages = {108},
  issn = {0004-637X, 1538-4357},
  doi = {10.3847/1538-4357/aa863a},
  urldate = {2025-02-06},
  langid = {english}
}

@article{Cochrane2019,
  title = {Predictions for the Spatial Distribution of the Dust Continuum Emission in \$\textbackslash boldsymbol \textbraceleft 1\textbackslash,\textbackslash lt\textbackslash, {{Z}}\textbackslash,\textbackslash lt\textbackslash, 5\textbraceright\$ Star-Forming Galaxies},
  author = {Cochrane, R K and Hayward, C C and {Angl{\'e}s-Alc{\'a}zar}, D and Lotz, J and Parsotan, T and Ma, X and Kere{\v s}, D and Feldmann, R and {Faucher-Gigu{\`e}re}, C A and Hopkins, P F},
  year = 2019,
  month = sep,
  journal = {MNRAS},
  volume = {488},
  number = {2},
  pages = {1779--1789},
  issn = {0035-8711},
  doi = {10.1093/mnras/stz1736},
  urldate = {2025-02-06}
}

@article{Costantin2023,
  title = {A {{Milky Way-like}} Barred Spiral Galaxy at a Redshift of 3},
  author = {Costantin, Luca and {P{\'e}rez-Gonz{\'a}lez}, Pablo G. and Guo, Yuchen and Buttitta, Chiara and Jogee, Shardha and Bagley, Micaela B. and Barro, Guillermo and Kartaltepe, Jeyhan S. and Koekemoer, Anton M. and Cabello, Cristina and Corsini, Enrico Maria and {M{\'e}ndez-Abreu}, Jairo and {de la Vega}, Alexander and Iyer, Kartheik G. and Bisigello, Laura and Cheng, Yingjie and Morelli, Lorenzo and Arrabal Haro, Pablo and Buitrago, Fernando and Cooper, M. C. and Dekel, Avishai and Dickinson, Mark and Finkelstein, Steven L. and Giavalisco, Mauro and Holwerda, Benne W. and {Huertas-Company}, Marc and Lucas, Ray A. and Papovich, Casey and Pirzkal, Nor and Seill{\'e}, Lise-Marie and {Vega-Ferrero}, Jes{\'u}s and Wuyts, Stijn and Yung, L. Y. Aaron},
  year = 2023,
  month = nov,
  journal = {Nature},
  volume = {623},
  pages = {499--501},
  issn = {0028-0836},
  doi = {10.1038/s41586-023-06636-x},
  urldate = {2025-09-10}
}

@article{Daddi2007,
  title = {Multiwavelength {{Study}} of {{Massive Galaxies}} at {{Z}}\textasciitilde 2. {{I}}. {{Star Formation}} and {{Galaxy Growth}}},
  author = {Daddi, E. and Dickinson, M. and Morrison, G. and Chary, R. and Cimatti, A. and Elbaz, D. and Frayer, D. and Renzini, A. and Pope, A. and Alexander, D. M. and Bauer, F. E. and Giavalisco, M. and Huynh, M. and Kurk, J. and Mignoli, M.},
  year = 2007,
  month = nov,
  journal = {ApJ},
  volume = {670},
  pages = {156--172},
  issn = {0004-637X},
  doi = {10.1086/521818},
  urldate = {2025-08-22}
}

@article{Daddi2010,
  title = {Very {{High Gas Fractions}} and {{Extended Gas Reservoirs}} in z = 1.5 {{Disk Galaxies}}},
  author = {Daddi, E. and Bournaud, F. and Walter, F. and Dannerbauer, H. and Carilli, C. L. and Dickinson, M. and Elbaz, D. and Morrison, G. E. and Riechers, D. and Onodera, M. and Salmi, F. and Krips, M. and Stern, D.},
  year = 2010,
  month = apr,
  journal = {ApJ},
  volume = {713},
  pages = {686--707},
  issn = {0004-637X},
  doi = {10.1088/0004-637X/713/1/686},
  urldate = {2025-08-19}
}

@article{Daddi2015,
  title = {{{CO}} Excitation of Normal Star-Forming Galaxies out to z = 1.5 as Regulated by the Properties of Their Interstellar Medium},
  author = {Daddi, E. and Dannerbauer, H. and Liu, D. and Aravena, M. and Bournaud, F. and Walter, F. and Riechers, D. and Magdis, G. and Sargent, M. and B{\'e}thermin, M. and Carilli, C. and Cibinel, A. and Dickinson, M. and Elbaz, D. and Gao, Y. and Gobat, R. and Hodge, J. and Krips, M.},
  year = 2015,
  month = may,
  journal = {A\&A},
  volume = {577},
  pages = {A46},
  issn = {0004-6361},
  doi = {10.1051/0004-6361/201425043},
  urldate = {2025-08-21}
}

@article{Danovich2015,
  title = {Four Phases of Angular-Momentum Buildup in High-z Galaxies: {{From}} Cosmic-Web Streams through an Extended Ring to Disc and Bulge},
  shorttitle = {Four Phases of Angular-Momentum Buildup in High-z Galaxies},
  author = {Danovich, Mark and Dekel, Avishai and Hahn, Oliver and Ceverino, Daniel and Primack, Joel},
  year = 2015,
  month = may,
  journal = {MNRAS},
  volume = {449},
  pages = {2087--2111},
  issn = {0035-8711},
  doi = {10.1093/mnras/stv270},
  urldate = {2025-09-10}
}

@article{Davies2009,
  title = {{{STELLAR AND MOLECULAR GAS KINEMATICS OF NGC}} 1097: {{INFLOW DRIVEN BY A NUCLEAR SPIRAL}}*},
  shorttitle = {{{STELLAR AND MOLECULAR GAS KINEMATICS OF NGC}} 1097},
  author = {Davies, R. I. and Maciejewski, W. and Hicks, E. K. S. and Tacconi, L. J. and Genzel, R. and Engel, H.},
  year = 2009,
  month = aug,
  journal = {ApJ},
  volume = {702},
  number = {1},
  pages = {114},
  publisher = {The American Astronomical Society},
  issn = {0004-637X},
  doi = {10.1088/0004-637X/702/1/114},
  urldate = {2025-09-10},
  langid = {english}
}

@article{Dekel2009a,
  title = {Formation of {{Massive Galaxies}} at {{High Redshift}}: {{Cold Streams}}, {{Clumpy Disks}}, and {{Compact Spheroids}}},
  shorttitle = {Formation of {{Massive Galaxies}} at {{High Redshift}}},
  author = {Dekel, Avishai and Sari, Re'em and Ceverino, Daniel},
  year = 2009,
  month = sep,
  journal = {ApJ},
  volume = {703},
  pages = {785--801},
  issn = {0004-637X},
  doi = {10.1088/0004-637X/703/1/785},
  urldate = {2025-08-26}
}

@article{Dekel2014,
  title = {Wet Disc Contraction to Galactic Blue Nuggets and Quenching to Red Nuggets},
  author = {Dekel, A. and Burkert, A.},
  year = 2014,
  month = feb,
  journal = {MNRAS},
  volume = {438},
  pages = {1870--1879},
  issn = {0035-8711},
  doi = {10.1093/mnras/stt2331},
  urldate = {2025-08-17}
}

@inproceedings{Dickinson2003,
  title = {The {{Great Observatories Origins Deep Survey}}},
  booktitle = {Mass {{Galaxies Low High Redshift}}},
  author = {Dickinson, Mark and Giavalisco, Mauro and {GOODS Team}},
  year = 2003,
  month = jan,
  pages = {324},
  address = {eprint: arXiv:astro-ph/0204213},
  doi = {10.1007/10899892_78},
  urldate = {2025-08-14}
}

@article{DiTeodoro2021,
  title = {Radial {{Motions}} and {{Radial Gas Flows}} in {{Local Spiral Galaxies}}},
  author = {Di Teodoro, Enrico M. and Peek, J. E. G.},
  year = 2021,
  month = dec,
  journal = {ApJ},
  volume = {923},
  number = {2},
  pages = {220},
  publisher = {The American Astronomical Society},
  issn = {0004-637X},
  doi = {10.3847/1538-4357/ac2cbd},
  urldate = {2025-09-10},
  langid = {english}
}

@article{Drory2007,
  title = {A {{Connection}} between {{Bulge Properties}} and the {{Bimodality}} of {{Galaxies}}},
  author = {Drory, Niv and Fisher, David B.},
  year = 2007,
  month = aug,
  journal = {ApJ},
  volume = {664},
  pages = {640--649},
  issn = {0004-637X},
  doi = {10.1086/519441},
  urldate = {2025-09-11}
}

@article{Dunne2022,
  title = {Dust, {{CO}}, and [{{C I}}]: {{Cross-calibration}} of Molecular Gas Mass Tracers in Metal-Rich Galaxies across Cosmic Time},
  shorttitle = {Dust, {{CO}}, and [{{C I}}]},
  author = {Dunne, L. and Maddox, S. J. and Papadopoulos, P. P. and Ivison, R. J. and Gomez, H. L.},
  year = 2022,
  month = nov,
  journal = {MNRAS},
  volume = {517},
  pages = {962--999},
  publisher = {OUP},
  issn = {0035-8711},
  doi = {10.1093/mnras/stac2098},
  urldate = {2025-02-12}
}

@article{Eales2012,
  title = {Can {{Dust Emission}} Be {{Used}} to {{Estimate}} the {{Mass}} of the {{Interstellar Medium}} in {{Galaxies}}---{{A Pilot Project}} with the {{Herschel Reference Survey}}},
  author = {Eales, Stephen and Smith, Matthew W. L. and Auld, Robbie and Baes, Maarten and Bendo, George J. and Bianchi, Simone and Boselli, Alessandro and Ciesla, Laure and Clements, David and Cooray, Asantha and Cortese, Luca and Davies, Jon and De Looze, Ilse and Galametz, Maud and Gear, Walter and Gentile, Gianfranco and Gomez, Haley and Fritz, Jacopo and Hughes, Tom and Madden, Suzanne and Magrini, Laura and Pohlen, Michael and Spinoglio, Luigi and Verstappen, Joris and Vlahakis, Catherine and Wilson, Christine D.},
  year = 2012,
  month = dec,
  journal = {ApJ},
  volume = {761},
  pages = {168},
  issn = {0004-637X},
  doi = {10.1088/0004-637X/761/2/168},
  urldate = {2025-08-17}
}

@article{Elbaz2007,
  title = {The Reversal of the Star Formation-Density Relation in the Distant Universe},
  author = {Elbaz, D. and Daddi, E. and Le Borgne, D. and Dickinson, M. and Alexander, D. M. and Chary, R. -R. and Starck, J. -L. and Brandt, W. N. and Kitzbichler, M. and MacDonald, E. and Nonino, M. and Popesso, P. and Stern, D. and Vanzella, E.},
  year = 2007,
  month = jun,
  journal = {A\&A},
  volume = {468},
  pages = {33--48},
  issn = {0004-6361},
  doi = {10.1051/0004-6361:20077525},
  urldate = {2025-08-22}
}

@article{ForsterSchreiber2011,
  title = {Constraints on the {{Assembly}} and {{Dynamics}} of {{Galaxies}}. {{II}}. {{Properties}} of {{Kiloparsec-scale Clumps}} in {{Rest-frame Optical Emission}} of z \textasciitilde{} 2 {{Star-forming Galaxies}}},
  author = {F{\"o}rster Schreiber, N. M. and Shapley, A. E. and Genzel, R. and Bouch{\'e}, N. and Cresci, G. and Davies, R. and Erb, D. K. and Genel, S. and Lutz, D. and Newman, S. and Shapiro, K. L. and Steidel, C. C. and Sternberg, A. and Tacconi, L. J.},
  year = 2011,
  month = sep,
  journal = {ApJ},
  volume = {739},
  pages = {45},
  issn = {0004-637X},
  doi = {10.1088/0004-637X/739/1/45},
  urldate = {2025-08-21}
}

@article{ForsterSchreiber2020,
  title = {Star-{{Forming Galaxies}} at {{Cosmic Noon}}},
  author = {F{\"o}rster Schreiber, Natascha M. and Wuyts, Stijn},
  year = 2020,
  journal = {ARA\&A},
  volume = {58},
  number = {1},
  pages = {661--725},
  doi = {10.1146/annurev-astro-032620-021910},
  urldate = {2021-02-26}
}

@article{Fudamoto2022,
  title = {Red {{Spiral Galaxies}} at {{Cosmic Noon Unveiled}} in the {{First JWST Image}}},
  author = {Fudamoto, Yoshinobu and Inoue, Akio K. and Sugahara, Yuma},
  year = 2022,
  month = oct,
  journal = {ApJL},
  volume = {938},
  number = {2},
  pages = {L24},
  issn = {2041-8205, 2041-8213},
  doi = {10.3847/2041-8213/ac982b},
  urldate = {2025-08-23},
  langid = {english}
}

@article{Geach2012,
  title = {Molecular and {{Atomic Line Surveys}} of {{Galaxies}}. {{I}}. {{The Dense}}, {{Star-Forming Gas Phase}} as a {{Beacon}}},
  author = {Geach, James E. and Papadopoulos, Padelis P.},
  year = 2012,
  month = oct,
  journal = {ApJ},
  volume = {757},
  pages = {156},
  issn = {0004-637X},
  doi = {10.1088/0004-637X/757/2/156},
  urldate = {2025-08-20}
}

@article{Genzel2008,
  title = {From {{Rings}} to {{Bulges}}: {{Evidence}} for {{Rapid Secular Galaxy Evolution}} at z \textasciitilde{} 2 from {{Integral Field Spectroscopy}} in the {{SINS Survey}}},
  shorttitle = {From {{Rings}} to {{Bulges}}},
  author = {Genzel, R. and Burkert, A. and Bouch{\'e}, N. and Cresci, G. and F{\"o}rster Schreiber, N. M. and Shapley, A. and Shapiro, K. and Tacconi, L. J. and Buschkamp, P. and Cimatti, A. and Daddi, E. and Davies, R. and Eisenhauer, F. and Erb, D. K. and Genel, S. and Gerhard, O. and Hicks, E. and Lutz, D. and Naab, T. and Ott, T. and Rabien, S. and Renzini, A. and Steidel, C. C. and Sternberg, A. and Lilly, S. J.},
  year = 2008,
  month = nov,
  journal = {ApJ},
  volume = {687},
  pages = {59--77},
  issn = {0004-637X},
  doi = {10.1086/591840},
  urldate = {2025-08-27}
}

@article{Genzel2014,
  title = {The {{SINS}}/{{zC-SINF Survey}} of z \textasciitilde{} 2 {{Galaxy Kinematics}}: {{Evidence}} for {{Gravitational Quenching}}},
  shorttitle = {The {{SINS}}/{{zC-SINF Survey}} of z \textasciitilde{} 2 {{Galaxy Kinematics}}},
  author = {Genzel, R. and F{\"o}rster Schreiber, N. M. and Lang, P. and Tacchella, S. and Tacconi, L. J. and Wuyts, S. and Bandara, K. and Burkert, A. and Buschkamp, P. and Carollo, C. M. and Cresci, G. and Davies, R. and Eisenhauer, F. and Hicks, E. K. S. and Kurk, J. and Lilly, S. J. and Lutz, D. and Mancini, C. and Naab, T. and Newman, S. and Peng, Y. and Renzini, A. and Shapiro Griffin, K. and Sternberg, A. and Vergani, D. and Wisnioski, E. and Wuyts, E. and Zamorani, G.},
  year = 2014,
  month = apr,
  journal = {ApJ},
  volume = {785},
  pages = {75},
  issn = {0004-637X},
  doi = {10.1088/0004-637X/785/1/75},
  urldate = {2025-08-26}
}

@article{Genzel2015,
  title = {Combined {{CO}} and {{Dust Scaling Relations}} of {{Depletion Time}} and {{Molecular Gas Fractions}} with {{Cosmic Time}}, {{Specific Star-formation Rate}}, and {{Stellar Mass}}},
  author = {Genzel, R. and Tacconi, L. J. and Lutz, D. and Saintonge, A. and Berta, S. and Magnelli, B. and Combes, F. and {Garc{\'i}a-Burillo}, S. and Neri, R. and Bolatto, A. and Contini, T. and Lilly, S. and Boissier, J. and Boone, F. and Bouch{\'e}, N. and Bournaud, F. and Burkert, A. and Carollo, M. and Colina, L. and Cooper, M. C. and Cox, P. and Feruglio, C. and F{\"o}rster Schreiber, N. M. and Freundlich, J. and {Gracia-Carpio}, J. and Juneau, S. and Kovac, K. and Lippa, M. and Naab, T. and Salome, P. and Renzini, A. and Sternberg, A. and Walter, F. and Weiner, B. and Weiss, A. and Wuyts, S.},
  year = 2015,
  month = feb,
  journal = {ApJ},
  volume = {800},
  pages = {20},
  publisher = {IOP},
  issn = {0004-637X},
  doi = {10.1088/0004-637X/800/1/20},
  urldate = {2025-06-11}
}

@article{Genzel2023,
  title = {Evidence for {{Large-scale}}, {{Rapid Gas Inflows}} in z 2 {{Star-forming Disks}}},
  author = {Genzel, R. and Jolly, J. -B. and Liu, D. and Price, S. H. and Lee, L. L. and F{\"o}rster Schreiber, N. M. and Tacconi, L. J. and {Herrera-Camus}, R. and Barfety, C. and Burkert, A. and Cao, Y. and Davies, R. I. and Dekel, A. and Lee, M. M. and Lutz, D. and Naab, T. and Neri, R. and Nestor Shachar, A. and Pastras, S. and Pulsoni, C. and Renzini, A. and Schuster, K. and Shimizu, T. T. and Stanley, F. and Sternberg, A. and {\"U}bler, H.},
  year = 2023,
  month = nov,
  journal = {ApJ},
  volume = {957},
  pages = {48},
  publisher = {IOP},
  issn = {0004-637X},
  doi = {10.3847/1538-4357/acef1a},
  urldate = {2025-03-14}
}

@article{Grogin2011,
  title = {{{CANDELS}}: {{The Cosmic Assembly Near-infrared Deep Extragalactic Legacy Survey}}},
  shorttitle = {{{CANDELS}}},
  author = {Grogin, Norman A. and Kocevski, Dale D. and Faber, S. M. and Ferguson, Henry C. and Koekemoer, Anton M. and Riess, Adam G. and Acquaviva, Viviana and Alexander, David M. and Almaini, Omar and Ashby, Matthew L. N. and Barden, Marco and Bell, Eric F. and Bournaud, Fr{\'e}d{\'e}ric and Brown, Thomas M. and Caputi, Karina I. and Casertano, Stefano and Cassata, Paolo and Castellano, Marco and Challis, Peter and Chary, Ranga-Ram and Cheung, Edmond and Cirasuolo, Michele and Conselice, Christopher J. and Roshan Cooray, Asantha and Croton, Darren J. and Daddi, Emanuele and Dahlen, Tomas and Dav{\'e}, Romeel and {de Mello}, Du{\'i}lia F. and Dekel, Avishai and Dickinson, Mark and Dolch, Timothy and Donley, Jennifer L. and Dunlop, James S. and Dutton, Aaron A. and Elbaz, David and Fazio, Giovanni G. and Filippenko, Alexei V. and Finkelstein, Steven L. and Fontana, Adriano and Gardner, Jonathan P. and Garnavich, Peter M. and Gawiser, Eric and Giavalisco, Mauro and Grazian, Andrea and Guo, Yicheng and Hathi, Nimish P. and H{\"a}ussler, Boris and Hopkins, Philip F. and Huang, Jia-Sheng and Huang, Kuang-Han and Jha, Saurabh W. and Kartaltepe, Jeyhan S. and Kirshner, Robert P. and Koo, David C. and Lai, Kamson and Lee, Kyoung-Soo and Li, Weidong and Lotz, Jennifer M. and Lucas, Ray A. and Madau, Piero and McCarthy, Patrick J. and McGrath, Elizabeth J. and McIntosh, Daniel H. and McLure, Ross J. and Mobasher, Bahram and Moustakas, Leonidas A. and Mozena, Mark and Nandra, Kirpal and Newman, Jeffrey A. and Niemi, Sami-Matias and Noeske, Kai G. and Papovich, Casey J. and Pentericci, Laura and Pope, Alexandra and Primack, Joel R. and Rajan, Abhijith and Ravindranath, Swara and Reddy, Naveen A. and Renzini, Alvio and Rix, Hans-Walter and Robaina, Aday R. and Rodney, Steven A. and Rosario, David J. and Rosati, Piero and Salimbeni, Sara and Scarlata, Claudia and Siana, Brian and Simard, Luc and Smidt, Joseph and Somerville, Rachel S. and Spinrad, Hyron and Straughn, Amber N. and Strolger, Louis-Gregory and Telford, Olivia and Teplitz, Harry I. and Trump, Jonathan R. and {van der Wel}, Arjen and Villforth, Carolin and Wechsler, Risa H. and Weiner, Benjamin J. and Wiklind, Tommy and Wild, Vivienne and Wilson, Grant and Wuyts, Stijn and Yan, Hao-Jing and Yun, Min S.},
  year = 2011,
  month = dec,
  journal = {ApJS},
  volume = {197},
  pages = {35},
  publisher = {IOP},
  issn = {0067-0049},
  doi = {10.1088/0067-0049/197/2/35},
  urldate = {2025-08-09}
}

@article{Gullberg2019,
  title = {An {{ALMA}} Survey of the {{SCUBA-2 Cosmology Legacy Survey UKIDSS}}/{{UDS}} Field: {{High-resolution}} Dust Continuum Morphologies and the Link between Sub-Millimetre Galaxies and Spheroid Formation},
  shorttitle = {An {{ALMA}} Survey of the {{SCUBA-2 Cosmology Legacy Survey UKIDSS}}/{{UDS}} Field},
  author = {Gullberg, B. and Smail, Ian and Swinbank, A. M. and Dudzevi{\v c}i{\=u}t{\.e}, U. and Stach, S. M. and Thomson, A. P. and Almaini, O. and Chen, C. C. and Conselice, C. and Cooke, E. A. and Farrah, D. and Ivison, R. J. and Maltby, D. and Micha{\l}owski, M. J. and Simpson, J. M. and Scott, D. and Wardlow, J. L. and Weiss, A.},
  year = 2019,
  month = dec,
  journal = {MNRAS},
  volume = {490},
  number = {4},
  pages = {4956--4974},
  issn = {0035-8711},
  doi = {10.1093/mnras/stz2835},
  urldate = {2025-06-11},
  langid = {english}
}

@article{Guo2012,
  title = {{{MULTI-WAVELENGTH VIEW OF KILOPARSEC-SCALE CLUMPS IN STAR-FORMING GALAXIES AT}} z {$\sim$} 2},
  author = {Guo, Yicheng and Giavalisco, Mauro and Ferguson, Henry C. and Cassata, Paolo and Koekemoer, Anton M.},
  year = 2012,
  month = sep,
  journal = {ApJ},
  volume = {757},
  number = {2},
  pages = {120},
  publisher = {The American Astronomical Society},
  issn = {0004-637X},
  doi = {10.1088/0004-637X/757/2/120},
  urldate = {2025-08-21},
  langid = {english}
}

@article{Guo2023,
  title = {First {{Look}} at z {$>$} 1 {{Bars}} in the {{Rest-frame Near-infrared}} with {{JWST Early CEERS Imaging}}},
  author = {Guo, Yuchen and Jogee, Shardha and Finkelstein, Steven L. and Chen, Zilei and Wise, Eden and Bagley, Micaela B. and Barro, Guillermo and Wuyts, Stijn and Kocevski, Dale D. and Kartaltepe, Jeyhan S. and McGrath, Elizabeth J. and Ferguson, Henry C. and Mobasher, Bahram and Giavalisco, Mauro and Lucas, Ray A. and Zavala, Jorge A. and Lotz, Jennifer M. and Grogin, Norman A. and {Huertas-Company}, Marc and {Vega-Ferrero}, Jes{\'u}s and Hathi, Nimish P. and Arrabal Haro, Pablo and Dickinson, Mark and Koekemoer, Anton M. and Papovich, Casey and Pirzkal, Nor and Yung, L. Y. Aaron and Backhaus, Bren E. and Bell, Eric F. and Calabr{\`o}, Antonello and Cleri, Nikko J. and Coogan, Rosemary T. and Cooper, M. C. and Costantin, Luca and Croton, Darren and Davis, Kelcey and Dekel, Avishai and Franco, Maximilien and Gardner, Jonathan P. and Holwerda, Benne W. and Hutchison, Taylor A. and Pandya, Viraj and {P{\'e}rez-Gonz{\'a}lez}, Pablo G. and Ravindranath, Swara and Rose, Caitlin and Trump, Jonathan R. and {de la Vega}, Alexander and Wang, Weichen},
  year = 2023,
  month = mar,
  journal = {ApJ},
  volume = {945},
  pages = {L10},
  issn = {0004-637X},
  doi = {10.3847/2041-8213/acacfb},
  urldate = {2025-09-10}
}

@article{Harrington2021,
  title = {Turbulent {{Gas}} in {{Lensed Planck-selected Starbursts}} at z {$\sim$} 1-3.5},
  author = {Harrington, Kevin C. and Weiss, Axel and Yun, Min S. and Magnelli, Benjamin and Sharon, C. E. and Leung, T. K. D. and Vishwas, A. and Wang, Q. D. and Frayer, D. T. and {Jim{\'e}nez-Andrade}, E. F. and Liu, D. and Garc{\'i}a, P. and {Romano-D{\'i}az}, E. and Frye, B. L. and Jarugula, S. and B{\u a}descu, T. and Berman, D. and Dannerbauer, H. and {D{\'i}az-S{\'a}nchez}, A. and Grassitelli, L. and Kamieneski, P. and Kim, W. J. and Kirkpatrick, A. and Lowenthal, J. D. and Messias, H. and Puschnig, J. and Stacey, G. J. and Torne, P. and Bertoldi, F.},
  year = 2021,
  month = feb,
  journal = {ApJ},
  volume = {908},
  pages = {95},
  issn = {0004-637X},
  doi = {10.3847/1538-4357/abcc01},
  urldate = {2025-08-20}
}

@article{Hodge2015,
  title = {The {{Kiloparsec-scale Star Formation Law}} at {{Redshift}} 4: {{Widespread}}, {{Highly Efficient Star Formation}} in the {{Dust-obscured Starburst Galaxy GN20}}},
  shorttitle = {The {{Kiloparsec-scale Star Formation Law}} at {{Redshift}} 4},
  author = {Hodge, J. A. and Riechers, D. and Decarli, R. and Walter, F. and Carilli, C. L. and Daddi, E. and Dannerbauer, H.},
  year = 2015,
  month = jan,
  journal = {ApJ},
  volume = {798},
  pages = {L18},
  publisher = {IOP},
  issn = {0004-637X},
  doi = {10.1088/2041-8205/798/1/L18},
  urldate = {2025-03-14}
}

@article{Hodge2020,
  title = {High-Redshift Star Formation in the {{Atacama}} Large Millimetre/Submillimetre Array Era},
  author = {Hodge, J. A. and {da Cunha}, E.},
  year = 2020,
  month = dec,
  journal = {R. Soc. open sci.},
  volume = {7},
  number = {12},
  pages = {200556},
  issn = {2054-5703},
  doi = {10.1098/rsos.200556},
  urldate = {2022-12-29},
  langid = {english}
}

@article{Hodge2025,
  title = {{{ALESS-JWST}}: {{Joint}} ({{Sub}})Kiloparsec {{JWST}} and {{ALMA Imaging}} of z \textasciitilde{} 3 {{Submillimeter Galaxies Reveals Heavily Obscured Bulge Formation Events}}},
  shorttitle = {{{ALESS-JWST}}},
  author = {Hodge, J. A. and {da Cunha}, E. and Kendrew, S. and Li, J. and Smail, I. and Westoby, B. A. and Nayak, O. and Swinbank, A. M. and Chen, C. -C. and Walter, F. and {van der Werf}, P. and Cracraft, M. and Battisti, A. and Brandt, W. N. and Calistro Rivera, G. and Chapman, S. C. and Cox, P. and Dannerbauer, H. and Decarli, R. and Frias Castillo, M. and Greve, T. R. and Knudsen, K. K. and Leslie, S. and Menten, K. M. and Rybak, M. and Schinnerer, E. and Wardlow, J. L. and Weiss, A.},
  year = 2025,
  month = jan,
  journal = {ApJ},
  volume = {978},
  pages = {165},
  publisher = {IOP},
  issn = {0004-637X},
  doi = {10.3847/1538-4357/ad9a52},
  urldate = {2025-05-20}
}

@article{Huang2025,
  title = {Large Gas Inflow Driven by a Matured Galactic Bar in the Early {{Universe}}},
  author = {Huang, Shuo and Kawabe, Ryohei and Umehata, Hideki and Kohno, Kotaro and Tamura, Yoichi and Saito, Toshiki},
  year = 2025,
  month = may,
  journal = {Nature},
  volume = {641},
  number = {8064},
  pages = {861--865},
  publisher = {Nature Publishing Group},
  issn = {1476-4687},
  doi = {10.1038/s41586-025-08914-2},
  urldate = {2025-09-10},
  copyright = {2025 The Author(s), under exclusive licence to Springer Nature Limited},
  langid = {english}
}

@article{Jiao2019,
  title = {Resolved {{Neutral Carbon Emission}} in {{Nearby Galaxies}}: [{{C I}}] {{Lines}} as {{Total Molecular Gas Tracers}}},
  shorttitle = {Resolved {{Neutral Carbon Emission}} in {{Nearby Galaxies}}},
  author = {Jiao, Qian and Zhao, Yinghe and Lu, Nanyao and Gao, Yu and Salak, Dragan and Zhu, Ming and Zhang, Zhi-Yu and Jiang, Xue-Jian and Tan, Qinghua},
  year = 2019,
  month = aug,
  journal = {ApJ},
  volume = {880},
  pages = {133},
  issn = {0004-637X},
  doi = {10.3847/1538-4357/ab29ed},
  urldate = {2022-11-27}
}

@article{Kaasinen2020,
  title = {A {{Comparison}} of the {{Stellar}}, {{CO}}, and {{Dust-continuum Emission}} from {{Three Star-forming HUDF Galaxies}} at z {$\sim$} 2},
  author = {Kaasinen, Melanie and Walter, Fabian and Novak, Mladen and Neeleman, Marcel and Smail, Ian and Boogaard, Leindert and {da Cunha}, Elisabete and Weiss, Axel and Liu, Daizhong and Decarli, Roberto and Popping, Gerg{\"o} and {Diaz-Santos}, Tanio and Cort{\'e}s, Paulo and Aravena, Manuel and {van der Werf}, Paul and Riechers, Dominik and Inami, Hanae and Hodge, Jacqueline A. and Rix, Hans-Walter and Cox, Pierre},
  year = 2020,
  month = aug,
  journal = {ApJ},
  volume = {899},
  pages = {37},
  publisher = {IOP},
  issn = {0004-637X},
  doi = {10.3847/1538-4357/aba438},
  urldate = {2025-03-14}
}

@article{Kennicutt1998,
  title = {{{STAR FORMATION IN GALAXIES ALONG THE HUBBLE SEQUENCE}}},
  author = {Kennicutt, Robert C.},
  year = 1998,
  month = sep,
  journal = {ARA\&A},
  volume = {36},
  number = {1},
  pages = {189--231},
  issn = {0066-4146, 1545-4282},
  doi = {10.1146/annurev.astro.36.1.189},
  urldate = {2020-08-26},
  langid = {english}
}

@article{Keres2005,
  title = {How Do Galaxies Get Their Gas?},
  author = {Kere{\v s}, Du{\v s}an and Katz, Neal and Weinberg, David H. and Dav{\'e}, Romeel},
  year = 2005,
  month = oct,
  journal = {Mon Not R Astron Soc},
  volume = {363},
  number = {1},
  pages = {2--28},
  issn = {0035-8711},
  doi = {10.1111/j.1365-2966.2005.09451.x},
  urldate = {2025-08-21}
}

@article{Koekemoer2011,
  title = {{{CANDELS}}: {{The Cosmic Assembly Near-infrared Deep Extragalactic Legacy Survey}}---{{The Hubble Space Telescope Observations}}, {{Imaging Data Products}}, and {{Mosaics}}},
  shorttitle = {{{CANDELS}}},
  author = {Koekemoer, Anton M. and Faber, S. M. and Ferguson, Henry C. and Grogin, Norman A. and Kocevski, Dale D. and Koo, David C. and Lai, Kamson and Lotz, Jennifer M. and Lucas, Ray A. and McGrath, Elizabeth J. and Ogaz, Sara and Rajan, Abhijith and Riess, Adam G. and Rodney, Steve A. and Strolger, Louis and Casertano, Stefano and Castellano, Marco and Dahlen, Tomas and Dickinson, Mark and Dolch, Timothy and Fontana, Adriano and Giavalisco, Mauro and Grazian, Andrea and Guo, Yicheng and Hathi, Nimish P. and Huang, Kuang-Han and {van der Wel}, Arjen and Yan, Hao-Jing and Acquaviva, Viviana and Alexander, David M. and Almaini, Omar and Ashby, Matthew L. N. and Barden, Marco and Bell, Eric F. and Bournaud, Fr{\'e}d{\'e}ric and Brown, Thomas M. and Caputi, Karina I. and Cassata, Paolo and Challis, Peter J. and Chary, Ranga-Ram and Cheung, Edmond and Cirasuolo, Michele and Conselice, Christopher J. and Roshan Cooray, Asantha and Croton, Darren J. and Daddi, Emanuele and Dav{\'e}, Romeel and {de Mello}, Duilia F. and {de Ravel}, Loic and Dekel, Avishai and Donley, Jennifer L. and Dunlop, James S. and Dutton, Aaron A. and Elbaz, David and Fazio, Giovanni G. and Filippenko, Alexei V. and Finkelstein, Steven L. and Frazer, Chris and Gardner, Jonathan P. and Garnavich, Peter M. and Gawiser, Eric and Gruetzbauch, Ruth and Hartley, Will G. and H{\"a}ussler, Boris and Herrington, Jessica and Hopkins, Philip F. and Huang, Jia-Sheng and Jha, Saurabh W. and Johnson, Andrew and Kartaltepe, Jeyhan S. and Khostovan, Ali A. and Kirshner, Robert P. and Lani, Caterina and Lee, Kyoung-Soo and Li, Weidong and Madau, Piero and McCarthy, Patrick J. and McIntosh, Daniel H. and McLure, Ross J. and McPartland, Conor and Mobasher, Bahram and Moreira, Heidi and Mortlock, Alice and Moustakas, Leonidas A. and Mozena, Mark and Nandra, Kirpal and Newman, Jeffrey A. and Nielsen, Jennifer L. and Niemi, Sami and Noeske, Kai G. and Papovich, Casey J. and Pentericci, Laura and Pope, Alexandra and Primack, Joel R. and Ravindranath, Swara and Reddy, Naveen A. and Renzini, Alvio and Rix, Hans-Walter and Robaina, Aday R. and Rosario, David J. and Rosati, Piero and Salimbeni, Sara and Scarlata, Claudia and Siana, Brian and Simard, Luc and Smidt, Joseph and Snyder, Diana and Somerville, Rachel S. and Spinrad, Hyron and Straughn, Amber N. and Telford, Olivia and Teplitz, Harry I. and Trump, Jonathan R. and Vargas, Carlos and Villforth, Carolin and Wagner, Cory R. and Wandro, Pat and Wechsler, Risa H. and Weiner, Benjamin J. and Wiklind, Tommy and Wild, Vivienne and Wilson, Grant and Wuyts, Stijn and Yun, Min S.},
  year = 2011,
  month = dec,
  journal = {ApJS},
  volume = {197},
  pages = {36},
  publisher = {IOP},
  issn = {0067-0049},
  doi = {10.1088/0067-0049/197/2/36},
  urldate = {2025-08-09}
}

@article{Laurikainen2010,
  title = {Photometric Scaling Relations of Lenticular and Spiral Galaxies},
  author = {Laurikainen, E. and Salo, H. and Buta, R. and Knapen, J. H. and Comer{\'o}n, S.},
  year = 2010,
  month = jun,
  journal = {Mon Not R Astron Soc},
  volume = {405},
  number = {2},
  pages = {1089--1118},
  issn = {0035-8711},
  doi = {10.1111/j.1365-2966.2010.16521.x},
  urldate = {2025-09-12}
}

@article{Leauthaud2012,
  title = {New {{Constraints}} on the {{Evolution}} of the {{Stellar-to-dark Matter Connection}}: {{A Combined Analysis}} of {{Galaxy-Galaxy Lensing}}, {{Clustering}}, and {{Stellar Mass Functions}} from z = 0.2 to z =1},
  shorttitle = {New {{Constraints}} on the {{Evolution}} of the {{Stellar-to-dark Matter Connection}}},
  author = {Leauthaud, Alexie and Tinker, Jeremy and Bundy, Kevin and Behroozi, Peter S. and Massey, Richard and Rhodes, Jason and George, Matthew R. and Kneib, Jean-Paul and Benson, Andrew and Wechsler, Risa H. and Busha, Michael T. and Capak, Peter and Cort{\^e}s, Marina and Ilbert, Olivier and Koekemoer, Anton M. and Le F{\`e}vre, Oliver and Lilly, Simon and McCracken, Henry J. and Salvato, Mara and Schrabback, Tim and Scoville, Nick and Smith, Tristan and Taylor, James E.},
  year = 2012,
  month = jan,
  journal = {ApJ},
  volume = {744},
  pages = {159},
  publisher = {IOP},
  issn = {0004-637X},
  doi = {10.1088/0004-637X/744/2/159},
  urldate = {2025-07-25}
}

@article{LeConte2024,
  title = {A {{JWST}} Investigation into the Bar Fraction at Redshifts 1 {$\leq$} z {$\leq$} 3},
  author = {Le Conte, Zoe A. and Gadotti, Dimitri A. and Ferreira, Leonardo and Conselice, Christopher J. and {de S{\'a}-Freitas}, Camila and Kim, Taehyun and Neumann, Justus and Fragkoudi, Francesca and Athanassoula, E. and Adams, Nathan J.},
  year = 2024,
  month = may,
  journal = {MNRAS},
  volume = {530},
  pages = {1984--2000},
  issn = {0035-8711},
  doi = {10.1093/mnras/stae921},
  urldate = {2025-09-10}
}

@article{Leroy2011,
  title = {The {{CO-to-H2 Conversion Factor}} from {{Infrared Dust Emission}} across the {{Local Group}}},
  author = {Leroy, Adam K. and Bolatto, Alberto and Gordon, Karl and Sandstrom, Karin and Gratier, Pierre and Rosolowsky, Erik and Engelbracht, Charles W. and Mizuno, Norikazu and Corbelli, Edvige and Fukui, Yasuo and Kawamura, Akiko},
  year = 2011,
  month = aug,
  journal = {ApJ},
  volume = {737},
  pages = {12},
  publisher = {IOP},
  issn = {0004-637X},
  doi = {10.1088/0004-637X/737/1/12},
  urldate = {2025-06-23}
}

@article{Lilly2013,
  title = {Gas {{Regulation}} of {{Galaxies}}: {{The Evolution}} of the {{Cosmic Specific Star Formation Rate}}, the {{Metallicity-Mass-Star-formation Rate Relation}}, and the {{Stellar Content}} of {{Halos}}},
  shorttitle = {Gas {{Regulation}} of {{Galaxies}}},
  author = {Lilly, Simon J. and Carollo, C. Marcella and Pipino, Antonio and Renzini, Alvio and Peng, Yingjie},
  year = 2013,
  month = aug,
  journal = {ApJ},
  volume = {772},
  pages = {119},
  publisher = {IOP},
  issn = {0004-637X},
  doi = {10.1088/0004-637X/772/2/119},
  urldate = {2025-03-14}
}

@article{Liu2023,
  title = {An 600 Pc {{View}} of the {{Strongly Lensed}}, {{Massive Main-sequence Galaxy J0901}}: {{A Baryon-dominated}}, {{Thick Turbulent Rotating Disk}} with a {{Clumpy Cold Gas Ring}} at z = 2.259},
  shorttitle = {An 600 Pc {{View}} of the {{Strongly Lensed}}, {{Massive Main-sequence Galaxy J0901}}},
  author = {Liu, Daizhong and F{\"o}rster Schreiber, N. M. and Genzel, R. and Lutz, D. and Price, S. H. and Lee, L. L. and Baker, Andrew J. and Burkert, A. and Coogan, R. T. and Davies, R. I. and Davies, R. L. and {Herrera-Camus}, R. and Kodama, Tadayuki and Lee, Minju M. and Nestor, A. and Pulsoni, C. and Renzini, A. and Sharon, Chelsea E. and Shimizu, T. T. and Tacconi, L. J. and Tadaki, Ken-ichi and {\"U}bler, H.},
  year = 2023,
  month = jan,
  journal = {ApJ},
  volume = {942},
  pages = {98},
  issn = {0004-637X},
  doi = {10.3847/1538-4357/aca46b},
  urldate = {2025-08-26}
}

@misc{Lopez-Coba2025,
  title = {Physical Properties of Gas Departing from Circular Rotation at 50 Pc Scales Using the {{PHANGS-MUSE}} Galaxies},
  author = {{L{\'o}pez-Cob{\'a}}, Carlos and Lin, Lihwai and Cruz Gonz{\'a}lez, Irene and S{\'a}nchez, Sebasti{\'a}n F. and Pan, Hsi-An and {Barrera-Ballesteros}, J. K. and Hsieh, Bau-Ching},
  year = 2025,
  month = aug,
  publisher = {arXiv},
  doi = {10.48550/arXiv.2508.18517},
  urldate = {2025-09-10}
}

@article{Madau2014,
  title = {Cosmic {{Star-Formation History}}},
  author = {Madau, Piero and Dickinson, Mark},
  year = 2014,
  month = aug,
  journal = {ARA\&A},
  volume = {52},
  pages = {415--486},
  issn = {0066-4146},
  doi = {10.1146/annurev-astro-081811-125615},
  urldate = {2025-03-14}
}

@article{Magdis2012,
  title = {The {{Evolving Interstellar Medium}} of {{Star-forming Galaxies}} since z = 2 as {{Probed}} by {{Their Infrared Spectral Energy Distributions}}},
  author = {Magdis, Georgios E. and Daddi, E. and B{\'e}thermin, M. and Sargent, M. and Elbaz, D. and Pannella, M. and Dickinson, M. and Dannerbauer, H. and {da Cunha}, E. and Walter, F. and Rigopoulou, D. and Charmandaris, V. and Hwang, H. S. and Kartaltepe, J.},
  year = 2012,
  month = nov,
  journal = {ApJ},
  volume = {760},
  pages = {6},
  issn = {0004-637X},
  doi = {10.1088/0004-637X/760/1/6},
  urldate = {2025-08-17}
}

@article{Masters2010,
       author = {{Masters}, Karen L. and {Mosleh}, Moein and {Romer}, A. Kathy and {Nichol}, Robert C. and {Bamford}, Steven P. and {Schawinski}, Kevin and {Lintott}, Chris J. and {Andreescu}, Dan and {Campbell}, Heather C. and {Crowcroft}, Ben and {Doyle}, Isabelle and {Edmondson}, Edward M. and {Murray}, Phil and {Raddick}, M. Jordan and {Slosar}, An{\v{z}}e and {Szalay}, Alexander S. and {Vandenberg}, Jan},
        title = "{Galaxy Zoo: passive red spirals}",
      journal = {MNRAS},
         year = 2010,
        month = jun,
       volume = {405},
       number = {2},
        pages = {783-799},
          doi = {10.1111/j.1365-2966.2010.16503.x},
archivePrefix = {arXiv},
       eprint = {0910.4113},
 primaryClass = {astro-ph.CO},
       adsurl = {https://ui.adsabs.harvard.edu/abs/2010MNRAS.405..783M}
}

@article{Mitsuhashi2024,
  title = {The {{ALMA-CRISTAL}} Survey: {{Widespread}} Dust-Obscured Star Formation in Typical Star-Forming Galaxies at z = 4--6},
  shorttitle = {The {{ALMA-CRISTAL}} Survey},
  author = {Mitsuhashi, Ikki and Tadaki, Ken-ichi and Ikeda, Ryota and {Herrera-Camus}, Rodrigo and Aravena, Manuel and De Looze, Ilse and F{\"o}rster Schreiber, Natascha M. and {Gonz{\'a}lez-L{\'o}pez}, Jorge and Spilker, Justin and Assef, Roberto J. and Bouwens, Rychard and {Barcos-Munoz}, Loreto and Birkin, Jack and Bowler, Rebecca A. A. and Calistro Rivera, Gabriela and Davies, Rebecca and Da Cunha, Elisabete and {D{\'i}az-Santos}, Tanio and Ferrara, Andrea and Fisher, Deanne B. and Lee, Lilian L. and Li, Juno and Lutz, Dieter and Rela{\~n}o, Monica and Naab, Thorsten and Palla, Marco and Posses, Ana and Solimano, Manuel and Tacconi, Linda and {\"U}bler, Hannah and {van der Giessen}, Stefan and Veilleux, Sylvain},
  year = 2024,
  month = oct,
  journal = {A\&A},
  volume = {690},
  pages = {A197},
  issn = {0004-6361},
  doi = {10.1051/0004-6361/202348782},
  urldate = {2025-06-11},
  langid = {english}
}

@article{Mowla2019,
  title = {A {{Mass-dependent Slope}} of the {{Galaxy Size-Mass Relation}} out to z {$\sim$} 3: {{Further Evidence}} for a {{Direct Relation}} between {{Median Galaxy Size}} and {{Median Halo Mass}}},
  shorttitle = {A {{Mass-dependent Slope}} of the {{Galaxy Size-Mass Relation}} out to z {$\sim$} 3},
  author = {Mowla, Lamiya and {van der Wel}, Arjen and {van Dokkum}, Pieter and Miller, Tim B.},
  year = 2019,
  month = feb,
  journal = {ApJ},
  volume = {872},
  pages = {L13},
  publisher = {IOP},
  issn = {0004-637X},
  doi = {10.3847/2041-8213/ab0379},
  urldate = {2025-07-23}
}

@article{Narayanan2014,
  title = {A Theory for the Excitation of {{CO}} in Star-Forming Galaxies},
  author = {Narayanan, Desika and Krumholz, Mark R.},
  year = 2014,
  month = aug,
  journal = {MNRAS},
  volume = {442},
  pages = {1411--1428},
  issn = {0035-8711},
  doi = {10.1093/mnras/stu834},
  urldate = {2025-08-21}
}

@article{Pantoni2021,
  title = {An {{ALMA}} View of 11 Dusty Star-Forming Galaxies at the Peak of Cosmic Star Formation History},
  author = {Pantoni, L. and Massardi, M. and Lapi, A. and Donevski, D. and D'Amato, Q. and Giulietti, M. and Pozzi, F. and Talia, M. and Vignali, C. and Cimatti, A. and Silva, L. and Bressan, A. and Ronconi, T.},
  year = 2021,
  month = nov,
  journal = {MNRAS},
  volume = {507},
  pages = {3998--4015},
  publisher = {OUP},
  issn = {0035-8711},
  doi = {10.1093/mnras/stab2346},
  urldate = {2025-03-14}
}

@article{Papadopoulos2004,
  title = {{{CI}} Lines as Tracers of Molecular Gas, and Their Prospects at High Redshifts},
  author = {Papadopoulos, P. P. and Thi, W. -F. and Viti, S.},
  year = 2004,
  month = jun,
  journal = {MNRAS},
  volume = {351},
  pages = {147--160},
  issn = {0035-8711},
  doi = {10.1111/j.1365-2966.2004.07762.x},
  urldate = {2025-08-20}
}

@article{Papadopoulos2010,
  title = {A {{Cosmic-ray-dominated Interstellar Medium}} in {{Ultra Luminous Infrared Galaxies}}: {{New Initial Conditions}} for {{Star Formation}}},
  shorttitle = {A {{Cosmic-ray-dominated Interstellar Medium}} in {{Ultra Luminous Infrared Galaxies}}},
  author = {Papadopoulos, Padelis P.},
  year = 2010,
  month = sep,
  journal = {ApJ},
  volume = {720},
  pages = {226--232},
  issn = {0004-637X},
  doi = {10.1088/0004-637X/720/1/226},
  urldate = {2025-08-20}
}

@article{Papadopoulos2012,
  title = {Molecular and {{Atomic Line Surveys}} of {{Galaxies}}. {{II}}. {{Unbiased Estimates}} of Their {{Star Formation Mode}}},
  author = {Papadopoulos, Padelis P. and Geach, James E.},
  year = 2012,
  month = oct,
  journal = {ApJ},
  volume = {757},
  pages = {157},
  publisher = {IOP},
  issn = {0004-637X},
  doi = {10.1088/0004-637X/757/2/157},
  urldate = {2025-02-12}
}

@article{Papadopoulos2012a,
  title = {The Molecular Gas in Luminous Infrared Galaxies - {{I}}. {{CO}} Lines, Extreme Physical Conditions and Their Drivers},
  author = {Papadopoulos, Padelis P. and {van der Werf}, Paul P. and Xilouris, E. M. and Isaak, K. G. and Gao, Yu and M{\"u}hle, S.},
  year = 2012,
  month = nov,
  journal = {MNRAS},
  volume = {426},
  pages = {2601--2629},
  publisher = {OUP},
  issn = {0035-8711},
  doi = {10.1111/j.1365-2966.2012.21001.x},
  urldate = {2025-02-12}
}

@article{Papadopoulos2022,
  title = {The Subthermal Excitation of the {{C I}} Lines in the Molecular Gas Reservoirs of Galaxies: {{Its}} Significance and Potential Utility},
  shorttitle = {The Subthermal Excitation of the {{C I}} Lines in the Molecular Gas Reservoirs of Galaxies},
  author = {Papadopoulos, P. and Dunne, L. and Maddox, S.},
  year = 2022,
  month = feb,
  journal = {MNRAS},
  volume = {510},
  pages = {725--733},
  issn = {0035-8711},
  doi = {10.1093/mnras/stab3194},
  urldate = {2025-08-20}
}

@article{Peroux2020,
  title = {The {{Cosmic Baryon}} and {{Metal Cycles}}},
  author = {P{\'e}roux, C{\'e}line and Howk, J. Christopher},
  year = 2020,
  journal = {ARA\&A},
  volume = {58},
  number = {1},
  pages = {363--406},
  doi = {10.1146/annurev-astro-021820-120014},
  urldate = {2021-01-29}
}

@article{Rodighiero2011,
  title = {The {{Lesser Role}} of {{Starbursts}} in {{Star Formation}} at z = 2},
  author = {Rodighiero, G. and Daddi, E. and Baronchelli, I. and Cimatti, A. and Renzini, A. and Aussel, H. and Popesso, P. and Lutz, D. and Andreani, P. and Berta, S. and Cava, A. and Elbaz, D. and Feltre, A. and Fontana, A. and F{\"o}rster Schreiber, N. M. and Franceschini, A. and Genzel, R. and Grazian, A. and Gruppioni, C. and Ilbert, O. and Le Floch, E. and Magdis, G. and Magliocchetti, M. and Magnelli, B. and Maiolino, R. and McCracken, H. and Nordon, R. and Poglitsch, A. and Santini, P. and Pozzi, F. and Riguccini, L. and Tacconi, L. J. and Wuyts, S. and Zamorani, G.},
  year = 2011,
  month = oct,
  journal = {ApJ},
  volume = {739},
  pages = {L40},
  publisher = {IOP},
  issn = {0004-637X},
  doi = {10.1088/2041-8205/739/2/L40},
  urldate = {2025-03-14}
}

@article{Saha2018,
  title = {Forming {{Lenticular Galaxies}} via {{Violent Disk Instability}}},
  author = {Saha, Kanak and Cortesi, Arianna},
  year = 2018,
  month = jul,
  journal = {ApJL},
  volume = {862},
  number = {1},
  pages = {L12},
  publisher = {The American Astronomical Society},
  issn = {2041-8205},
  doi = {10.3847/2041-8213/aad23a},
  urldate = {2025-09-12},
  langid = {english}
}

@article{Sargent2012,
  title = {The {{Contribution}} of {{Starbursts}} and {{Normal Galaxies}} to {{Infrared Luminosity Functions}} at z {$<$} 2},
  author = {Sargent, M. T. and B{\'e}thermin, M. and Daddi, E. and Elbaz, D.},
  year = 2012,
  month = mar,
  journal = {ApJ},
  volume = {747},
  pages = {L31},
  publisher = {IOP},
  issn = {0004-637X},
  doi = {10.1088/2041-8205/747/2/L31},
  urldate = {2025-03-14}
}

@article{Schmidt2016,
  title = {Radial Gas Motions in {{The H I Nearby Galaxy Survey}} ({{THINGS}})},
  author = {Schmidt, Tobias M. and Bigiel, Frank and Klessen, Ralf S. and {de Blok}, W. J. G.},
  year = 2016,
  month = apr,
  journal = {MNRAS},
  volume = {457},
  pages = {2642--2664},
  issn = {0035-8711},
  doi = {10.1093/mnras/stw011},
  urldate = {2025-09-10}
}

@article{Scoville2014,
  title = {The {{Evolution}} of {{Interstellar Medium Mass Probed}} by {{Dust Emission}}: {{ALMA Observations}} at z = 0.3-2},
  shorttitle = {The {{Evolution}} of {{Interstellar Medium Mass Probed}} by {{Dust Emission}}},
  author = {Scoville, N. and Aussel, H. and Sheth, K. and Scott, K. S. and Sanders, D. and Ivison, R. and Pope, A. and Capak, P. and Vanden Bout, P. and Manohar, S. and Kartaltepe, J. and Robertson, B. and Lilly, S.},
  year = 2014,
  month = mar,
  journal = {ApJ},
  volume = {783},
  number = {2},
  pages = {84},
  issn = {0004-637X},
  doi = {10.1088/0004-637X/783/2/84},
  urldate = {2023-07-31},
  langid = {english}
}

@article{Scoville2016,
  title = {{{ISM Masses}} and the {{Star}} Formation {{Law}} at {{Z}} = 1 to 6: {{ALMA Observations}} of {{Dust Continuum}} in 145 {{Galaxies}} in the {{COSMOS Survey Field}}},
  shorttitle = {{{ISM Masses}} and the {{Star}} Formation {{Law}} at {{Z}} = 1 to 6},
  author = {Scoville, N. and Sheth, K. and Aussel, H. and Vanden Bout, P. and Capak, P. and Bongiorno, A. and Casey, C. M. and Murchikova, L. and Koda, J. and {\'A}lvarez-M{\'a}rquez, J. and Lee, N. and Laigle, C. and McCracken, H. J. and Ilbert, O. and Pope, A. and Sanders, D. and Chu, J. and Toft, S. and Ivison, R. J. and Manohar, S.},
  year = 2016,
  month = apr,
  journal = {ApJ},
  volume = {820},
  pages = {83},
  issn = {0004-637X},
  doi = {10.3847/0004-637X/820/2/83},
  urldate = {2023-07-31}
}

@article{Scoville2017,
  title = {Evolution of {{Interstellar Medium}}, {{Star Formation}}, and {{Accretion}} at {{High Redshift}}},
  author = {Scoville, N. and Lee, N. and Vanden Bout, P. and {Diaz-Santos}, T. and Sanders, D. and Darvish, B. and Bongiorno, A. and Casey, C. M. and Murchikova, L. and Koda, J. and Capak, P. and Vlahakis, Catherine and Ilbert, O. and Sheth, K. and {Morokuma-Matsui}, K. and Ivison, R. J. and Aussel, H. and Laigle, C. and McCracken, H. J. and Armus, L. and Pope, A. and Toft, S. and Masters, D.},
  year = 2017,
  month = mar,
  journal = {ApJ},
  volume = {837},
  pages = {150},
  issn = {0004-637X},
  doi = {10.3847/1538-4357/aa61a0},
  urldate = {2025-08-18}
}

@article{Smail2023,
  title = {Hidden {{Giants}} in {{JWST}}'s {{PEARLS}}: {{An Ultramassive}} z = 4.26 {{Submillimeter Galaxy}} That {{Is Invisible}} to {{HST}}},
  shorttitle = {Hidden {{Giants}} in {{JWST}}'s {{PEARLS}}},
  author = {Smail, Ian and Dudzevi{\v c}i{\=u}t{\.e}, Ugn{\.e} and Gurwell, Mark and Fazio, Giovanni G. and Willner, S. P. and Swinbank, A. M. and Arumugam, Vinodiran and Summers, Jake and Cohen, Seth H. and Jansen, Rolf A. and Windhorst, Rogier A. and Meena, Ashish and Zitrin, Adi and Keel, William C. and Cheng, Cheng and Coe, Dan and Conselice, Christopher J. and D'Silva, Jordan C. J. and Driver, Simon P. and Frye, Brenda and Grogin, Norman A. and Koekemoer, Anton M. and Marshall, Madeline A. and Nonino, Mario and Pirzkal, Nor and Robotham, Aaron and Rutkowski, Michael J. and Ryan, Jr., Russell E. and Tompkins, Scott and Willmer, Christopher N. A. and Yan, Haojing and Broadhurst, Thomas J. and Diego, Jos{\'e} M. and Kamieneski, Patrick and Yun, Min},
  year = 2023,
  month = nov,
  journal = {ApJ},
  volume = {958},
  pages = {36},
  publisher = {IOP},
  issn = {0004-637X},
  doi = {10.3847/1538-4357/acf931},
  urldate = {2025-06-16}
}

@article{Sormani2023,
  title = {Fuelling the Nuclear Ring of {{NGC}} 1097},
  author = {Sormani, Mattia C. and Barnes, Ashley T. and Sun, Jiayi and Stuber, Sophia K. and Schinnerer, Eva and Emsellem, Eric and Leroy, Adam K. and Glover, Simon C. O. and Henshaw, Jonathan D. and Meidt, Sharon E. and Neumann, Justus and Querejeta, Miguel and Williams, Thomas G. and Bigiel, Frank and Eibensteiner, Cosima and Fragkoudi, Francesca and Levy, Rebecca C. and Grasha, Kathryn and Klessen, Ralf S. and Kruijssen, J. M. Diederik and Neumayer, Nadine and Pinna, Francesca and Rosolowsky, Erik W. and Smith, Rowan J. and Teng, Yu-Hsuan and Tress, Robin G. and Watkins, Elizabeth J.},
  year = 2023,
  month = aug,
  journal = {MNRAS},
  volume = {523},
  pages = {2918--2927},
  issn = {0035-8711},
  doi = {10.1093/mnras/stad1554},
  urldate = {2025-09-10}
}

@article{Speagle2014,
  title = {A {{Highly Consistent Framework}} for the {{Evolution}} of the {{Star-Forming}} "{{Main Sequence}}" from z \textasciitilde{} 0-6},
  author = {Speagle, J. S. and Steinhardt, C. L. and Capak, P. L. and Silverman, J. D.},
  year = 2014,
  month = oct,
  journal = {ApJS},
  volume = {214},
  pages = {15},
  publisher = {IOP},
  issn = {0067-0049},
  doi = {10.1088/0067-0049/214/2/15},
  urldate = {2025-03-14}
}

@article{Spilker2015,
  title = {Sub-Kiloparsec {{Imaging}} of {{Cool Molecular Gas}} in {{Two Strongly Lensed Dusty}}, {{Star-forming Galaxies}}},
  author = {Spilker, J. S. and Aravena, M. and Marrone, D. P. and B{\'e}thermin, M. and Bothwell, M. S. and Carlstrom, J. E. and Chapman, S. C. and Collier, J. D. and {de Breuck}, C. and Fassnacht, C. D. and Galvin, T. and Gonzalez, A. H. and {Gonz{\'a}lez-L{\'o}pez}, J. and Grieve, K. and Hezaveh, Y. and Ma, J. and Malkan, M. and O'Brien, A. and Rotermund, K. M. and Strandet, M. and Vieira, J. D. and Weiss, A. and Wong, G. F.},
  year = 2015,
  month = oct,
  journal = {ApJ},
  volume = {811},
  pages = {124},
  publisher = {IOP},
  issn = {0004-637X},
  doi = {10.1088/0004-637X/811/2/124},
  urldate = {2025-03-14}
}

@article{Tacchella2016,
  title = {The Confinement of Star-Forming Galaxies into a Main Sequence through Episodes of Gas Compaction, Depletion and Replenishment},
  author = {Tacchella, Sandro and Dekel, Avishai and Carollo, C. Marcella and Ceverino, Daniel and DeGraf, Colin and Lapiner, Sharon and Mandelker, Nir and Primack Joel, R.},
  year = 2016,
  month = apr,
  journal = {MNRAS},
  volume = {457},
  pages = {2790--2813},
  publisher = {OUP},
  issn = {0035-8711},
  doi = {10.1093/mnras/stw131},
  urldate = {2025-03-14}
}

@article{Tacconi2010,
  title = {High Molecular Gas Fractions in Normal Massive Star-Forming Galaxies in the Young {{Universe}}},
  author = {Tacconi, L. J. and Genzel, R. and Neri, R. and Cox, P. and Cooper, M. C. and Shapiro, K. and Bolatto, A. and Bouch{\'e}, N. and Bournaud, F. and Burkert, A. and Combes, F. and Comerford, J. and Davis, M. and F{\"o}rster Schreiber, N. M. and {Garcia-Burillo}, S. and {Gracia-Carpio}, J. and Lutz, D. and Naab, T. and Omont, A. and Shapley, A. and Sternberg, A. and Weiner, B.},
  year = 2010,
  month = feb,
  journal = {Nature},
  volume = {463},
  pages = {781--784},
  issn = {0028-0836},
  doi = {10.1038/nature08773},
  urldate = {2025-08-19}
}

@article{Tacconi2018,
  title = {{{PHIBSS}}: {{Unified Scaling Relations}} of {{Gas Depletion Time}} and {{Molecular Gas Fractions}}},
  shorttitle = {{{PHIBSS}}},
  author = {Tacconi, L. J. and Genzel, R. and Saintonge, A. and Combes, F. and {Garc{\'i}a-Burillo}, S. and Neri, R. and Bolatto, A. and Contini, T. and F{\"o}rster Schreiber, N. M. and Lilly, S. and Lutz, D. and Wuyts, S. and Accurso, G. and Boissier, J. and Boone, F. and Bouch{\'e}, N. and Bournaud, F. and Burkert, A. and Carollo, M. and Cooper, M. and Cox, P. and Feruglio, C. and Freundlich, J. and {Herrera-Camus}, R. and Juneau, S. and Lippa, M. and Naab, T. and Renzini, A. and Salome, P. and Sternberg, A. and Tadaki, K. and {\"U}bler, H. and Walter, F. and Weiner, B. and Weiss, A.},
  year = 2018,
  month = feb,
  journal = {ApJ},
  volume = {853},
  pages = {179},
  publisher = {IOP},
  issn = {0004-637X},
  doi = {10.3847/1538-4357/aaa4b4},
  urldate = {2025-08-09}
}

@article{Tacconi2020,
  title = {The {{Evolution}} of the {{Star-Forming Interstellar Medium Across Cosmic Time}}},
  author = {Tacconi, Linda J. and Genzel, Reinhard and Sternberg, Amiel},
  year = 2020,
  journal = {ARA\&A},
  volume = {58},
  number = {1},
  pages = {157--203},
  doi = {10.1146/annurev-astro-082812-141034},
  urldate = {2021-01-28}
}

@article{Tadaki2017,
  title = {Rotating {{Starburst Cores}} in {{Massive Galaxies}} at z = 2.5},
  author = {Tadaki, Ken-ichi and Kodama, Tadayuki and Nelson, Erica J. and Belli, Sirio and F{\"o}rster Schreiber, Natascha M. and Genzel, Reinhard and Hayashi, Masao and {Herrera-Camus}, Rodrigo and Koyama, Yusei and Lang, Philipp and Lutz, Dieter and Shimakawa, Rhythm and Tacconi, Linda J. and {\"U}bler, Hannah and Wisnioski, Emily and Wuyts, Stijn and Hatsukade, Bunyo and Lippa, Magdalena and Nakanishi, Kouichiro and Ikarashi, Soh and Kohno, Kotaro and Suzuki, Tomoko L. and Tamura, Yoichi and Tanaka, Ichi},
  year = 2017,
  month = jun,
  journal = {ApJ},
  volume = {841},
  pages = {L25},
  publisher = {IOP},
  issn = {0004-637X},
  doi = {10.3847/2041-8213/aa7338},
  urldate = {2025-03-14}
}

@article{Tadaki2017a,
  title = {{{BULGE-FORMING GALAXIES WITH AN EXTENDED ROTATING DISK AT}} z {$\sim$} 2},
  author = {Tadaki, Ken-ichi and Genzel, Reinhard and Kodama, Tadayuki and Wuyts, Stijn and Wisnioski, Emily and Schreiber, Natascha M. F{\"o}rster and Burkert, Andreas and Lang, Philipp and Tacconi, Linda J. and Lutz, Dieter and Belli, Sirio and Davies, Richard I. and Hatsukade, Bunyo and Hayashi, Masao and {Herrera-Camus}, Rodrigo and Ikarashi, Soh and Inoue, Shigeki and Kohno, Kotaro and Koyama, Yusei and Mendel, J. Trevor and Nakanishi, Kouichiro and Shimakawa, Rhythm and Suzuki, Tomoko L. and Tamura, Yoichi and Tanaka, Ichi and {\"U}bler, Hannah and Wilman, Dave J.},
  year = 2017,
  month = jan,
  journal = {ApJ},
  volume = {834},
  number = {2},
  pages = {135},
  issn = {0004-637X, 1538-4357},
  doi = {10.3847/1538-4357/834/2/135},
  urldate = {2025-08-23},
  langid = {english}
}

@article{Tadaki2020,
  title = {Structural {{Evolution}} in {{Massive Galaxies}} at z {$\sim$} 2},
  author = {Tadaki, Ken-ichi and Belli, Sirio and Burkert, Andreas and Dekel, Avishai and F{\"o}rster Schreiber, Natascha M. and Genzel, Reinhard and Hayashi, Masao and {Herrera-Camus}, Rodrigo and Kodama, Tadayuki and Kohno, Kotaro and Koyama, Yusei and Lee, Minju M. and Lutz, Dieter and Mowla, Lamiya and Nelson, Erica J. and Renzini, Alvio and Suzuki, Tomoko L. and Tacconi, Linda J. and {\"U}bler, Hannah and Wisnioski, Emily and Wuyts, Stijn},
  year = 2020,
  month = sep,
  journal = {ApJ},
  volume = {901},
  number = {1},
  pages = {74},
  issn = {0004-637X},
  doi = {10.3847/1538-4357/abaf4a},
  urldate = {2025-06-11},
  langid = {english}
}

@article{Toft2007,
  title = {Hubble {{Space Telescope}} and {{Spitzer Imaging}} of {{Red}} and {{Blue Galaxies}} at z \textasciitilde{} 2.5: {{A Correlation}} between {{Size}} and {{Star Formation Activity}} from {{Compact Quiescent Galaxies}} to {{Extended Star-forming Galaxies}}},
  shorttitle = {Hubble {{Space Telescope}} and {{Spitzer Imaging}} of {{Red}} and {{Blue Galaxies}} at z \textasciitilde{} 2.5},
  author = {Toft, S. and {van Dokkum}, P. and Franx, M. and Labbe, I. and F{\"o}rster Schreiber, N. M. and Wuyts, S. and Webb, T. and Rudnick, G. and Zirm, A. and Kriek, M. and {van der Werf}, P. and Blakeslee, J. P. and Illingworth, G. and Rix, H. -W. and Papovich, C. and Moorwood, A.},
  year = 2007,
  month = dec,
  journal = {ApJ},
  volume = {671},
  pages = {285--302},
  publisher = {IOP},
  issn = {0004-637X},
  doi = {10.1086/521810},
  urldate = {2025-06-24}
}

@article{Trapp2022,
  title = {Gas Infall and Radial Transport in Cosmological Simulations of Milky Way-Mass Discs},
  author = {Trapp, Cameron W. and Kere{\v s}, Du{\v s}an and Chan, Tsang Keung and Escala, Ivanna and Hummels, Cameron and Hopkins, Philip F. and {Faucher-Gigu{\`e}re}, Claude-Andr{\'e} and Murray, Norman and Quataert, Eliot and Wetzel, Andrew},
  year = 2022,
  month = jan,
  journal = {MNRAS},
  volume = {509},
  pages = {4149--4170},
  issn = {0035-8711},
  doi = {10.1093/mnras/stab3251},
  urldate = {2025-09-10}
}

@article{Umehata2025,
  title = {{{ADF22-WEB}}: {{A}} Giant Barred Spiral Starburst Galaxy in the z = 3.1 {{SSA22}} Protocluster Core},
  shorttitle = {{{ADF22-WEB}}},
  author = {Umehata, Hideki and Steidel, Charles C. and Smail, Ian and Swinbank, Mark and Monson, Erik B. and Rosario, David and Lehmer, Bret D. and Nakanishi, Kouichiro and Kubo, Mariko and Iono, Daisuke and Alexander, David M. and Kohno, Kotaro and Tamura, Yoichi and Ivison, Rob J. and Saito, Toshiki and Mitsuhashi, Ikki and Huang, Shuo and Matsuda, Yuichi},
  year = 2025,
  month = apr,
  journal = {PASJ},
  volume = {77},
  pages = {432--445},
  publisher = {OUP},
  issn = {0004-6264},
  doi = {10.1093/pasj/psaf010},
  urldate = {2025-06-19}
}

@article{Valentino2018,
  title = {A {{Survey}} of {{Atomic Carbon}} [{{C I}}] in {{High-redshift Main-sequence Galaxies}}},
  author = {Valentino, Francesco and Magdis, Georgios E. and Daddi, Emanuele and Liu, Daizhong and Aravena, Manuel and Bournaud, Fr{\'e}d{\'e}ric and Cibinel, Anna and Cormier, Diane and Dickinson, Mark E. and Gao, Yu and Jin, Shuowen and Juneau, St{\'e}phanie and Kartaltepe, Jeyhan and Lee, Min-Young and Madden, Suzanne C. and Puglisi, Annagrazia and Sanders, David and Silverman, John},
  year = 2018,
  month = dec,
  journal = {ApJ},
  volume = {869},
  pages = {27},
  issn = {0004-637X},
  doi = {10.3847/1538-4357/aaeb88},
  urldate = {2025-08-20}
}

@article{Valentino2020,
  title = {The {{Properties}} of the {{Interstellar Medium}} of {{Galaxies}} across {{Time}} as {{Traced}} by the {{Neutral Atomic Carbon}} [{{C I}}]},
  author = {Valentino, Francesco and Magdis, Georgios E. and Daddi, Emanuele and Liu, Daizhong and Aravena, Manuel and Bournaud, Fr{\'e}d{\'e}ric and Cortzen, Isabella and Gao, Yu and Jin, Shuowen and Juneau, St{\'e}phanie and Kartaltepe, Jeyhan S. and Kokorev, Vasily and Lee, Min-Young and Madden, Suzanne C. and Narayanan, Desika and Popping, Gerg{\"o} and Puglisi, Annagrazia},
  year = 2020,
  month = feb,
  journal = {ApJ},
  volume = {890},
  pages = {24},
  publisher = {IOP},
  issn = {0004-637X},
  doi = {10.3847/1538-4357/ab6603},
  urldate = {2025-02-12}
}

@article{Valentino2020a,
  title = {{{CO}} Emission in Distant Galaxies on and above the Main Sequence},
  author = {Valentino, F. and Daddi, E. and Puglisi, A. and Magdis, G. E. and Liu, D. and Kokorev, V. and Cortzen, I. and Madden, S. and Aravena, M. and {G{\'o}mez-Guijarro}, C. and Lee, M. -Y. and Le Floc'h, E. and Gao, Y. and Gobat, R. and Bournaud, F. and Dannerbauer, H. and Jin, S. and Dickinson, M. E. and Kartaltepe, J. and Sanders, D.},
  year = 2020,
  month = sep,
  journal = {A\&A},
  volume = {641},
  pages = {A155},
  issn = {0004-6361},
  doi = {10.1051/0004-6361/202038322},
  urldate = {2025-08-21}
}

@article{vanderWel2014,
  title = {{{3D-HST}}+{{CANDELS}}: {{The Evolution}} of the {{Galaxy Size-Mass Distribution}} since z = 3},
  shorttitle = {{{3D-HST}}+{{CANDELS}}},
  author = {{van der Wel}, A. and Franx, M. and {van Dokkum}, P. G. and Skelton, R. E. and Momcheva, I. G. and Whitaker, K. E. and Brammer, G. B. and Bell, E. F. and Rix, H. -W. and Wuyts, S. and Ferguson, H. C. and Holden, B. P. and Barro, G. and Koekemoer, A. M. and Chang, Yu-Yen and McGrath, E. J. and H{\"a}ussler, B. and Dekel, A. and Behroozi, P. and Fumagalli, M. and Leja, J. and Lundgren, B. F. and Maseda, M. V. and Nelson, E. J. and Wake, D. A. and Patel, S. G. and Labb{\'e}, I. and Faber, S. M. and Grogin, N. A. and Kocevski, D. D.},
  year = 2014,
  month = jun,
  journal = {ApJ},
  volume = {788},
  pages = {28},
  publisher = {IOP},
  issn = {0004-637X},
  doi = {10.1088/0004-637X/788/1/28},
  urldate = {2025-06-28}
}

@article{vanDokkum2008,
  title = {Confirmation of the {{Remarkable Compactness}} of {{Massive Quiescent Galaxies}} at z \textasciitilde{} 2.3: {{Early-Type Galaxies Did}} Not {{Form}} in a {{Simple Monolithic Collapse}}},
  shorttitle = {Confirmation of the {{Remarkable Compactness}} of {{Massive Quiescent Galaxies}} at z \textasciitilde{} 2.3},
  author = {{van Dokkum}, Pieter G. and Franx, Marijn and Kriek, Mariska and Holden, Bradford and Illingworth, Garth D. and Magee, Daniel and Bouwens, Rychard and Marchesini, Danilo and Quadri, Ryan and Rudnick, Greg and Taylor, Edward N. and Toft, Sune},
  year = 2008,
  month = apr,
  journal = {ApJ},
  volume = {677},
  pages = {L5},
  issn = {0004-637X},
  doi = {10.1086/587874},
  urldate = {2025-08-26}
}

@article{Walter2011,
  title = {A {{Survey}} of {{Atomic Carbon}} at {{High Redshift}}},
  author = {Walter, F. and Wei{\ss}, A. and Downes, D. and Decarli, R. and Henkel, C.},
  year = 2011,
  month = mar,
  journal = {ApJ},
  volume = {730},
  pages = {18},
  issn = {0004-637X},
  doi = {10.1088/0004-637X/730/1/18},
  urldate = {2025-08-20}
}

@article{Walter2020,
  title = {The {{Evolution}} of the {{Baryons Associated}} with {{Galaxies Averaged}} over {{Cosmic Time}} and {{Space}}},
  author = {Walter, Fabian and Carilli, Chris and Neeleman, Marcel and Decarli, Roberto and Popping, Gerg{\"o} and Somerville, Rachel S. and Aravena, Manuel and Bertoldi, Frank and Boogaard, Leindert and Cox, Pierre and {da Cunha}, Elisabete and Magnelli, Benjamin and Obreschkow, Danail and Riechers, Dominik and Rix, Hans-Walter and Smail, Ian and Weiss, Axel and Assef, Roberto J. and Bauer, Franz and Bouwens, Rychard and Contini, Thierry and Cortes, Paulo C. and Daddi, Emanuele and {Diaz-Santos}, Tanio and {Gonz{\'a}lez-L{\'o}pez}, Jorge and Hennawi, Joseph and Hodge, Jacqueline A. and Inami, Hanae and Ivison, Rob and Oesch, Pascal and Sargent, Mark and {van der Werf}, Paul and Wagg, Jeff and Yung, L. Y. Aaron},
  year = 2020,
  month = oct,
  journal = {ApJ},
  volume = {902},
  pages = {111},
  issn = {0004-637X},
  doi = {10.3847/1538-4357/abb82e},
  urldate = {2021-02-01}
}

@article{Wong2004,
  title = {A {{Search}} for {{Kinematic Evidence}} of {{Radial Gas Flows}} in {{Spiral Galaxies}}},
  author = {Wong, Tony and Blitz, Leo and Bosma, Albert},
  year = 2004,
  month = apr,
  journal = {ApJ},
  volume = {605},
  number = {1},
  pages = {183},
  issn = {0004-637X},
  doi = {10.1086/382215},
  urldate = {2025-09-10},
  langid = {english}
}

@article{Xiao2025,
  title = {{{PANORAMIC}}: {{Discovery}} of an Ultra-Massive Grand-Design Spiral Galaxy at {\emph{z}} {$\sim$} 5.2},
  shorttitle = {{{PANORAMIC}}},
  author = {Xiao, Mengyuan and Williams, Christina C. and Oesch, Pascal A. and Elbaz, David and {Dessauges-Zavadsky}, Miroslava and {Marques-Chaves}, Rui and Bing, Longji and Ji, Zhiyuan and Weibel, Andrea and Bezanson, Rachel and Brammer, Gabriel and Casey, Caitlin and Cloonan, Aidan P. and Daddi, Emanuele and Dayal, Pratika and Faisst, Andreas L. and Franx, Marijn and Glazebrook, Karl and Hutter, Anne and Kartaltepe, Jeyhan S. and Labbe, Ivo and Lagache, Guilaine and Lim, Seunghwan and Magnelli, Benjamin and Martinez, Felix and Maseda, Michael V. and Nanayakkara, Themiya and Schaerer, Daniel and Whitaker, Katherine E.},
  year = 2025,
  month = apr,
  journal = {A\&A},
  volume = {696},
  pages = {A156},
  issn = {0004-6361, 1432-0746},
  doi = {10.1051/0004-6361/202453487},
  urldate = {2025-08-23},
  langid = {english}
}

@article{Zolotov2015,
  title = {Compaction and Quenching of High-z Galaxies in Cosmological Simulations: {{Blue}} and Red Nuggets},
  shorttitle = {Compaction and Quenching of High-z Galaxies in Cosmological Simulations},
  author = {Zolotov, Adi and Dekel, Avishai and Mandelker, Nir and Tweed, Dylan and Inoue, Shigeki and DeGraf, Colin and Ceverino, Daniel and Primack, Joel R. and Barro, Guillermo and Faber, Sandra M.},
  year = 2015,
  month = jul,
  journal = {MNRAS},
  volume = {450},
  pages = {2327--2353},
  publisher = {OUP},
  issn = {0035-8711},
  doi = {10.1093/mnras/stv740},
  urldate = {2025-06-11}
}

@article{Fabian2012,
  title = {Observational {{Evidence}} of {{Active Galactic Nuclei Feedback}}},
  author = {Fabian, A.C.},
  year = 2012,
  month = sep,
  journal = {Annu. Rev. Astron. Astrophys.},
  volume = {50},
  number = {1},
  pages = {455--489},
  issn = {0066-4146, 1545-4282},
  doi = {10.1146/annurev-astro-081811-125521},
  url = {https://www.annualreviews.org/doi/10.1146/annurev-astro-081811-125521},
  urldate = {2025-01-17},
  langid = {english}
}

@article{Marti-Vidal2014,
  title = {{{UVMULTIFIT}}: {{A}} Versatile Tool for Fitting Astronomical Radio Interferometric Data},
  shorttitle = {{{UVMULTIFIT}}},
  author = {{Mart{\'i}-Vidal}, I. and Vlemmings, W. H. T. and Muller, S. and Casey, S.},
  year = 2014,
  month = mar,
  journal = {A\&A},
  volume = {563},
  pages = {A136},
  publisher = {EDP Sciences},
  issn = {0004-6361, 1432-0746},
  doi = {10.1051/0004-6361/201322633},
  url = {https://www.aanda.org/articles/aa/abs/2014/03/aa22633-13/aa22633-13.html},
  urldate = {2025-09-19},
  copyright = {\copyright{} ESO, 2014},
  langid = {english}
}

@article{Kurtovic2024,
  title = {Recovering the Gas Properties of Protoplanetary Disks through Parametric Visibility Modeling: {{MHO}} 6},
  shorttitle = {Recovering the Gas Properties of Protoplanetary Disks through Parametric Visibility Modeling},
  author = {Kurtovic, N. T. and Pinilla, P.},
  year = 2024,
  month = jul,
  journal = {A\&A},
  volume = {687},
  pages = {A188},
  issn = {0004-6361},
  doi = {10.1051/0004-6361/202449667},
  url = {https://ui.adsabs.harvard.edu/abs/2024A&A...687A.188K},
  urldate = {2025-09-19}
}

@article{Spilker2016,
  title = {{{ALMA Imaging}} and {{Gravitational Lens Models}} of {{South Pole Telescope}}---{{Selected Dusty}}, {{Star-Forming Galaxies}} at {{High Redshifts}}},
  author = {Spilker, J. S. and Marrone, D. P. and Aravena, M. and B{\'e}thermin, M. and Bothwell, M. S. and Carlstrom, J. E. and Chapman, S. C. and Crawford, T. M. and {de Breuck}, C. and Fassnacht, C. D. and Gonzalez, A. H. and Greve, T. R. and Hezaveh, Y. and Litke, K. and Ma, J. and Malkan, M. and Rotermund, K. M. and Strandet, M. and Vieira, J. D. and Weiss, A. and Welikala, N.},
  year = 2016,
  month = aug,
  journal = {ApJ},
  volume = {826},
  pages = {112},
  issn = {0004-637X},
  doi = {10.3847/0004-637X/826/2/112},
  url = {https://ui.adsabs.harvard.edu/abs/2016ApJ...826..112S},
  urldate = {2025-09-19}
}

@article{Tazzari2018,
  title = {{{GALARIO}}: A {{GPU}} Accelerated Library for Analysing Radio Interferometer Observations},
  shorttitle = {{{GALARIO}}},
  author = {Tazzari, Marco and Beaujean, Frederik and Testi, Leonardo},
  year = 2018,
  month = jun,
  journal = {MNRAS},
  volume = {476},
  pages = {4527--4542},
  issn = {0035-8711},
  doi = {10.1093/mnras/sty409},
  url = {https://ui.adsabs.harvard.edu/abs/2018MNRAS.476.4527T},
  urldate = {2025-09-19}
}

@article{Rodriguez-Gomez2019,
  title = {The Optical Morphologies of Galaxies in the {{IllustrisTNG}} Simulation: A Comparison to {{Pan-STARRS}} Observations},
  shorttitle = {The Optical Morphologies of Galaxies in the {{IllustrisTNG}} Simulation},
  author = {{Rodriguez-Gomez}, Vicente and Snyder, Gregory F. and Lotz, Jennifer M. and Nelson, Dylan and Pillepich, Annalisa and Springel, Volker and Genel, Shy and Weinberger, Rainer and Tacchella, Sandro and Pakmor, R{\"u}diger and Torrey, Paul and Marinacci, Federico and Vogelsberger, Mark and Hernquist, Lars and Thilker, David A.},
  year = 2019,
  month = mar,
  journal = {MNRAS},
  volume = {483},
  pages = {4140--4159},
  issn = {0035-8711},
  doi = {10.1093/mnras/sty3345},
  url = {https://ui.adsabs.harvard.edu/abs/2019MNRAS.483.4140R},
  urldate = {2025-09-21}
}

@article{Liu2019,
  title = {Automated {{Mining}} of the {{ALMA Archive}} in the {{COSMOS Field}} ({{A3COSMOS}}). {{II}}. {{Cold Molecular Gas Evolution}} out to {{Redshift}} 6},
  author = {Liu, Daizhong and Schinnerer, E. and Groves, B. and Magnelli, B. and Lang, P. and Leslie, S. and {Jim{\'e}nez-Andrade}, E. and Riechers, D. A. and Popping, G. and Magdis, Georgios E. and Daddi, E. and Sargent, M. and Gao, Yu and Fudamoto, Y. and Oesch, P. A. and Bertoldi, F.},
  year = 2019,
  month = dec,
  journal = {ApJ},
  volume = {887},
  pages = {235},
  issn = {0004-637X},
  doi = {10.3847/1538-4357/ab578d},
  url = {https://ui.adsabs.harvard.edu/abs/2019ApJ...887..235L},
  urldate = {2025-09-21}
}

@article{Riechers2020,
  title = {{{VLA-ALMA Spectroscopic Survey}} in the {{Hubble Ultra Deep Field}} ({{VLASPECS}}): {{Total Cold Gas Masses}} and {{CO Line Ratios}} for z = 2-3 {{Main-sequence Galaxies}}},
  shorttitle = {{{VLA-ALMA Spectroscopic Survey}} in the {{Hubble Ultra Deep Field}} ({{VLASPECS}})},
  author = {Riechers, Dominik A. and Boogaard, Leindert A. and Decarli, Roberto and {Gonz{\'a}lez-L{\'o}pez}, Jorge and Smail, Ian and Walter, Fabian and Aravena, Manuel and Carilli, Christopher L. and Cortes, Paulo C. and Cox, Pierre and {D{\'i}az-Santos}, Tanio and Hodge, Jacqueline A. and Inami, Hanae and Ivison, Rob J. and Kaasinen, Melanie and Wagg, Jeff and Wei{\ss}, Axel and {van der Werf}, Paul},
  year = 2020,
  month = jun,
  journal = {ApJ},
  volume = {896},
  pages = {L21},
  issn = {0004-637X},
  doi = {10.3847/2041-8213/ab9595},
  url = {https://ui.adsabs.harvard.edu/abs/2020ApJ...896L..21R},
  urldate = {2025-09-21}
}

@article{Boogaard2023,
  title = {A {{NOEMA Molecular Line Scan}} of the {{Hubble Deep Field North}}: {{Improved Constraints}} on the {{CO Luminosity Functions}} and {{Cosmic Density}} of {{Molecular Gas}}},
  shorttitle = {A {{NOEMA Molecular Line Scan}} of the {{Hubble Deep Field North}}},
  author = {Boogaard, Leindert A. and Decarli, Roberto and Walter, Fabian and Wei{\ss}, Axel and Popping, Gerg{\"o} and Neri, Roberto and Aravena, Manuel and Riechers, Dominik and Ellis, Richard S. and Carilli, Chris and Cox, Pierre and Pety, J{\'e}r{\^o}me},
  year = 2023,
  month = mar,
  journal = {ApJ},
  volume = {945},
  pages = {111},
  issn = {0004-637X},
  doi = {10.3847/1538-4357/acb4f0},
  url = {https://ui.adsabs.harvard.edu/abs/2023ApJ...945..111B},
  urldate = {2025-09-21}
}

@article{Bollo2025,
  title = {{{ALMACAL}}: {{XIII}}. {{Evolution}} of the {{CO}} Luminosity Function and the Molecular Gas Mass Density out to z {$\sim$} 6},
  shorttitle = {{{ALMACAL}}},
  author = {Bollo, Victoria and P{\'e}roux, C{\'e}line and Zwaan, Martin and Hamanowicz, Aleksandra and Chen, Jianhang and Weng, Simon and Lagos, Claudia del P. and Bravo, Mat{\'i}as and Ivison, Rob J. and Biggs, Andrew},
  year = 2025,
  month = mar,
  journal = {A\&A},
  volume = {695},
  pages = {A163},
  issn = {0004-6361},
  doi = {10.1051/0004-6361/202453223},
  url = {https://ui.adsabs.harvard.edu/abs/2025A&A...695A.163B},
  urldate = {2025-09-21}
}

@article{Kennicutt2012,
  title = {Star {{Formation}} in the {{Milky Way}} and {{Nearby Galaxies}}},
  author = {Kennicutt, Robert C. and Evans, Neal J.},
  year = 2012,
  month = sep,
  journal = {ARA\&A},
  volume = {50},
  number = {1},
  pages = {531--608},
  issn = {0066-4146, 1545-4282},
  doi = {10.1146/annurev-astro-081811-125610},
  url = {http://www.annualreviews.org/doi/10.1146/annurev-astro-081811-125610},
  urldate = {2020-08-26},
  langid = {english}
}

@article{Wang2025a,
  title = {A Giant Disk Galaxy Two Billion Years after the {{Big Bang}}},
  author = {Wang, Weichen and Cantalupo, Sebastiano and Pensabene, Antonio and Galbiati, Marta and Travascio, Andrea and Steidel, Charles C. and Maseda, Michael V. and Pezzulli, Gabriele and {de Beer}, Stephanie and Fossati, Matteo and Fumagalli, Michele and Gallego, Sofia G. and Lazeyras, Titouan and Mackenzie, Ruari and Matthee, Jorryt and Nanayakkara, Themiya and Quadri, Giada},
  year = 2025,
  month = may,
  journal = {Nat. Astron.},
  volume = {9},
  pages = {710--719},
  issn = {2397-3366},
  doi = {10.1038/s41550-025-02500-2},
  url = {https://ui.adsabs.harvard.edu/abs/2025NatAs...9..710W},
  urldate = {2025-09-24}
}

@misc{Jiang2025,
  title = {Formation and {{Environmental Context}} of {{Giant Bulgeless Disk Galaxies}} in the {{Early Universe}}: {{Insights}} from {{Cosmological Simulations}}},
  shorttitle = {Formation and {{Environmental Context}} of {{Giant Bulgeless Disk Galaxies}} in the {{Early Universe}}},
  author = {Jiang, Fangzhou and Liang, Jinning and Jin, Bingcheng and Gao, Zeyu and Wang, Weichen and Cantalupo, Sebastiano and Shen, Xuejian and Ho, Luis C. and Peng, Yingjie and Wang, Jing},
  year = 2025,
  month = apr,
  publisher = {arXiv},
  doi = {10.48550/arXiv.2504.01070},
  url = {https://ui.adsabs.harvard.edu/abs/2025arXiv250401070J},
  urldate = {2025-09-24}
}

@article{vanDokkum2010,
  title = {The {{Growth}} of {{Massive Galaxies Since}} z = 2},
  author = {{van Dokkum}, Pieter G. and Whitaker, Katherine E. and Brammer, Gabriel and Franx, Marijn and Kriek, Mariska and Labb{\'e}, Ivo and Marchesini, Danilo and Quadri, Ryan and Bezanson, Rachel and Illingworth, Garth D. and Muzzin, Adam and Rudnick, Gregory and Tal, Tomer and Wake, David},
  year = 2010,
  month = feb,
  journal = {ApJ},
  volume = {709},
  pages = {1018--1041},
  issn = {0004-637X},
  doi = {10.1088/0004-637X/709/2/1018},
  url = {https://ui.adsabs.harvard.edu/abs/2010ApJ...709.1018V},
  urldate = {2025-09-27}
}

@article{Whitaker2014,
  title = {Constraining the {{Low-mass Slope}} of the {{Star Formation Sequence}} at 0.5 {$<$} z {$<$} 2.5},
  author = {Whitaker, Katherine E. and Franx, Marijn and Leja, Joel and {van Dokkum}, Pieter G. and Henry, Alaina and Skelton, Rosalind E. and Fumagalli, Mattia and Momcheva, Ivelina G. and Brammer, Gabriel B. and Labb{\'e}, Ivo and Nelson, Erica J. and Rigby, Jane R.},
  year = 2014,
  month = nov,
  journal = {ApJ},
  volume = {795},
  pages = {104},
  publisher = {IOP},
  issn = {0004-637X},
  doi = {10.1088/0004-637X/795/2/104},
  url = {https://ui.adsabs.harvard.edu/abs/2014ApJ...795..104W},
  urldate = {2025-11-04}
}

@article{Lutz2011,
  title = {{{PACS Evolutionary Probe}} ({{PEP}}) - {{A Herschel}} Key Program},
  author = {Lutz, D. and Poglitsch, A. and Altieri, B. and Andreani, P. and Aussel, H. and Berta, S. and Bongiovanni, A. and Brisbin, D. and Cava, A. and Cepa, J. and Cimatti, A. and Daddi, E. and {Dominguez-Sanchez}, H. and Elbaz, D. and F{\"o}rster Schreiber, N. M. and Genzel, R. and Grazian, A. and Gruppioni, C. and Harwit, M. and Le Floc'h, E. and Magdis, G. and Magnelli, B. and Maiolino, R. and Nordon, R. and P{\'e}rez Garc{\'i}a, A. M. and Popesso, P. and Pozzi, F. and Riguccini, L. and Rodighiero, G. and Saintonge, A. and Sanchez Portal, M. and Santini, P. and Shao, L. and Sturm, E. and Tacconi, L. J. and Valtchanov, I. and Wetzstein, M. and Wieprecht, E.},
  year = 2011,
  month = aug,
  journal = {A\&A},
  volume = {532},
  pages = {A90},
  publisher = {EDP},
  issn = {0004-6361},
  doi = {10.1051/0004-6361/201117107},
  url = {https://ui.adsabs.harvard.edu/abs/2011A&A...532A..90L},
  urldate = {2025-11-04}
}

@article{Calzetti2000,
  title = {The {{Dust Content}} and {{Opacity}} of {{Actively Star-forming Galaxies}}},
  author = {Calzetti, Daniela and Armus, Lee and Bohlin, Ralph C. and Kinney, Anne L. and Koornneef, Jan and {Storchi-Bergmann}, Thaisa},
  year = 2000,
  month = apr,
  journal = {ApJ},
  volume = {533},
  pages = {682--695},
  publisher = {IOP},
  issn = {0004-637X},
  doi = {10.1086/308692},
  url = {https://ui.adsabs.harvard.edu/abs/2000ApJ...533..682C},
  urldate = {2025-11-04}
}

@article{Dale2014,
  title = {A {{Two-parameter Model}} for the {{Infrared}}/{{Submillimeter}}/{{Radio Spectral Energy Distributions}} of {{Galaxies}} and {{Active Galactic Nuclei}}},
  author = {Dale, Daniel A. and Helou, George and Magdis, Georgios E. and Armus, Lee and {D{\'i}az-Santos}, Tanio and Shi, Yong},
  year = 2014,
  month = mar,
  journal = {ApJ},
  volume = {784},
  pages = {83},
  publisher = {IOP},
  issn = {0004-637X},
  doi = {10.1088/0004-637X/784/1/83},
  url = {https://ui.adsabs.harvard.edu/abs/2014ApJ...784...83D},
  urldate = {2025-11-04}
}

@article{Bruzual2003,
  title = {Stellar Population Synthesis at the Resolution of 2003},
  author = {Bruzual, G. and Charlot, S.},
  year = 2003,
  month = oct,
  journal = {MNRAS},
  volume = {344},
  pages = {1000--1028},
  publisher = {OUP},
  issn = {0035-8711},
  doi = {10.1046/j.1365-8711.2003.06897.x},
  url = {https://ui.adsabs.harvard.edu/abs/2003MNRAS.344.1000B},
  urldate = {2025-11-04}
}

@article{Bigiel2011,
  title = {A {{Constant Molecular Gas Depletion Time}} in {{Nearby Disk Galaxies}}},
  author = {Bigiel, F. and Leroy, A. K. and Walter, F. and Brinks, E. and {de Blok}, W. J. G. and Kramer, C. and Rix, H. W. and Schruba, A. and Schuster, K.-F. and Usero, A. and Wiesemeyer, H. W.},
  year = 2011,
  month = apr,
  journal = {ApJ},
  volume = {730},
  pages = {L13},
  publisher = {IOP},
  issn = {0004-637X},
  doi = {10.1088/2041-8205/730/2/L13},
  url = {https://ui.adsabs.harvard.edu/abs/2011ApJ...730L..13B},
  urldate = {2025-11-09}
}

@article{Muraoka2019,
  title = {{{CO Multi-line Imaging}} of {{Nearby Galaxies}} ({{COMING}}). {{VI}}. {{Radial}} Variations in Star Formation Efficiency},
  author = {Muraoka, Kazuyuki and Sorai, Kazuo and Miyamoto, Yusuke and Yoda, Moe and {Morokuma-Matsui}, Kana and Kobayashi, Masato I. N. and Kuroda, Mayu and Kaneko, Hiroyuki and Kuno, Nario and Takeuchi, Tsutomu T. and Nakanishi, Hiroyuki and Watanabe, Yoshimasa and Tanaka, Takahiro and Yasuda, Atsushi and Yajima, Yoshiyuki and Shibata, Shugo and Salak, Dragan and Espada, Daniel and Matsumoto, Naoko and Noma, Yuto and Kita, Shoichiro and Komatsuzaki, Ryusei and Kajikawa, Ayumi and Yashima, Yu and Pan, Hsi-An and Oi, Nagisa and Seta, Masumichi and Nakai, Naomasa},
  year = 2019,
  month = dec,
  journal = {PASJ},
  volume = {71},
  pages = {S15},
  publisher = {OUP},
  issn = {0004-6264},
  doi = {10.1093/pasj/psz015},
  url = {https://ui.adsabs.harvard.edu/abs/2019PASJ...71S..15M},
  urldate = {2025-11-09}
}

@article{Tadaki2023,
  title = {Spatial {{Extent}} of {{Molecular Gas}}, {{Dust}}, and {{Stars}} in {{Massive Galaxies}} at z {$\sim$} 2.2-2.5 {{Determined}} with {{ALMA}} and {{JWST}}},
  author = {Tadaki, Ken-ichi and Kodama, Tadayuki and Koyama, Yusei and Suzuki, Tomoko L. and Mitsuhashi, Ikki and Ikeda, Ryota},
  year = 2023,
  month = nov,
  journal = {ApJ},
  volume = {957},
  pages = {L15},
  publisher = {IOP},
  issn = {0004-637X},
  doi = {10.3847/2041-8213/ad03f2},
  url = {https://ui.adsabs.harvard.edu/abs/2023ApJ...957L..15T},
  urldate = {2025-11-09}
}

@article{Lee2021,
  title = {Revisited {{Cold Gas Content}} with {{Atomic Carbon}} [{{C I}}] in z = 2.5 {{Protocluster Galaxies}}},
  author = {Lee, Minju M. and Tanaka, Ichi and Iono, Daisuke and Kawabe, Ryohei and Kodama, Tadayuki and Kohno, Kotaro and Saito, Toshiki and Tamura, Yoichi},
  year = 2021,
  month = mar,
  journal = {ApJ},
  volume = {909},
  pages = {181},
  publisher = {IOP},
  issn = {0004-637X},
  doi = {10.3847/1538-4357/abdbb5},
  url = {https://ui.adsabs.harvard.edu/abs/2021ApJ...909..181L},
  urldate = {2026-01-02}
}

@article{Genel2010,
  title = {The {{Growth}} of {{Dark Matter Halos}}: {{Evidence}} for {{Significant Smooth Accretion}}},
  shorttitle = {The {{Growth}} of {{Dark Matter Halos}}},
  author = {Genel, Shy and Bouch{\'e}, Nicolas and Naab, Thorsten and Sternberg, Amiel and Genzel, Reinhard},
  year = 2010,
  month = aug,
  journal = {ApJ},
  volume = {719},
  pages = {229--239},
  publisher = {IOP},
  issn = {0004-637X},
  doi = {10.1088/0004-637X/719/1/229},
  url = {https://ui.adsabs.harvard.edu/abs/2010ApJ...719..229G},
  urldate = {2026-01-02}
}

@article{Ju2025,
  title = {{{MSA-3D}}: {{Metallicity Gradients}} in {{Galaxies}} at z {$\sim$} 1 with {{JWST}}/{{NIRSpec Slit-stepping Spectroscopy}}},
  shorttitle = {{{MSA-3D}}},
  author = {Ju, Mengting and Wang, Xin and Jones, Tucker and Bari{\v s}i{\'c}, Ivana and Nanayakkara, Themiya and Bundy, Kevin and {Faucher-Gigu{\`e}re}, Claude-Andr{\'e} and Feng, Shuai and Glazebrook, Karl and Henry, Alaina and Malkan, Matthew A. and Obreschkow, Danail and Roy, Namrata and Sanders, Ryan L. and Sun, Xunda and Treu, Tommaso and Zhou, Qianqiao},
  year = 2025,
  month = jan,
  journal = {ApJ},
  volume = {978},
  pages = {L39},
  publisher = {IOP},
  issn = {0004-637X},
  doi = {10.3847/2041-8213/ada150},
  url = {https://ui.adsabs.harvard.edu/abs/2025ApJ...978L..39J},
  urldate = {2026-01-02}
}

@article{ForsterSchreiber2018,
  title = {The {{SINS}}/{{zC-SINF Survey}} of z {$\sim$} 2 {{Galaxy Kinematics}}: {{SINFONI Adaptive Optics-assisted Data}} and {{Kiloparsec-scale Emission-line Properties}}},
  shorttitle = {The {{SINS}}/{{zC-SINF Survey}} of z {$\sim$} 2 {{Galaxy Kinematics}}},
  author = {F{\"o}rster Schreiber, N. M. and Renzini, A. and Mancini, C. and Genzel, R. and Bouch{\'e}, N. and Cresci, G. and Hicks, E. K. S. and Lilly, S. J. and Peng, Y. and Burkert, A. and Carollo, C. M. and Cimatti, A. and Daddi, E. and Davies, R. I. and Genel, S. and Kurk, J. D. and Lang, P. and Lutz, D. and Mainieri, V. and McCracken, H. J. and Mignoli, M. and Naab, T. and Oesch, P. and Pozzetti, L. and Scodeggio, M. and Shapiro Griffin, K. and Shapley, A. E. and Sternberg, A. and Tacchella, S. and Tacconi, L. J. and Wuyts, S. and Zamorani, G.},
  year = 2018,
  month = oct,
  journal = {ApJS},
  volume = {238},
  pages = {21},
  publisher = {IOP},
  issn = {0067-0049},
  doi = {10.3847/1538-4365/aadd49},
  url = {https://ui.adsabs.harvard.edu/abs/2018ApJS..238...21F},
  urldate = {2026-01-02}
}

@article{Wang2017,
  title = {The {{Grism Lens-Amplified Survey}} from {{Space}} ({{GLASS}}). {{X}}. {{Sub-kiloparsec Resolution Gas-phase Metallicity Maps}} at {{Cosmic Noon}} behind the {{Hubble Frontier Fields Cluster MACS1149}}.6+2223},
  author = {Wang, Xin and Jones, Tucker A. and Treu, Tommaso and Morishita, Takahiro and Abramson, Louis E. and Brammer, Gabriel B. and Huang, Kuang-Han and Malkan, Matthew A. and Schmidt, Kasper B. and Fontana, Adriano and Grillo, Claudio and Henry, Alaina L. and Karman, Wouter and Kelly, Patrick L. and Mason, Charlotte A. and Mercurio, Amata and Rosati, Piero and Sharon, Keren and Trenti, Michele and Vulcani, Benedetta},
  year = 2017,
  month = mar,
  journal = {ApJ},
  volume = {837},
  number = {1},
  pages = {89},
  issn = {0004-637X, 1538-4357},
  doi = {10.3847/1538-4357/aa603c},
  url = {https://iopscience.iop.org/article/10.3847/1538-4357/aa603c},
  urldate = {2026-01-02}
}

@article{Wuyts2016,
  title = {{{THE EVOLUTION OF METALLICITY AND METALLICITY GRADIENTS FROM}} z = 2.7 {{TO}} 0.6 {{WITH KMOS}}{{{\textsuperscript{3D}}}}},
  author = {Wuyts, Eva and Wisnioski, Emily and Fossati, Matteo and Schreiber, Natascha M. F{\"o}rster and Genzel, Reinhard and Davies, Ric and Mendel, J. Trevor and Naab, Thorsten and R{\"o}ttgers, Bernhard and Wilman, David J. and Wuyts, Stijn and Bandara, Kaushala and Beifiori, Alessandra and Belli, Sirio and Bender, Ralf and Brammer, Gabriel B. and Burkert, Andreas and Chan, Jeffrey and Galametz, Audrey and Kulkarni, Sandesh K. and Lang, Philipp and Lutz, Dieter and Momcheva, Ivelina G. and Nelson, Erica J. and Rosario, David and Saglia, Roberto P. and Seitz, Stella and Tacconi, Linda J. and Tadaki, Ken-ichi and {\"U}bler, Hannah and Dokkum, Pieter Van},
  year = 2016,
  month = aug,
  journal = {ApJ},
  volume = {827},
  number = {1},
  pages = {74},
  issn = {0004-637X, 1538-4357},
  doi = {10.3847/0004-637X/827/1/74},
  url = {https://iopscience.iop.org/article/10.3847/0004-637X/827/1/74},
  urldate = {2026-01-02}
}

@article{Tsukui2025,
  title = {The Emergence of Galactic Thin and Thick Discs across Cosmic History},
  author = {Tsukui, Takafumi and Wisnioski, Emily and {Bland-Hawthorn}, Joss and Freeman, Ken},
  year = 2025,
  month = jul,
  journal = {MNRAS},
  volume = {540},
  pages = {3493--3522},
  publisher = {OUP},
  issn = {0035-8711},
  doi = {10.1093/mnras/staf604},
  url = {https://ui.adsabs.harvard.edu/abs/2025MNRAS.540.3493T},
  urldate = {2026-01-03}
}

@article{Liu2025,
  title = {Deciphering {{Gas Dynamics}} and {{Star Formation}} in a z = 1.1 {{Main-sequence Spiral Galaxy}} with {{ALMA}} and {{JWST}}},
  author = {Liu, Zhaoran and Kodama, Tadayuki and Morishita, Takahiro and Lee, Kianhong and Sun, Fengwu and Kubo, Mariko and Cai, Zheng and Wu, Yunjing and Li, Zihao},
  year = 2025,
  month = feb,
  journal = {ApJ},
  volume = {980},
  pages = {69},
  publisher = {IOP},
  issn = {0004-637X},
  doi = {10.3847/1538-4357/ada937},
  url = {https://ui.adsabs.harvard.edu/abs/2025ApJ...980...69L},
  urldate = {2025-08-11}
}

@article{Casey2014,
  title = {Dusty Star-Forming Galaxies at High Redshift},
  author = {Casey, Caitlin M. and Narayanan, Desika and Cooray, Asantha},
  year = 2014,
  month = aug,
  journal = {PhR},
  volume = {541},
  number = {2},
  pages = {45--161},
  issn = {03701573},
  doi = {10.1016/j.physrep.2014.02.009},
  url = {https://linkinghub.elsevier.com/retrieve/pii/S0370157314000477},
  urldate = {2020-09-13},
  langid = {english}
}

@article{Dudzeviciute2020,
  title = {An {{ALMA}} Survey of the {{SCUBA-2 CLS UDS}} Field: Physical Properties of 707 Sub-Millimetre Galaxies},
  shorttitle = {An {{ALMA}} Survey of the {{SCUBA-2 CLS UDS}} Field},
  author = {Dudzevi{\v c}i{\=u}t{\.e}, U and Smail, Ian and Swinbank, A M and Stach, S M and Almaini, O and {da~Cunha}, E and An, Fang Xia and Arumugam, V and Birkin, J and Blain, A W and Chapman, S C and Chen, C-C and Conselice, C J and Coppin, K E K and Dunlop, J S and Farrah, D and Geach, J E and Gullberg, B and Hartley, W G and Hodge, J A and Ivison, R J and Maltby, D T and Scott, D and Simpson, C J and Simpson, J M and Thomson, A P and Walter, F and Wardlow, J L and Weiss, A and {van~der~Werf}, P},
  year = 2020,
  month = may,
  journal = {MNRAS},
  volume = {494},
  number = {3},
  pages = {3828--3860},
  issn = {0035-8711},
  doi = {10.1093/mnras/staa769},
  url = {https://doi.org/10.1093/mnras/staa769},
  urldate = {2021-11-12}
}

@article{Gadotti2026,
  title = {Robust Galaxy Image Decompositions with Differential Evolution Optimization and the Problem of Classical Bulges in and beyond the Nearby {{Universe}}},
  author = {Gadotti, Dimitri A.},
  year = 2026,
  month = feb,
  journal = {MNRAS},
  volume = {545},
  pages = {staf2072},
  publisher = {OUP},
  issn = {0035-8711},
  doi = {10.1093/mnras/staf2072},
  url = {https://ui.adsabs.harvard.edu/abs/2026MNRAS.545f2072G},
  urldate = {2026-03-19}
}

@article{Carilli2013,
  title = {Cool {{Gas}} in {{High-Redshift Galaxies}}},
  author = {Carilli, C. L. and Walter, F.},
  year = 2013,
  month = aug,
  journal = {ARA\&A},
  volume = {51},
  pages = {105--161},
  issn = {0066-4146},
  doi = {10.1146/annurev-astro-082812-140953},
  url = {https://ui.adsabs.harvard.edu/abs/2013ARA&A..51..105C},
  urldate = {2026-04-02}
}

@article{Arriagada-Neira2025,
  title = {Deep Kiloparsec View of the Molecular Gas in a Massive Star-Forming Galaxy at Cosmic Noon},
  author = {{Arriagada-Neira}, Sebasti{\'a}n and {Herrera-Camus}, Rodrigo and Villanueva, Vicente and F{\"o}rster Schreiber, Natascha M. and Lee, Minju and Bolatto, Alberto and Chen, Jianhang and Genzel, Reinhard and Liu, Daizhong and Renzini, Alvio and Tacconi, Linda J. and Tozzi, Giulia and {\"U}bler, Hannah},
  year = 2025,
  month = apr,
  journal = {A\&A},
  volume = {696},
  pages = {A83},
  publisher = {EDP},
  issn = {0004-6361},
  doi = {10.1051/0004-6361/202452652},
  url = {https://ui.adsabs.harvard.edu/abs/2025A&A...696A..83A},
  urldate = {2026-04-02}
}

@article{Pastras2025,
  title = {{{NOEMA3D}}: {{A}} First Kiloparsec Resolution Study of a z {$\sim$} 1.5 Main Sequence Barred Galaxy Channeling Gas into a Growing Bulge},
  shorttitle = {{{NOEMA3D}}},
  author = {Pastras, Stavros and Genzel, Reinhard and Tacconi, Linda J. and Schuster, Karl and Neri, Roberto and F{\"o}rster Schreiber, Natascha M. and Naab, Thorsten and Barfety, Capucine and Burkert, Andreas and Cao, Yixian and Chen, Jianhang and Combes, Fran{\c c}oise and Davies, Ric and Eisenhauer, Frank and Espejo Salcedo, Juan M. and {Garc{\'i}a-Burillo}, Santiago and {Herrera-Camus}, Rodrigo and Jolly, Jean-Baptiste and Lee, Lilian L. and Lee, Minju M. and Liu, Daizhong and Lutz, Dieter and Nestor Shachar, Amit and Parlanti, Eleonora and Price, Sedona H. and Pulsoni, Claudia and Renzini, Alvio and Scaloni, Letizia and Shimizu, Taro T. and Springel, Volker and Sternberg, Amiel and Sturm, Eckhard and Tozzi, Giulia and Wuyts, Stijn and {\"U}bler, Hannah},
  year = 2025,
  month = dec,
  journal = {A\&A},
  volume = {704},
  pages = {A329},
  publisher = {EDP},
  issn = {0004-6361},
  doi = {10.1051/0004-6361/202555430},
  url = {https://ui.adsabs.harvard.edu/abs/2025A&A...704A.329P},
  urldate = {2026-04-16}
}

@article{Claeyssens2023,
  title = {Star Formation at the Smallest Scales: A {{JWST}} Study of the Clump Populations in {{SMACS0723}}},
  shorttitle = {Star Formation at the Smallest Scales},
  author = {Claeyssens, Ad{\'e}la{\"i}de and Adamo, Angela and Richard, Johan and Mahler, Guillaume and Messa, Matteo and {Dessauges-Zavadsky}, Miroslava},
  year = 2023,
  month = apr,
  journal = {Mon. Not. R. Astron. Soc.},
  volume = {520},
  pages = {2180--2203},
  publisher = {OUP},
  issn = {0035-8711},
  doi = {10.1093/mnras/stac3791},
  url = {https://ui.adsabs.harvard.edu/abs/2023MNRAS.520.2180C},
  urldate = {2026-04-16}
}

@article{Kalita2025,
  title = {Near-{{IR}} Clumps and Their Properties in High-z Galaxies with {{JWST}}/{{NIRCam}}},
  author = {Kalita, Boris S. and Silverman, John D. and Daddi, Emanuele and Mercier, Wilfried and Ho, Luis C. and Ding, Xuheng},
  year = 2025,
  month = feb,
  journal = {MNRAS},
  volume = {537},
  pages = {402--418},
  publisher = {OUP},
  issn = {0035-8711},
  doi = {10.1093/mnras/staf031},
  url = {https://ui.adsabs.harvard.edu/abs/2025MNRAS.537..402K},
  urldate = {2026-04-16}
}

@misc{LeConte2026,
  title = {A Nuclear Disc at {{Cosmic Noon}}: Evidence of Early Bar-Driven Galaxy Evolution},
  shorttitle = {A Nuclear Disc at {{Cosmic Noon}}},
  author = {Le Conte, Zoe A. and Gadotti, Dimitri A. and Harvey, Thomas and Ferreira, Leonardo and Conselice, Christopher J. and Kim, Taehyun and {de S{\'a}-Freitas}, Camila and Fragkoudi, Francesca and Neumann, Justus and Athanassoula, E.},
  year = 2026,
  month = jan,
  publisher = {arXiv},
  doi = {10.48550/arXiv.2601.18871},
  url = {https://ui.adsabs.harvard.edu/abs/2026arXiv260118871L},
  urldate = {2026-04-16}
}

@misc{Lee2026,
  title = {The {{ALPINE-CRISTAL-JWST Survey}}: {{Gas-phase}} Abundance Gradients of Main Sequence Star-Forming Galaxies and Their Kinematics at \$4 {$<$} z {$<$} 6\$},
  shorttitle = {The {{ALPINE-CRISTAL-JWST Survey}}},
  author = {Lee, Lilian L. and F{\"o}rster Schreiber, Natascha M. and Fujimoto, Seiji and Faisst, Andreas L. and {Herrera-Camus}, Rodrigo and Genzel, Reinhard and Tacconi, Linda J. and Lutz, Dieter and Renzini, Alvio and Sanders, Ryan and Wisnioski, Emily and Wuyts, Stijn and Parlanti, Eleonora and Jones, Gareth and {\"U}bler, Hannah and Liu, Daizhong and Chen, Jianhang and Davies, Ric I. and Tozzi, Giulia and Burkert, Andreas and Price, Sedona H. and Aravena, Manuel and Boquien, M{\'e}d{\'e}ric and B{\'e}thermin, Matthieu and {da Cunha}, Elisabete and Davies, Rebecca L. and De Looze, Ilse and {Dessauges-Zavadsky}, Miroslava and Ferrara, Andrea and Fisher, Deanne B. and Gillman, Steven and Ginolfi, Michele and Ibar, Edo and Koekemoer, Anton M. and Molina, Juan and Naab, Thorsten and Rela{\~n}o, M{\'o}nica and Riechers, Dominik A. and Sanders, David B. and Spilker, Justin S. and Vallini, Livia and Zamorani, Giovanni and Nanni, Ambra and Dam, Poulomi and {Diaz-Santos}, Tanio and {G{\'o}mez-Espinoza}, Diego and Hadi, Ali and Ikeda, Ryota and Posses, Ana and Romano, Michael and Sternberg, Amiel and Villanueva, Vicente and Wang, Wuji},
  year = 2026,
  month = mar,
  publisher = {arXiv},
  doi = {10.48550/arXiv.2603.13493},
  url = {https://ui.adsabs.harvard.edu/abs/2026arXiv260313493L},
  urldate = {2026-04-16}
}

@misc{Fujimoto2025,
  title = {The {{ALPINE-CRISTAL-JWST Survey}}: {{NIRSpec IFU Data Processing}} and {{Spatially-resolved Views}} of {{Chemical Enrichment}} in {{Normal Galaxies}} at Z=4-6},
  shorttitle = {The {{ALPINE-CRISTAL-JWST Survey}}},
  author = {Fujimoto, Seiji and Faisst, Andreas L. and Tsujita, Akiyoshi and Kohandel, Mahsa and Lee, Lilian L. and {\"U}bler, Hannah and Loiacono, Federica and Nezhad, Negin and Pallottini, Andrea and Aravena, Manuel and Assef, Roberto J. and Battisti, Andrew J. and B{\'e}thermin, Matthieu and Boquien, M{\'e}d{\'e}ric and {da Cunha}, Elisabete and Ferrara, Andrea and Franco, Maximilien and Ginolfi, Michele and Hadi, Ali and Haghjoo, Aryana and {Herrera-Camus}, Rodrigo and Inami, Hanae and Koekemoer, Anton M. and Lemaux, Brian C. and Li, Yuan and Liu, Lun-Jun and Molina, Juan and Nanni, Ambra and Pozzi, Francesca and Relano, Monica and Romano, Michael and Sanders, David B. and F{\"o}rster Schreiber, Natascha M. and Silverman, John and Spilker, Justin and Telikova, Kseniia and Villanueva, Vicente and Vallini, Livia and Wang, Wuji and Zamorani, Giovanni},
  year = 2025,
  month = oct,
  publisher = {arXiv},
  doi = {10.48550/arXiv.2510.16116},
  url = {https://ui.adsabs.harvard.edu/abs/2025arXiv251016116F},
  urldate = {2026-04-16}
}

@article{Tacchella2015,
  title = {Evidence for Mature Bulges and an Inside-out Quenching Phase 3 Billion Years after the {{Big Bang}}},
  author = {Tacchella, S. and Carollo, C. M. and Renzini, A. and F{\"o}rster Schreiber, N. M. and Lang, P. and Wuyts, S. and Cresci, G. and Dekel, A. and Genzel, R. and Lilly, S. J. and Mancini, C. and Newman, S. and Onodera, M. and Shapley, A. and Tacconi, L. and Woo, J. and Zamorani, G.},
  year = 2015,
  month = apr,
  journal = {Science},
  volume = {348},
  pages = {314--317},
  issn = {0036-8075},
  doi = {10.1126/science.1261094},
  url = {https://ui.adsabs.harvard.edu/abs/2015Sci...348..314T},
  urldate = {2026-04-16}
}

@article{Schreiber2015,
  title = {The {{Herschel}} View of the Dominant Mode of Galaxy Growth from z = 4 to the Present Day},
  author = {Schreiber, C. and Pannella, M. and Elbaz, D. and B{\'e}thermin, M. and Inami, H. and Dickinson, M. and Magnelli, B. and Wang, T. and Aussel, H. and Daddi, E. and Juneau, S. and Shu, X. and Sargent, M. T. and Buat, V. and Faber, S. M. and Ferguson, H. C. and Giavalisco, M. and Koekemoer, A. M. and Magdis, G. and Morrison, G. E. and Papovich, C. and Santini, P. and Scott, D.},
  year = 2015,
  month = mar,
  journal = {A\&A},
  volume = {575},
  pages = {A74},
  publisher = {EDP},
  issn = {0004-6361},
  doi = {10.1051/0004-6361/201425017},
  url = {https://ui.adsabs.harvard.edu/abs/2015A&A...575A..74S},
  urldate = {2026-04-17}
}

@article{Prajapati2026,
  title = {Vz-{{GAL}}: {{Probing Cold Molecular Gas}} in {{Dusty Star-forming Galaxies}} at z = 1-6},
  shorttitle = {Vz-{{GAL}}},
  author = {Prajapati, Prachi and Riechers, Dominik and Cox, Pierre and Weiss, Axel and Saintonge, Am{\'e}lie and Jones, Bethany and Bakx, Tom J. L. C. and Berta, Stefano and {van der Werf}, Paul and Neri, Roberto and Butler, Kirsty M. and Cooray, Asantha and Ismail, Diana and Baker, Andrew J. and Borsato, Edoardo and Harris, Andrew and Ivison, Rob and Lehnert, Matthew and Marchetti, Lucia and Messias, Hugo and Omont, Alain and Vlahakis, Catherine and Yang, Chentao},
  year = 2026,
  month = feb,
  journal = {ApJS},
  volume = {282},
  pages = {40},
  publisher = {IOP},
  issn = {0067-0049},
  doi = {10.3847/1538-4365/ae27d4},
  url = {https://ui.adsabs.harvard.edu/abs/2026ApJS..282...40P},
  urldate = {2026-04-17}
}

@article{King2015,
  title = {Powerful {{Outflows}} and {{Feedback}} from {{Active Galactic Nuclei}}},
  author = {King, Andrew and Pounds, Ken},
  year = 2015,
  month = aug,
  journal = {ARA\&A},
  volume = {53},
  pages = {115--154},
  issn = {0066-4146},
  doi = {10.1146/annurev-astro-082214-122316},
  url = {https://ui.adsabs.harvard.edu/abs/2015ARA&A..53..115K},
  urldate = {2026-04-17}
}

@article{Ubler2024,
  title = {{{GA-NIFS}}: {{NIRSpec}} Reveals Evidence for Non-Circular Motions and {{AGN}} Feedback in {{GN20}}},
  shorttitle = {{{GA-NIFS}}},
  author = {{\"U}bler, Hannah and D'Eugenio, Francesco and Perna, Michele and Arribas, Santiago and Jones, Gareth C. and Bunker, Andrew J. and Carniani, Stefano and Charlot, St{\'e}phane and Maiolino, Roberto and {Rodr{\'i}guez del Pino}, Bruno and Willott, Chris J. and B{\"o}ker, Torsten and Cresci, Giovanni and Kumari, Nimisha and Lamperti, Isabella and Parlanti, Eleonora and Scholtz, Jan and Venturi, Giacomo},
  year = 2024,
  month = oct,
  journal = {MNRAS},
  volume = {533},
  pages = {4287--4299},
  publisher = {OUP},
  issn = {0035-8711},
  doi = {10.1093/mnras/stae1993},
  url = {https://ui.adsabs.harvard.edu/abs/2024MNRAS.533.4287U},
  urldate = {2026-04-17}
}

@article{Boogaard2026,
  title = {Resolving the {{Dusty Star-forming Galaxy GN20}} at z = 4.055 with {{NOEMA}} and {{JWST}}: {{A Similar Distribution}} of {{Stars}}, {{Gas}}, and {{Dust Despite Distinct Apparent Profiles}}},
  shorttitle = {Resolving the {{Dusty Star-forming Galaxy GN20}} at z = 4.055 with {{NOEMA}} and {{JWST}}},
  author = {Boogaard, Leindert A. and Walter, Fabian and Wei{\ss}, Axel and Colina, Luis and Hodge, Jacqueline and Bik, Arjan and Crespo G{\'o}mez, Alejandro and Daddi, Emanuele and Magdis, Georgios E. and Meyer, Romain A. and {\"O}stlin, G{\"o}ran},
  year = 2026,
  month = jan,
  journal = {ApJ},
  volume = {996},
  pages = {19},
  publisher = {IOP},
  issn = {0004-637X},
  doi = {10.3847/1538-4357/ae14eb},
  url = {https://ui.adsabs.harvard.edu/abs/2026ApJ...996...19B},
  urldate = {2026-04-17}
}

@article{Somerville2015,
  title = {Physical {{Models}} of {{Galaxy Formation}} in a {{Cosmological Framework}}},
  author = {Somerville, Rachel S. and Dav{\'e}, Romeel},
  year = 2015,
  month = aug,
  journal = {Annu. Rev. Astron. Astrophys.},
  volume = {53},
  pages = {51--113},
  issn = {0066-4146},
  doi = {10.1146/annurev-astro-082812-140951},
  url = {https://ui.adsabs.harvard.edu/abs/2015ARA&A..53...51S},
  urldate = {2026-04-17}
}

@article{Naab2017,
  title = {Theoretical {{Challenges}} in {{Galaxy Formation}}},
  author = {Naab, Thorsten and Ostriker, Jeremiah P.},
  year = 2017,
  month = aug,
  journal = {ARA\&A},
  volume = {55},
  pages = {59--109},
  issn = {0066-4146},
  doi = {10.1146/annurev-astro-081913-040019},
  url = {https://ui.adsabs.harvard.edu/abs/2017ARA&A..55...59N},
  urldate = {2026-04-17}
}

@article{Moster2020,
  title = {{{EMERGE}} - Empirical Constraints on the Formation of Passive Galaxies},
  author = {Moster, Benjamin P. and Naab, Thorsten and White, Simon D. M.},
  year = 2020,
  month = dec,
  journal = {MNRAS},
  volume = {499},
  pages = {4748--4767},
  publisher = {OUP},
  issn = {0035-8711},
  doi = {10.1093/mnras/staa3019},
  url = {https://ui.adsabs.harvard.edu/abs/2020MNRAS.499.4748M},
  urldate = {2026-04-17}
}

@article{Henriquez-Brocal2022,
  title = {Molecular Gas Properties of {{Q1700-MD94}}: {{A}} Massive Main-Sequence Galaxy at z {$\approx$} 2},
  shorttitle = {Molecular Gas Properties of {{Q1700-MD94}}},
  author = {{Henr{\'i}quez-Brocal}, K. and {Herrera-Camus}, R. and Tacconi, L. and Genzel, R. and Bolatto, A. and Bovino, S. and Demarco, R. and F{\"o}rster Schreiber, N. and Lee, M. and Lutz, D. and Rubio, M.},
  year = 2022,
  month = jan,
  journal = {A\&A},
  volume = {657},
  pages = {L15},
  publisher = {EDP},
  issn = {0004-6361},
  doi = {10.1051/0004-6361/202141870},
  url = {https://ui.adsabs.harvard.edu/abs/2022A&A...657L..15H},
  urldate = {2026-04-20}
}

@article{vandeVoort2011,
  title = {The Rates and Modes of Gas Accretion on to Galaxies and Their Gaseous Haloes},
  author = {{van de Voort}, Freeke and Schaye, Joop and Booth, C. M. and Haas, Marcel R. and Dalla Vecchia, Claudio},
  year = 2011,
  month = jul,
  journal = {MNRAS},
  volume = {414},
  pages = {2458--2478},
  publisher = {OUP},
  issn = {0035-8711},
  doi = {10.1111/j.1365-2966.2011.18565.x},
  url = {https://ui.adsabs.harvard.edu/abs/2011MNRAS.414.2458V},
  urldate = {2026-04-20}
}

\begin{appendix} 
\section{Curve-of-growth profile and size comparison between different filters}
\label{appendix:size_measurements}

\begin{figure*}[htpb]
  \centering
  \includegraphics[width=0.8\textwidth]{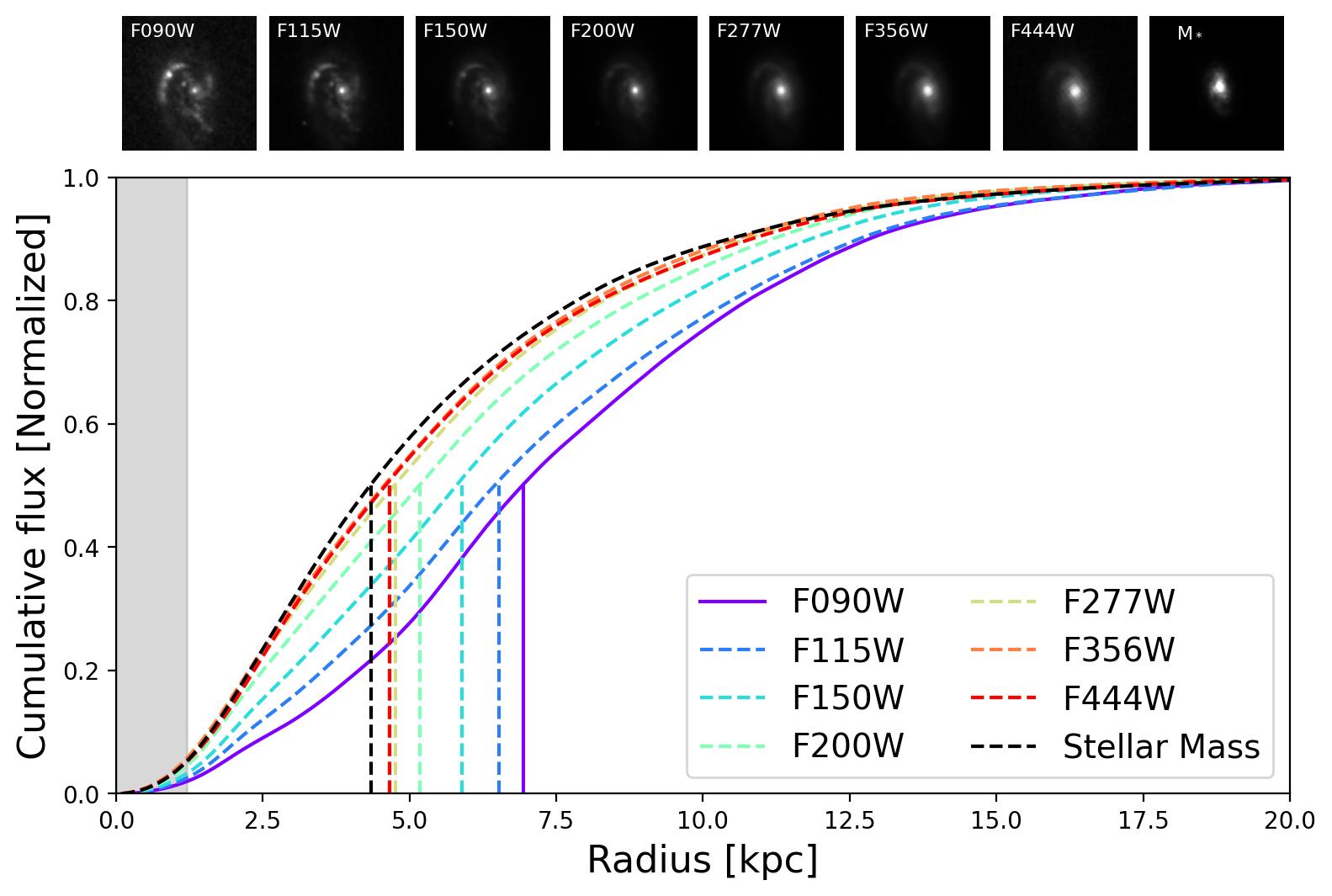}
  \caption{Curve-of-growth method used to measure the stellar size. \textit{First row}: Stellar images from different JWST/NIRCam broadband filters. The last image shows the stellar-mass map derived from spatially resolved SED fitting. The results of the curve-of-growth analysis for each filter are shown as colored curves. The curve of growth derived from the stellar mass map is shown by the dashed dark line, which is close to that of F444W but smaller than those of all available filters. The gray-shaded region indicates the FWHM of the PSF for F444W.}
  \label{appendixfig:size_measurements}
\end{figure*}

We applied the curve-of-growth method to derive the cumulative radial profile and the effective radius for different broadband filters, as well as the projected dust mass map from the spatially resolved fitting.  
The results from G4\_23011 are shown in Fig.~\ref{appendixfig:size_measurements}.  
In general, the effective radius decreases with increasing rest-frame wavelength.  
Based on our observations, the reddest broadband filter (F444W) exhibits the closest cumulative radial profile to that of the mass profile.  
This comparison supports the use of rest-frame near-infrared broadband images as a reliable approximation for the stellar mass distribution.

\section{UV modeling}
\label{appendix:uv_modelling}

The visibility fitting results for all targets are presented in Appendix Fig.~\ref{appendixfig:uvfitting_co43} and \ref{appendixfig:uvfitting_co43_page2} for the group-1 targets, and Fig.~\ref{appendixfig:uvfitting_co32} for the group-2 targets. For each individual system, only the converged fits are shown.
The best-fit parameters for each target are provided in Table~\ref{appendixtab:uvmodelling}. Parameters without error bars were held fixed during the fitting process.

\begin{figure*}[p]
  \centering
  \includegraphics[width=0.48\textwidth]{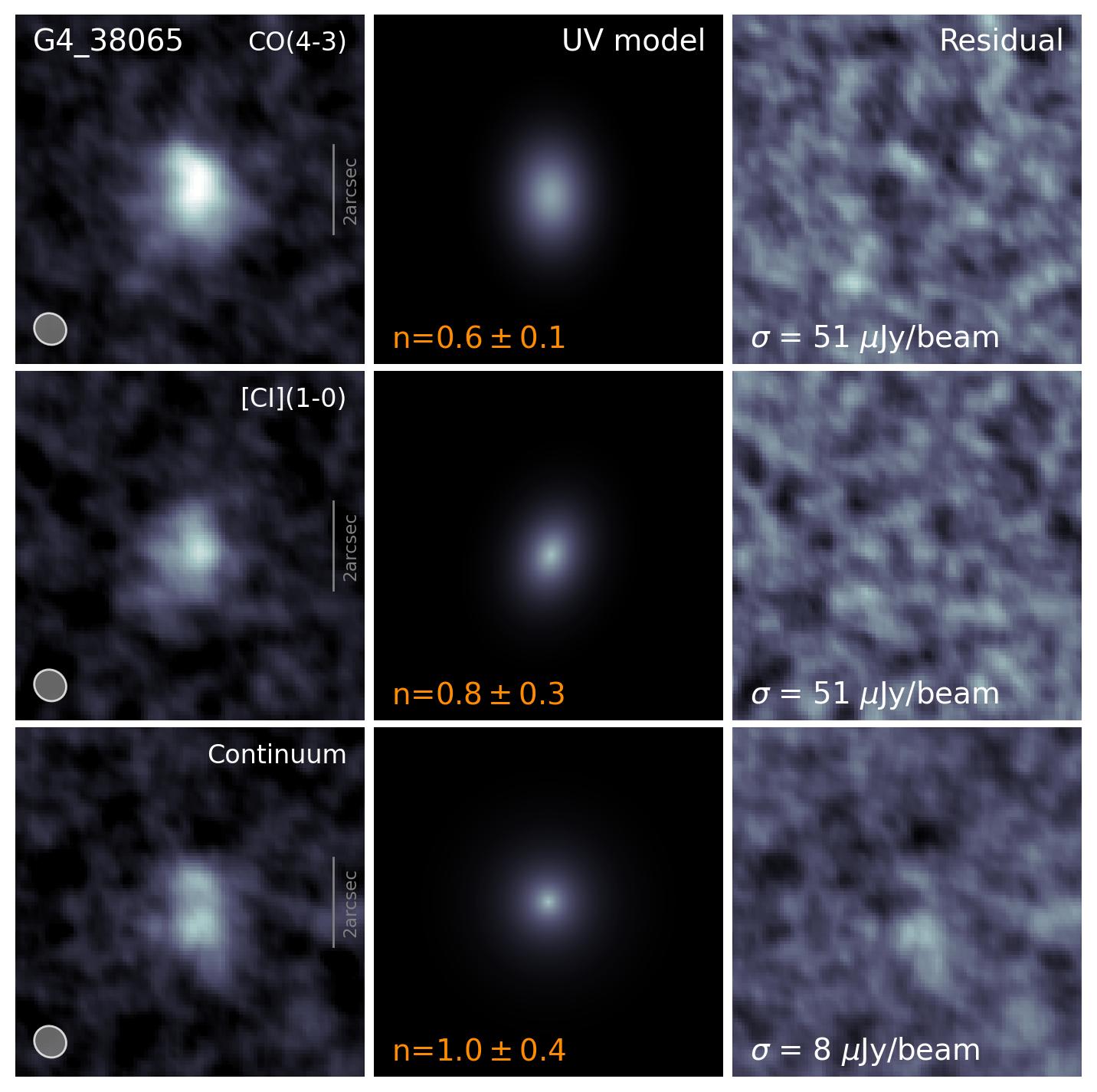}
  \includegraphics[width=0.48\textwidth]{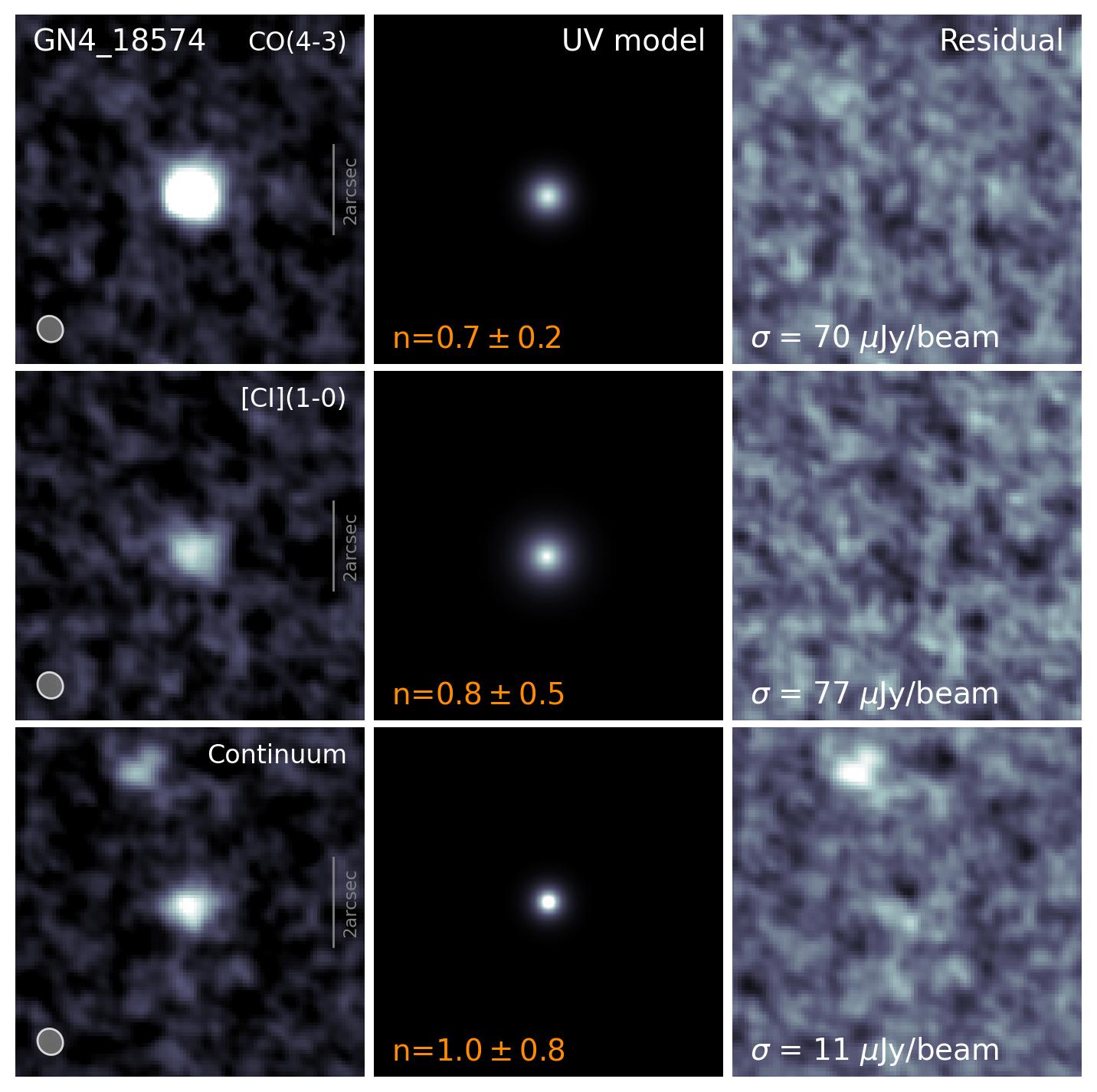}
  \includegraphics[width=0.48\textwidth]{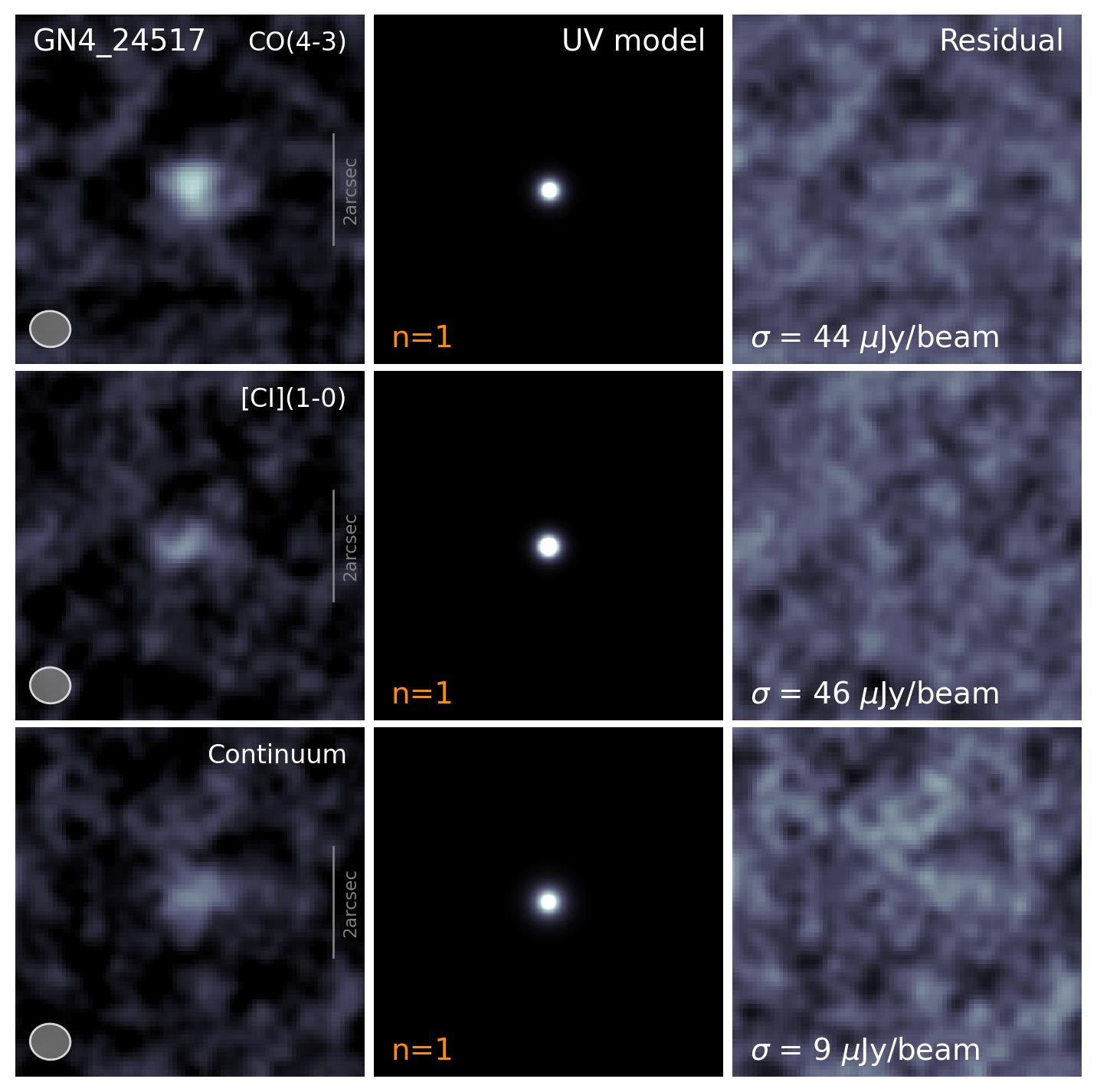}
  \includegraphics[width=0.48\textwidth]{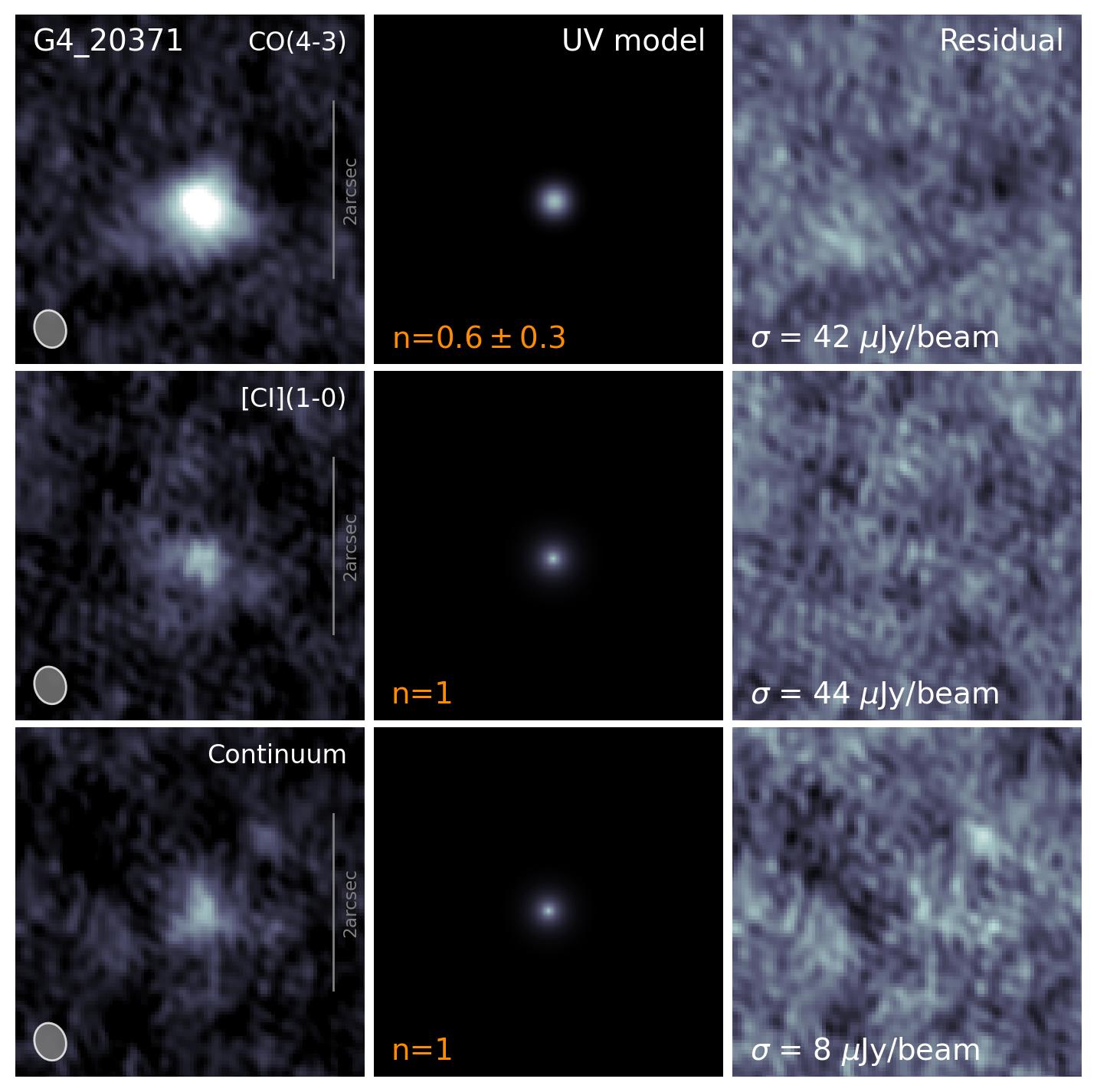}
  \caption{UV fitting for \firstgroup{} targets. \textit{First column}: Total intensity map, with the beam size indicated by the gray ellipse. \textit{Second column}: Best-fit model, with the best-fit S\'ersic index displayed in the bottom-left corner. The value without error indicates a fixed value during fitting. \textit{Last column}: Residual map, with the RMS value in the bottom-left corner.}
  \label{appendixfig:uvfitting_co43}
\end{figure*}

\begin{figure}[htpb]
  \centering
  \includegraphics[width=0.48\textwidth]{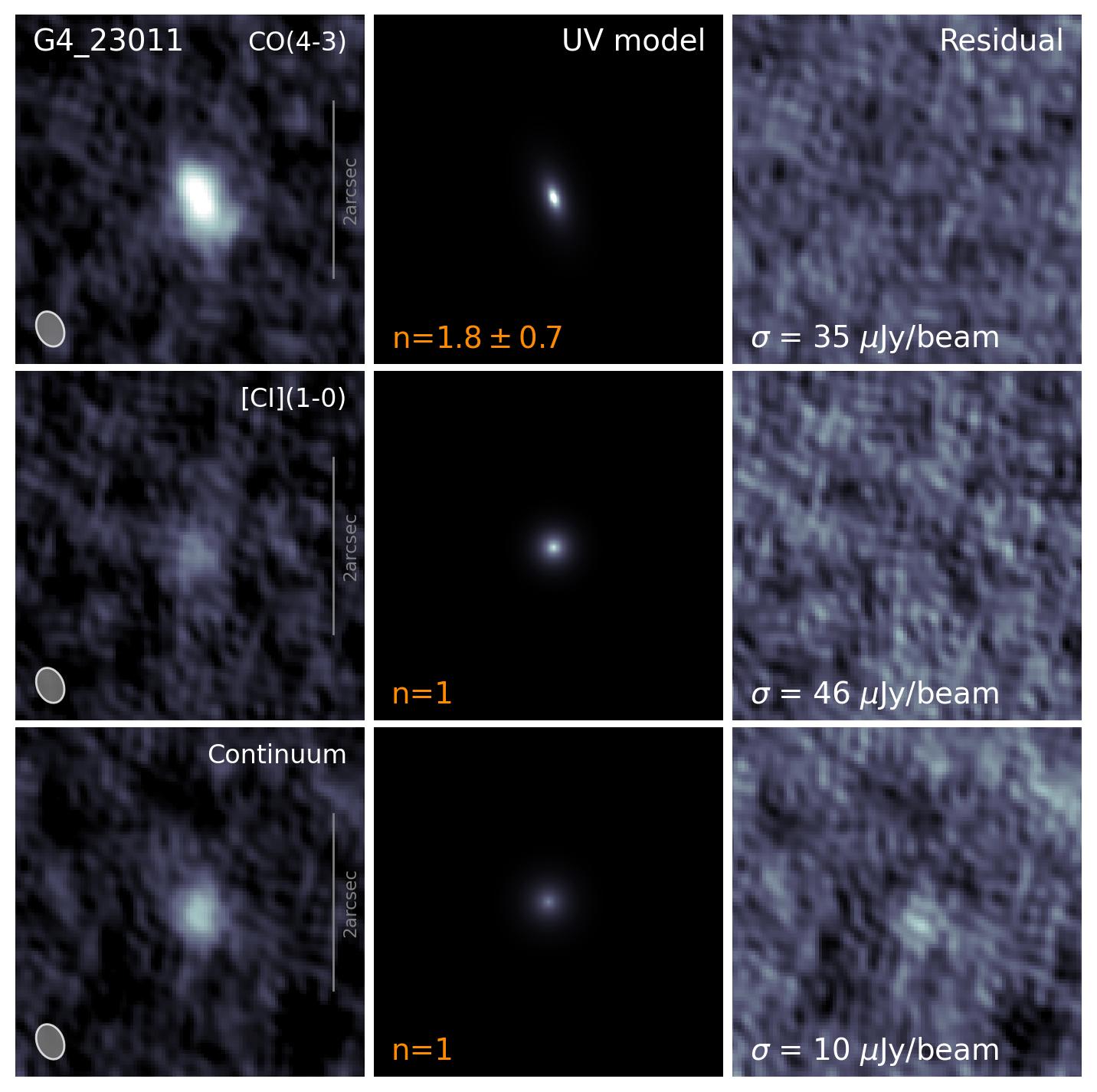}
  \includegraphics[width=0.48\textwidth]{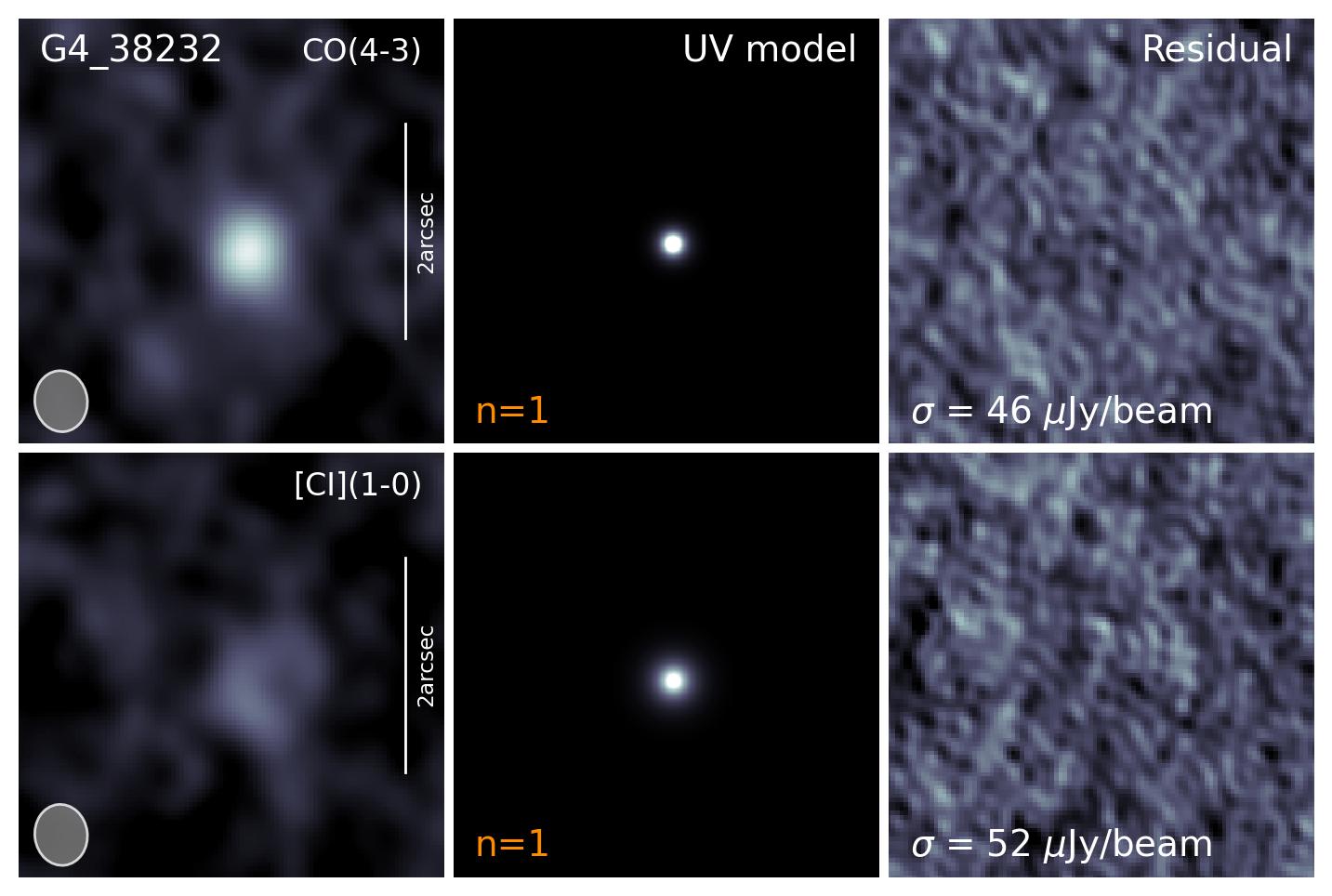}
  \caption{Figure~\ref{appendixfig:uvfitting_co43} continued.}
  \label{appendixfig:uvfitting_co43_page2}
\end{figure}

\begin{figure}[htpb]
  \centering
  \includegraphics[width=0.48\textwidth]{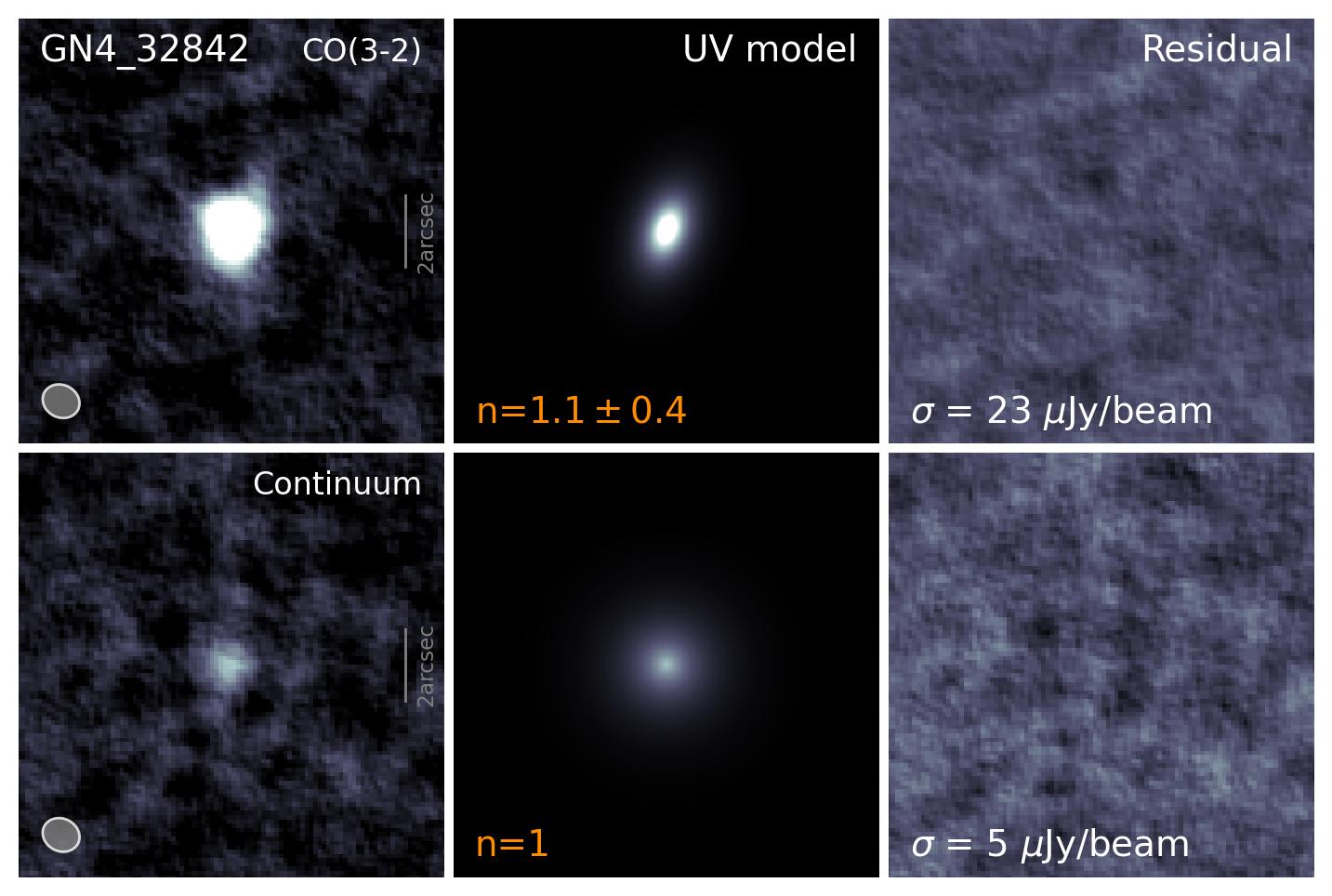}
  \includegraphics[width=0.48\textwidth]{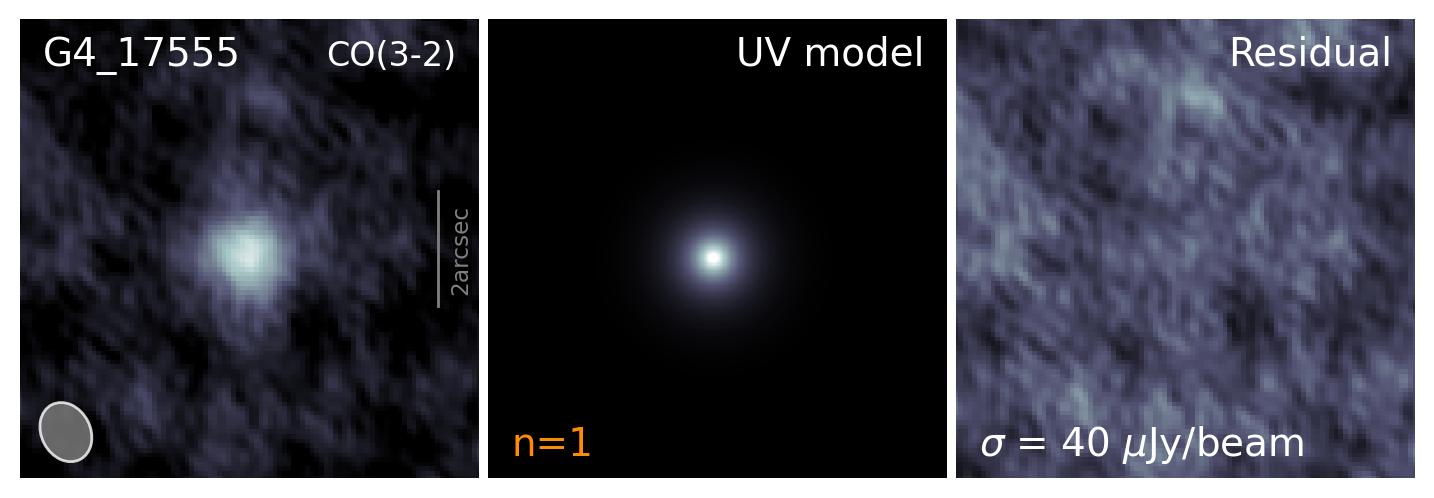}
  \includegraphics[width=0.48\textwidth]{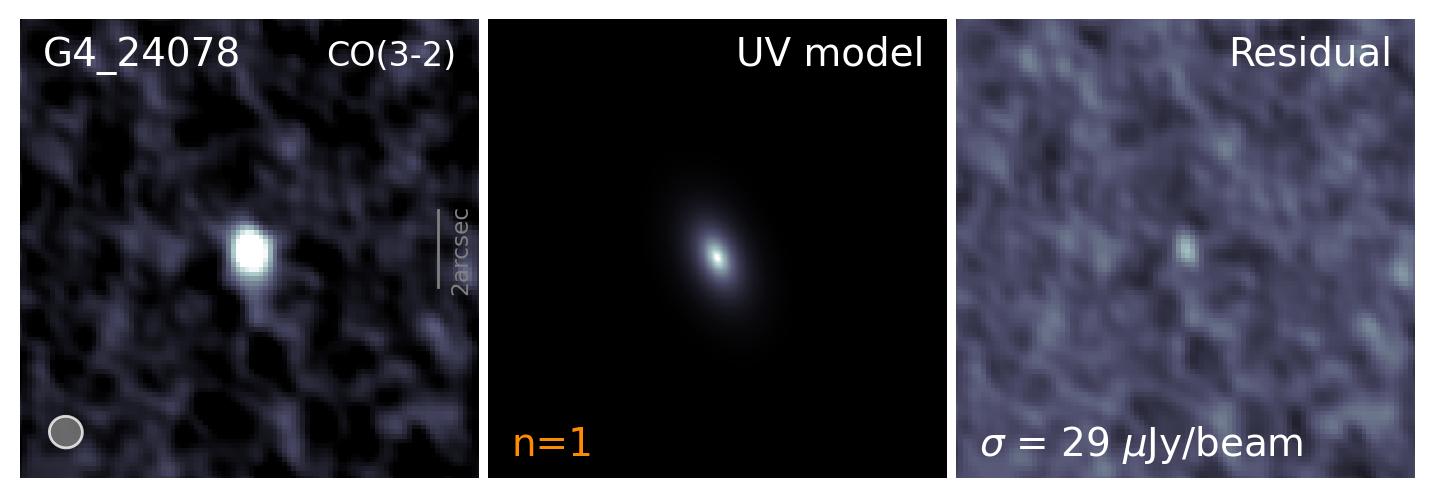}
  \includegraphics[width=0.48\textwidth]{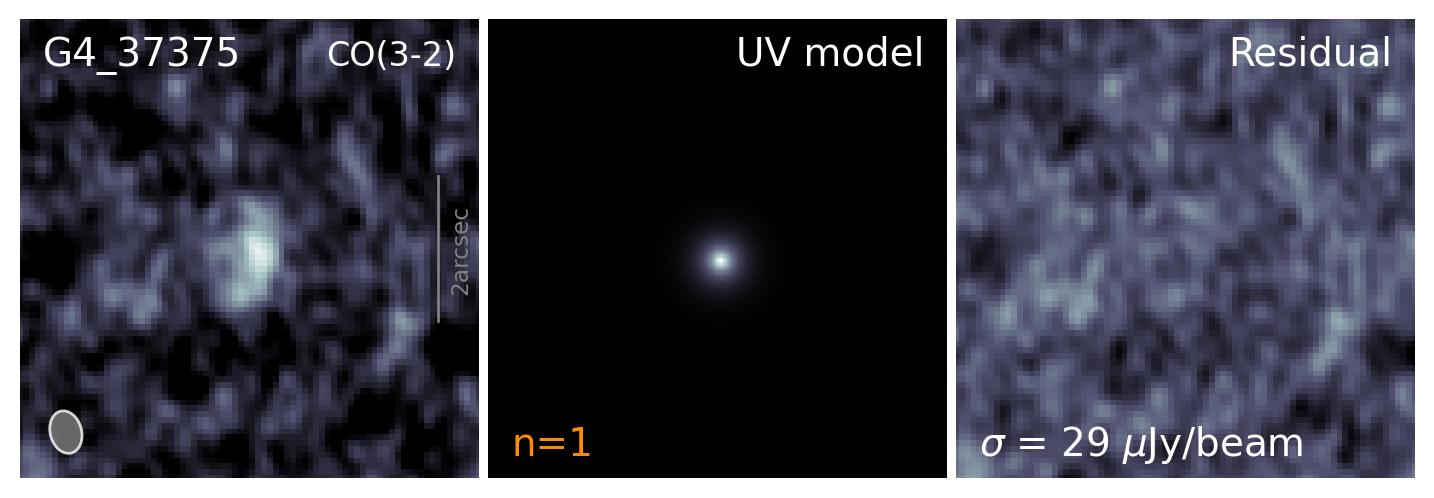}
  \caption{Same as Fig.\ref{appendixfig:uvfitting_co43} but for \secondgroup{} targets.}
  \label{appendixfig:uvfitting_co32}
\end{figure}

\begin{table*}
  \centering
  \caption{Size measurements based on visibility modeling.}
  \label{appendixtab:uvmodelling}
  \begin{tabular}{ccccccccc}
    \hline\hline
    Name & Tracer & Amplitude & Re & n & x0 & y0 & ellip & theta \\
         & & $\times10^{-8}$\,Jy  & arcsec &  & arcsec & arcsec & & deg \\ 
    \hline
G4\_38065 & CO(4$-$3) & 28.0$\pm$3.9 & 8.66$\pm$0.72 & 0.6$\pm$0.1 & 0.054 & -0.146 & 0.3$\pm$0.1 & 92$\pm$9 \\
         & [CI](1$-$0) & 16.6$\pm$4.9 & 8.02$\pm$1.37 & 0.8$\pm$0.3 & 0.067 & -0.219 & 0.3$\pm$0.2 & 72$\pm$19 \\
         & Continuum & 1.2$\pm$0.5 & 9.18$\pm$2.37 & 1.0$\pm$0.4 & -0.00$\pm$0.10 & 0.00$\pm$0.10 & 0.0$\pm$0.2 & 90$\pm$209 \\
\hline
GN4\_18574 & CO(4$-$3) & 128.7$\pm$17.7 & 3.47$\pm$0.20 & 0.7$\pm$0.2 & -0.022 & -0.162 & 0.0 & 0$\pm$0 \\
          & [CI](1$-$0) & 39.3$\pm$15.7 & 4.56$\pm$0.79 & 0.8$\pm$0.5 & -0.046 & -0.256 & 0.0 & 0$\pm$0 \\
          & Continuum & 7.9$\pm$3.8 & 2.64$\pm$0.54 & 1.0$\pm$0.8 & -0.004 & -0.004 & 0.0 & 0$\pm$0 \\
\hline
GN4\_24517 & CO(4$-$3) & 35.4$\pm$24.0 & 1.83$\pm$0.76 & 1.0 & 0.037 & -0.036 & 0.0 & 0$\pm$0 \\
          & [CI](1$-$0) & 36.7$\pm$32.1 & 1.67$\pm$0.89 & 1.0 & 0.008 & -0.027 & 0.0 & 0$\pm$0 \\
          & Continuum & 3.1$\pm$2.6 & 2.50$\pm$1.35 & 1.0 & 0.00$\pm$0.12 & 0.00$\pm$0.11 & 0.0 & 0$\pm$0 \\
\hline
G4\_20371 & CO(4$-$3) & 94.8$\pm$20.6 & 2.59$\pm$0.23 & 0.6$\pm$0.3 & 0.143 & -0.273 & 0.0 & 0$\pm$0 \\
         & [CI](1$-$0) & 24.7$\pm$9.1 & 3.45$\pm$0.92 & 1.0 & 0.118 & -0.308 & 0.0 & 0$\pm$0 \\
         & Continuum & 4.0$\pm$1.7 & 3.28$\pm$0.97 & 1.0 & 0.008 & -0.208 & 0.0 & 0$\pm$0 \\
\hline
G4\_23011 & CO(4$-$3) & 35.7$\pm$18.7 & 5.24$\pm$1.49 & 1.8$\pm$0.7 & 0.123 & -0.203 & 0.5$\pm$0.1 & 107$\pm$9 \\
         & [CI](1$-$0) & 25.5$\pm$10.7 & 3.23$\pm$0.96 & 1.0 & 0.128 & -0.048 & 0.0 & 0$\pm$0 \\
         & Continuum & 4.4$\pm$1.4 & 4.14$\pm$0.99 & 1.0 & 0.008 & -0.008 & 0.0 & 0$\pm$0 \\
\hline
G4\_38232 & CO(4$-$3) & 91.7$\pm$34.6 & 1.68$\pm$0.42 & 1.0 & 0.124 & -0.254 & 0.0 & 0$\pm$0 \\
 & [CI](1$-$0) & 30.2$\pm$14.0 & 3.01$\pm$1.01 & 1.0 & 0.133 & -0.307 & 0.0 & 0$\pm$0 \\
\hline
GN4\_32842 & CO(3$-$2) & 15.6$\pm$4.0 & 6.04$\pm$0.78 & 1.1$\pm$0.4 & 0.013 & 0.024 & 0.4$\pm$0.1 & 71$\pm$11 \\
          & Continuum & 0.4$\pm$0.2 & 8.47$\pm$2.72 & 1.0 & -0.00$\pm$0.22 & -0.00$\pm$0.20 & 0.0 & 0$\pm$0 \\
\hline
G4\_17555 & CO(3$-$2) & 7.7$\pm$2.5 & 5.79$\pm$1.30 & 1.0 & -0.07$\pm$0.11 & -0.17$\pm$0.11 & 0.0 & 0$\pm$0 \\
\hline
G4\_24078 & CO(3$-$2) & 10.0$\pm$2.7 & 5.67$\pm$1.26 & 1.0 & -0.006 & -0.168 & 0.4$\pm$0.2 & 113$\pm$18 \\
\hline
G4\_37375 & CO(3$-$2) & 10.0$\pm$5.2 & 3.48$\pm$1.16 & 1.0 & 0.06$\pm$0.11 & -0.21$\pm$0.11 & 0.0 & 0$\pm$0 \\
\hline

  \end{tabular}

\raggedright\small{\textbf{Notes:} The values without errorbar are fixed parameters during the modeling.}
  
\end{table*}

\section{Dust temperature}
\label{appendix:dust_temperature}

\begin{figure*}[htpb]
  \centering
  \includegraphics[width=0.96\textwidth]{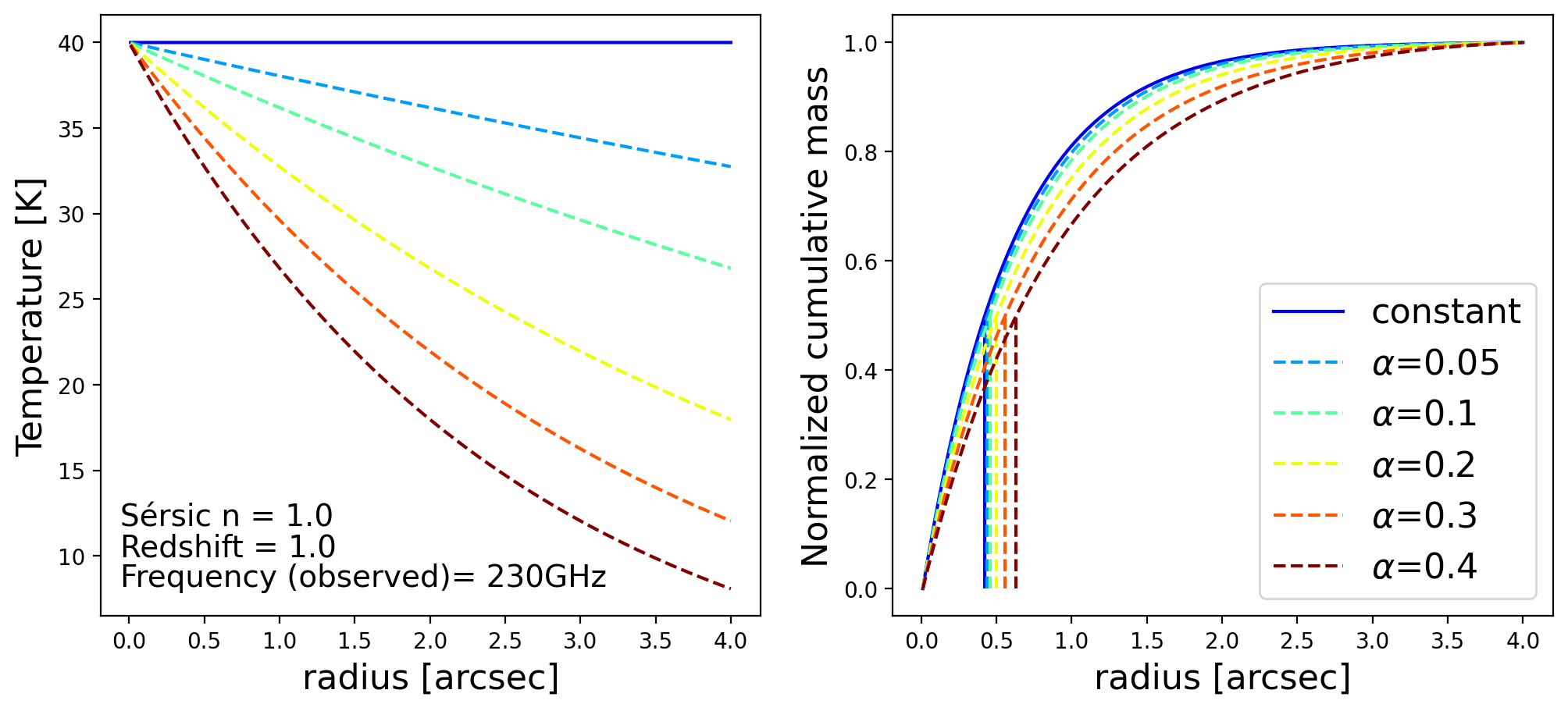}
  \caption{Dust-mass size with different temperature gradients. \textit{Left:} Input dust temperature profile. \textit{Right} Corresponding normalized mass profile in the same color. The vertical line in the second plot represents the half-mass radius.}
  \label{fig:dust-size-temperature}
\end{figure*}

Dust temperatures vary across galaxies, and are typically higher near star-forming regions.
To test how much varying dust temperature profiles affect final size measurements, we measured the sizes of an ideal galaxy with different temperature profiles.
We parameterized the dust radial temperature as an exponential profile $T_\text{dust}(r)\propto r^{-\alpha}$. 
The galaxy was constructed with a 2D S\'ersic profile with \(n=1\) to approximate the general extended disk morphology of the \noema{} galaxies.
We find that steeper temperature profiles predict larger disk sizes, indicating that the constant-temperature assumption tends to underestimate the true size of the dust distribution.
However, the overall effect is generally small; even with a steep slope of 0.4, the size difference remains within 0.3 dex, typically smaller than the measured uncertainties.
Due to the lack of observational constraints on the temperature gradient of cosmic noon galaxies, we did not apply corrections for it.
Nevertheless, the generally underestimated dust size does not change the conclusions of our work.

\section{Morphological analysis}

We applied morphological analysis using \textsc{statmorph} to the cold ISM maps from CO, [C\,I], and the dust continuum. The results are shown in Fig.~\ref{appendix:2d_morphology}. For most metrics, measurements across different tracers scatter around the 1:1 line. The strongest deviation from the one-to-one correlation is observed in M20 between CO and [C\,I], where CO shows a systematically larger M20, indicating that CO emission is more extended than [C\,I]. This appears contradictory to the general consensus that [C\,I] traces more diffuse molecular gas than CO. However, given the limited sample size and the small magnitude of the offset, a larger sample with improved statistics is needed to better quantify their morphological differences.

\label{appendix:2d_morphology}

\begin{figure*}
\begin{center}
  \includegraphics[width=0.95\textwidth]{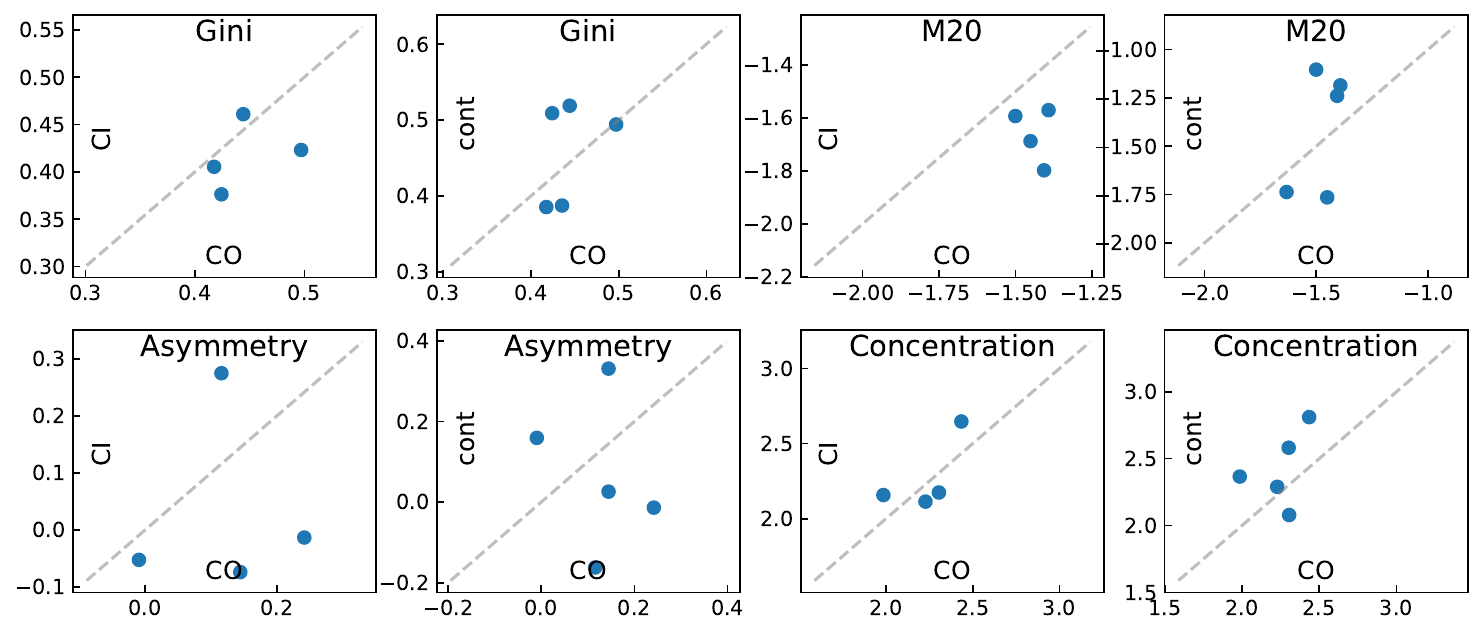}
\end{center}
\caption{Morphology analysis based on \textsc{statmorph} \citep{Rodriguez-Gomez2019}.  The computed morphology parameter is indicated at the top side of each plot, and the two tracers' names are indicated near each axis. The dashed line shows the one-to-one correlation. For most morphological parameters, the measured values from different tracers are scattered around the 1:1 line. Compared to \ci{} and dust, CO shows systematically larger asymmetry, largely due to its higher sensitivity in the outskirts of the galaxy.}
\label{appendixfig:2d-morpholgies}
\end{figure*} 

\section{Radial gas fraction and depletion timescale}

\begin{figure*}
  \centering
  \includegraphics[width=1.0\textwidth]{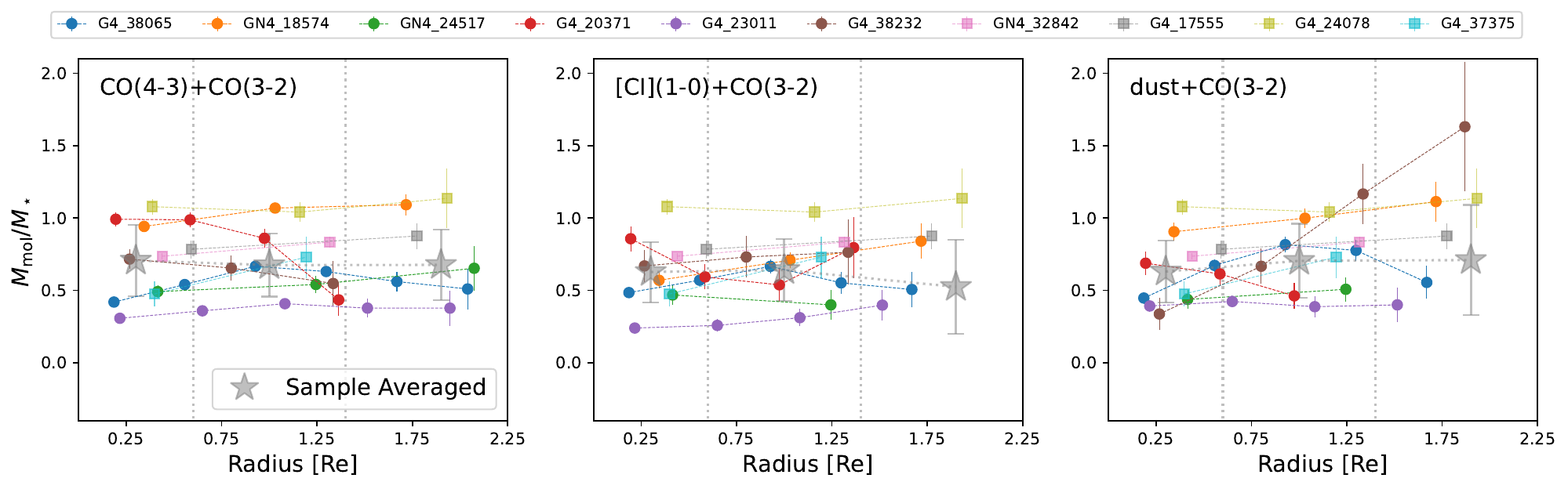}
  \includegraphics[width=1.0\textwidth]{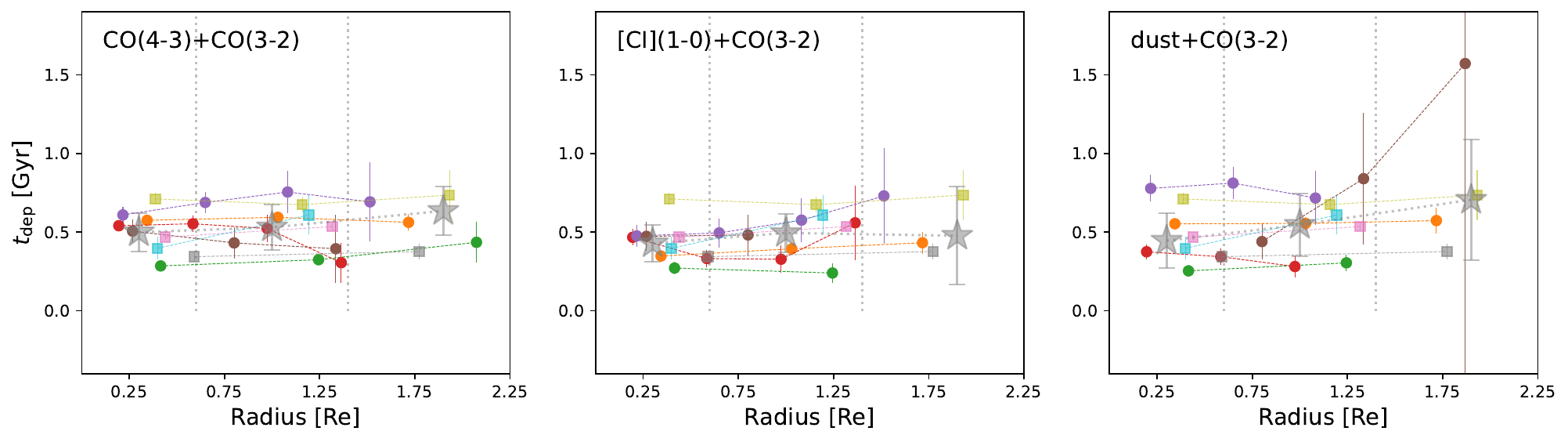}
  \caption{Radial distribution of the molecular gas fraction and its depletion timescale, derived from different available molecular gas tracers. Each column shows the results for a specified tracer, indicated by the text in the upper-left corner. The radial profile of each galaxy is represented by different colors. The \co{} targets are marked with filled circles, while the CO(3$-$2) targets are shown as color-filled squares. The normalization factor is the MS value for the same stellar mass. Overall, the various gas tracers provide consistent radial profiles of the molecular gas distribution.}
  \label{appendixfig:fgas_tdep_radial}
\end{figure*}

We calculated the radial gas mass profile using all three cold gas tracers, thereby allowing us to independently derive the radial gas fraction and the depletion timescale.  
The results from the different tracers are shown in Fig.~\ref{appendixfig:fgas_tdep_radial}.  
As noted in the main text, there is considerable scatter from galaxy to galaxy, but most exhibit only moderate radial variation.  
The sample-averaged radial profile also shows a generally flat trend.

\end{appendix}
\end{document}